\documentclass{jfm}
\usepackage{amsmath,graphicx}

\newcommand{\Pe}{\mathrm{P\hspace{-1pt}e}}
\newcommand{\Peo}{\mathrm{P\hspace{-1pt}e}_{\hspace{-0.5pt}o}}
\newcommand{\Nu}{\mathrm{N\hspace{-1pt}u}}
\newcommand{\p}[3][]{\frac{\partial^{#1}#2}{\partial{#3}^{#1}}}
\renewcommand{\Re}{\mathrm{Re}\,}
\renewcommand{\Im}{\mathrm{Im}\,}

\newcommand{\erfc}{\mathrm{erfc}}
\newcommand{\erf}{\mathrm{erf}}
\newcommand{\frsq}[2]{\frac{#1}{\sqrt{#2}}}

\newcommand{\quadtext}[2][\quad]{{#1}\text{#2}{#1}}

\newcommand{\intdt}[2][t]{\int\hspace{-2pt}d{#1}_{#2}}

\newcommand{\ub}{\boldsymbol{u}}
\newcommand{\del}{\boldsymbol{\nabla}}
\newcommand{\nhat}{\boldsymbol{\hat{n}}}
\newcommand{\xhat}{\boldsymbol{\hat{x}}}
\newcommand{\OP}{\mathcal{L}}

\title[Advection-diffusion in two dimensions]{Steady advection-diffusion around finite absorbers in two-dimensional potential flows%
  \footnote{
    The computer programming codes used for the results in this paper
    are available at \texttt{http://www.advection-diffusion.net}.
  }
}

\author[J. Choi, D. Margetis, T. M. Squires, and M. Z. Bazant]{J\ls A\ls E\ls H\ls Y\ls U\ls K\ns  C\ls H\ls O\ls I$^1$, \ns D\ls I\ls O\ls N\ls I\ls S\ls I\ls O\ls S \ns M\ls A\ls R\ls G\ls E\ls T\ls I\ls S$^1$, \ns T\ls O\ls D\ls D\ns M.\ns S\ls Q\ls U\ls I\ls R\ls E\ls S$^2$,\ns \and M\ls A\ls R\ls T\ls I\ls N\ns Z.\ns B\ls A\ls Z\ls A\ls N\ls T$^1$}

\affiliation{$^1$Department of Mathematics\\ 
Massachusetts Institute of Technology, Cambridge, MA 02139 \\
$^2$Departments of Applied and Computational Mathematics and Physics,\\
California Institute of Technology, Pasadena, CA 91125 }

\pubyear{2003} \volume{?} \pagerange{?--?}
\date{\today}
\begin{document}

\maketitle

\begin{abstract}
We consider perhaps the simplest, nontrivial problem in
advection-diffusion --- a finite absorber of arbitrary cross section
in a steady two-dimensional potential flow of concentrated fluid.
This problem has been studied extensively in the theory of
solidification from a flowing melt, and it also arises in
advection-diffusion-limited aggregation. In both cases, the
fundamental object is the flux to a circular disk, obtained by
conformal mapping from more complicated shapes. Here, we construct an
accurate numerical solution by an efficient method that involves
mapping to the interior of the disk and using a spectral method in
polar coordinates. The method combines exact asymptotics and an
adaptive mesh to handle boundary layers. Starting from a well-known
integral equation in streamline coordinates, we also derive high-order
asymptotic expansions for high and low P\'eclet numbers
($\Pe$). Remarkably, the ``high'' $\Pe$ expansion remains accurate
even for such low $\Pe$ as $10^{-3}$.  The two expansions overlap well
near $\Pe = 0.1$, allowing the construction of an analytical
connection formula that is uniformly accurate for all $\Pe$ and angles
on the disk with a maximum relative error of 1.75\%.  We also obtain
an analytical formula for the Nusselt number ($\Nu$) as a function of
$\Pe$ with a maximum relative error of 0.53\% for all possible
geometries after conformal mapping.  Considering the concentration
disturbance around a disk, we find that the crossover from a diffusive
cloud (at low $\Pe$) to an advective wake (at high $\Pe$) occurs at
$\Pe \approx 60$.
\end{abstract}


\section{Introduction}
\label{sec:introduction} 

The transfer of mass, heat, or other passive scalars in fluid flows is
a major theme in transport science (\cite{leal}, 1992). The canonical model
problem involves a uniform background flow of speed, $U_\infty$, and
concentration, $C_\infty$, past an absorbing object of characteristic
size, $L$. Given the steady, incompressible flow field, $\ub$ (scaled
to $U_\infty$), the steady tracer concentration around the object, $c$
(scaled to $C_\infty$), satisfies the (dimensionless) linear
advection-diffusion equation,
\begin{equation}
\Peo \, \ub \cdot \del c = \del^2 c,\label{eq:PDE-c}
\end{equation}
where $\Pe_o = U_\infty L/D$ is the {\it bare} P\'eclet number, which
measures the relative importance of advection compared to diffusion
(with a diffusivity $D$).  The partial differential equation (PDE) (\ref{eq:PDE-c})
must be solved subject to the boundary conditions (BCs) $c=0$ on the
object and $c=1$ far away (at $\infty$), to obtain the dimensionless
normal flux density,
\begin{equation}
\sigma = \nhat\cdot \del c,\label{eq:flux-def} 
\end{equation}
everywhere on the surface of the object. (Since the problem is linear,
we may also consider the equivalent problem of a source object at
$c=1$ relative to a depleted background fluid at $c=0$.)

Although similarity solutions exist for infinite leading edges (see
below), in the usual case of a {\it finite} absorber, the mathematical
problem is intractable.  It can even be difficult to solve numerically
because $\Peo$ appears as a singular perturbation in the PDE in both
limits, $\Peo \rightarrow \infty$ and $\Peo \rightarrow 0$.  The
classical approach, therefore, has been to employ asymptotic analysis
to obtain approximate solutions, usually relating the total integrated
flux, or Nusselt number, $\Nu$, to the P\'eclet number, $\Pe$. Early
studies of this type focused on spheres (\cite{acrivos62}, 1962) and
more complicated shapes (\cite{brenner63}, 1963) in Stokes flows, and
later studies dealt with heat or mass transfer in steady shear flow
(\cite{phillips90}, 1990). Related work continues to the present day,
e.g. in the context of nutrient uptake by single-cell organisms
(\cite{magar03}, 2003).

\begin{figure}
  \centering \includegraphics[width=0.5\linewidth]{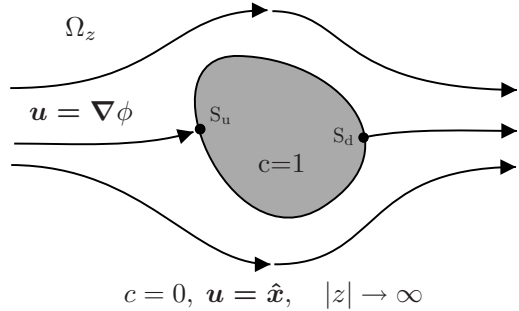}
  \caption{\label{fig:domain} The fundamental problem of advection-diffusion:
    A finite absorber (source) in a steady flow of uniformly concentrated
    (depleted) fluid in the complex $z$ plane;
    $\Omega_z$ is the region exterior to the object.}
\end{figure}

The $\Nu (\Pe)$ relation contains useful global information, but one
sometimes requires a complete solution to the problem, including the
local flux profile on the absorber. In this paper, we focus on a well
known special case, ideally suited for mathematical analysis and
physical interpretation: steady
advection-diffusion in a two-dimensional, irrotational flow.  The
velocity field is described by the flow potential, $\phi$, as
$\ub=\del\phi$ and the dimensionless boundary-value problem (BVP) is:
\begin{gather}
  \label{eq:z_pde} \Peo \del\phi \cdot \del c = \del^2 c ,\quad \del^2
  \phi = 0,\quad (x,y)\ \in\ \Omega_z \\ 
 \label{eq:z_bc1}  c=1,\quad \nhat\cdot\del\phi=0,\quad (x,y)\ \in \
  \partial\Omega_z \quadtext{and} c \to 0,\quad x^2+y^2\to \infty;\\
 \label{eq:z_bc2} \del\phi \to \xhat,\quad x^2+y^2\to \infty,
\end{gather}
where $\Omega_z$ is the flow region, exterior to the object's 
boundary, $\partial \Omega_z$, as shown in
Fig.~\ref{fig:domain}. Note that in (\ref{eq:z_bc1}) we use boundary
conditions for desorption ($c=1$ on the object and $c=0$ far away),
which are somewhat more convenient that those of adsorption ($c=0$ on
the object and $c=1$ far away). The two problems are mathematically
equivalent by linearity. 

The key to analyzing (\ref{eq:z_pde}) is to view points in the plane
as complex numbers, $z = x+iy$. For example, this enables a
transformation to streamline coordinates, $\Phi = \phi + i \psi$,
which reduces the problem to a thin absorbing strip in a uniform flow
(\cite{boussinesq05}, 1905). For a more general perspective in terms of
conformal mapping, see Bazant (2004).  Using complex-variable
techniques, the general system of PDEs (\ref{eq:z_pde}) has been
studied recently with applications to tracer dispersion
(\cite{koplik94}, 1994) and heat transfer (\cite{morega94}, 1994) in
porous media, as well as vorticity diffusion in strained wakes
(\cite{eames99}, 1999; \cite{hunt02}, 2002).

Complex analysis becomes particularly useful when the interface,
$\partial \Omega_z(t)$, is a moving free boundary, driven by the local
flux density, $\sigma_z$. For a broad class of transport-limited
growth phenomena, the interfacial dynamics, whether deterministic or
stochastic, can be formulated in terms of a time-dependent conformal
map from a simple, static domain to the evolving, physical domain
(\cite{bazant03}, 2003).  For continuous growth by
advection-diffusion, this approach was introduced by
\cite{wijngaarden66} (1966) and \cite{maksimov76} (1976),  who solved
(\ref{eq:z_pde})--(\ref{eq:z_bc2}) in streamline coordinates.
Maksimov's method has been used extensively by Kornev and his
collaborators to describe solidification and freezing from a flowing
melt (\cite{chugunov86}, 1986; \cite{kornev88}, 1988;
\cite{kornev94}, 1994; \cite{alimovetal94}, 1994; \cite{alimovetal98}, 1998;
\cite{cummingskornev99}; 1999; \cite{cummings99}; 1999). These studies
significantly extend the conformal-map dynamics for Laplacian growth
by pure diffusion, without advection (\cite{polubarinova45} 1945;
\cite{galin45}, 1945; \cite{howison92}, 1992).

Here, we are motivated by a new, discrete growth model,
Advection-Diffusion-Limited Aggregation (ADLA), which describes the
growth of fractal aggregates in a fluid flow via a stochastic
conformal map (\cite{bazant03}, 2003). In that case, the BVP
(\ref{eq:z_pde})-(\ref{eq:z_bc2}) must be solved for a circular
absorber for all values of $\Peo$.  The normal flux distribution,
$\sigma_z$, determines the probability measure for growth
events.  The detailed description of the growth measure in this paper
shows how dynamics crosses over from diffusion-dominated to
advection-dominated growth regimes, as a function of the
time-dependent P\'eclet number. These two ``fixed points'' of the
dynamics are related to special similarity solutions (\cite{bazant04},
2004), which correspond to the asymptotic regimes of high and low
$\Peo$ analyzed in this paper.

Beyond such applications, the BVP (\ref{eq:z_pde})--(\ref{eq:z_bc2}),
for an arbitrary single-connected domain, $\Omega_z$, merits serious
mathematical study in its own right.  It is perhaps the simplest
advection-diffusion problem with a nontrivial dependence on
the P\'eclet number.  It may also be the most complicated
advection-diffusion problem for which a nearly exact analytical
solution is possible, as we show here.

The paper is organized as follows.  In Sec.~\ref{sec:background} we
set the stage for our analysis by reviewing two key properties of the
BVP (\ref{eq:z_pde})--(\ref{eq:z_bc2}): (i) It can be recast as a
singular integral equation in streamline coordinates, and (ii)
conformal mapping can be applied to work in other convenient
geometries for numerical solution and mathematical analysis.  In
Sec.~\ref{sec:numerical_soln}, we present an efficient, new numerical
method to solve the BVP, after conformal mapping to the interior of a
circular disk.  In Sec.~\ref{sec:highPe_asymptot}, we derive accurate
asymptotic expansions for $\sigma$ for high P\'eclet numbers by
applying an exact iterative procedure to the integral equation.  In
Sec.~\ref{sec:lowPe_asymptot} approximate formulae for $\sigma$ are
derived when $\Pe$ is sufficiently small via approximating the kernel
of the integral equation and solving the resulting equation exactly by
known methods. In Sec.~\ref{sec:connection} an accurate {\it ad hoc}
connection formula is given for $\sigma$ by combining the formulae for
high and low $\Pe$, and it is also integrated to obtain the $\Nu(\Pe)$
relation.  Finally, in Sec.~\ref{sec:discussion} we conclude with a
discussion of some of the implications and applications of our results,
as well as by posing a few challenges for future work.

\section{Mathematical Preliminaries}
\label{sec:background}

\subsection{Streamline Coordinates}
\label{subsec:streamline}

In his analysis of high Reynolds number flows, \cite{boussinesq05}
(1905) discovered a hodograph transformation (exchanging dependent and
independent variables) which greatly simplifies (\ref{eq:z_pde}).  It
is well known that the velocity potential is the real part of an
analytic complex potential, $\Phi = \phi + i\psi$, where the harmonic
conjugate, $\psi$, is the stream function (\cite{batchelor}
1967). Using the Cauchy-Riemann equations, it is easy to show that the
concentration profile, $c$, satisfies the simplified PDE
\begin{equation}
  \label{eq:stream_pde} \Peo \p{c}{\phi} = \p[2]{c}{\phi} + \p[2]{c}{\psi},
\end{equation}
in ``streamline coordinates'', $(\phi,\psi)$. The physical
significance of this equation is that advection (the left hand side)
only occurs along streamlines, while diffusion (the right
hand side) also occurs in the transverse direction, along isopotential
lines. 

\begin{figure}

  \centering \raisebox{1.7in}{(a)}
  \includegraphics[height=1.9in]{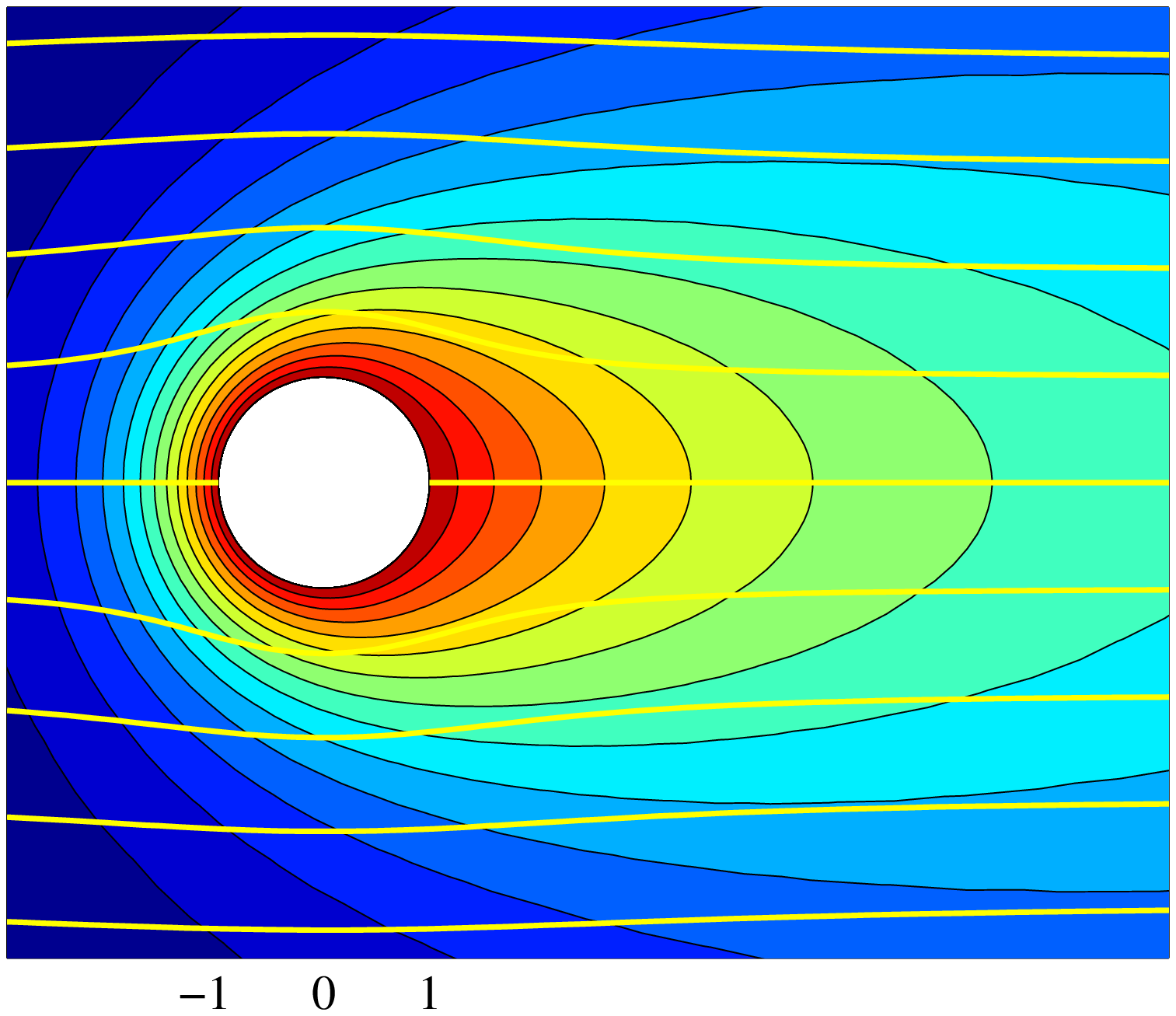} \quad
  \raisebox{1.7in}{(b)} \includegraphics[height=1.9in]{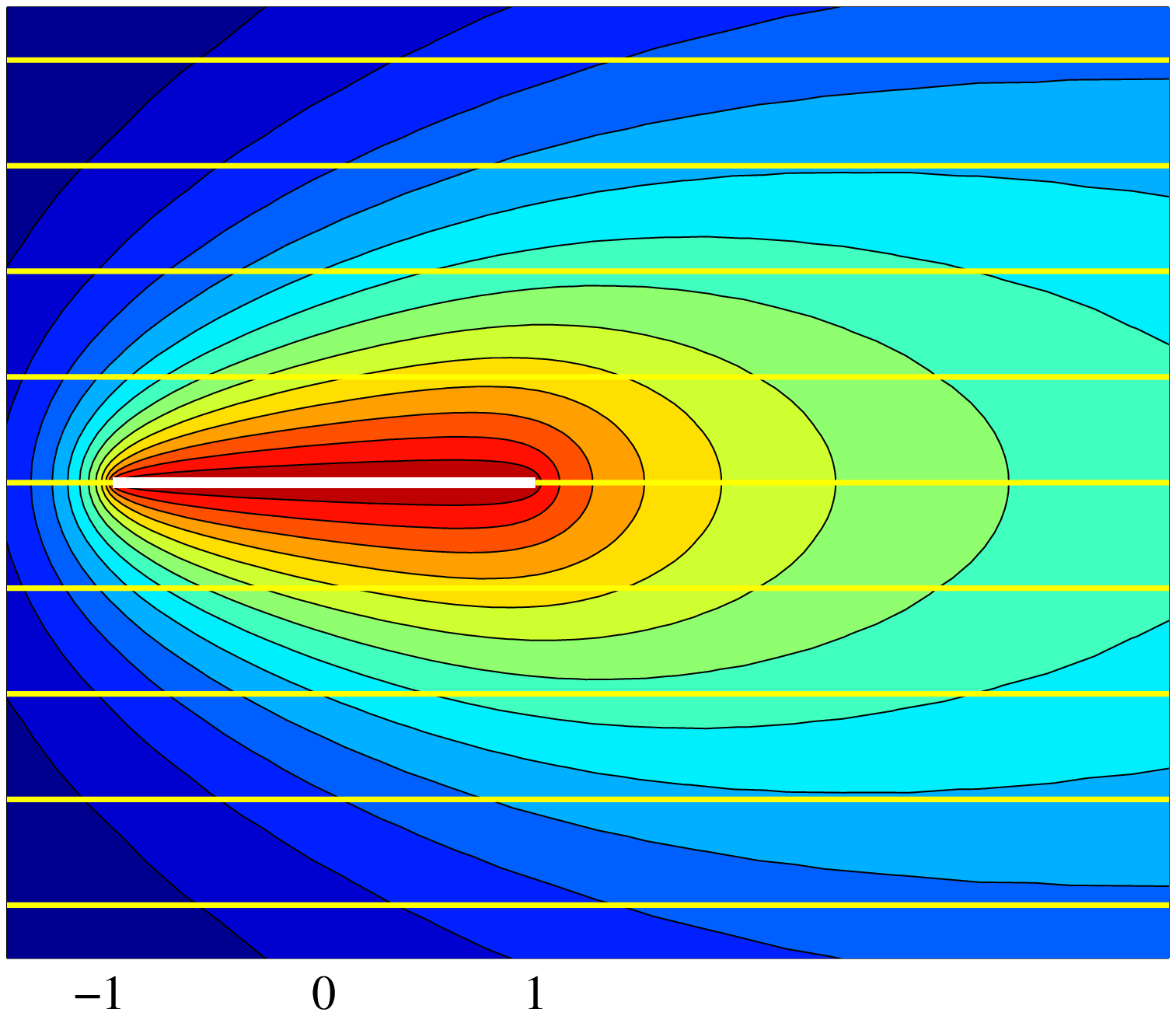}
  \\[2ex] \raisebox{1.7in}{(c)}
  \includegraphics[height=1.9in]{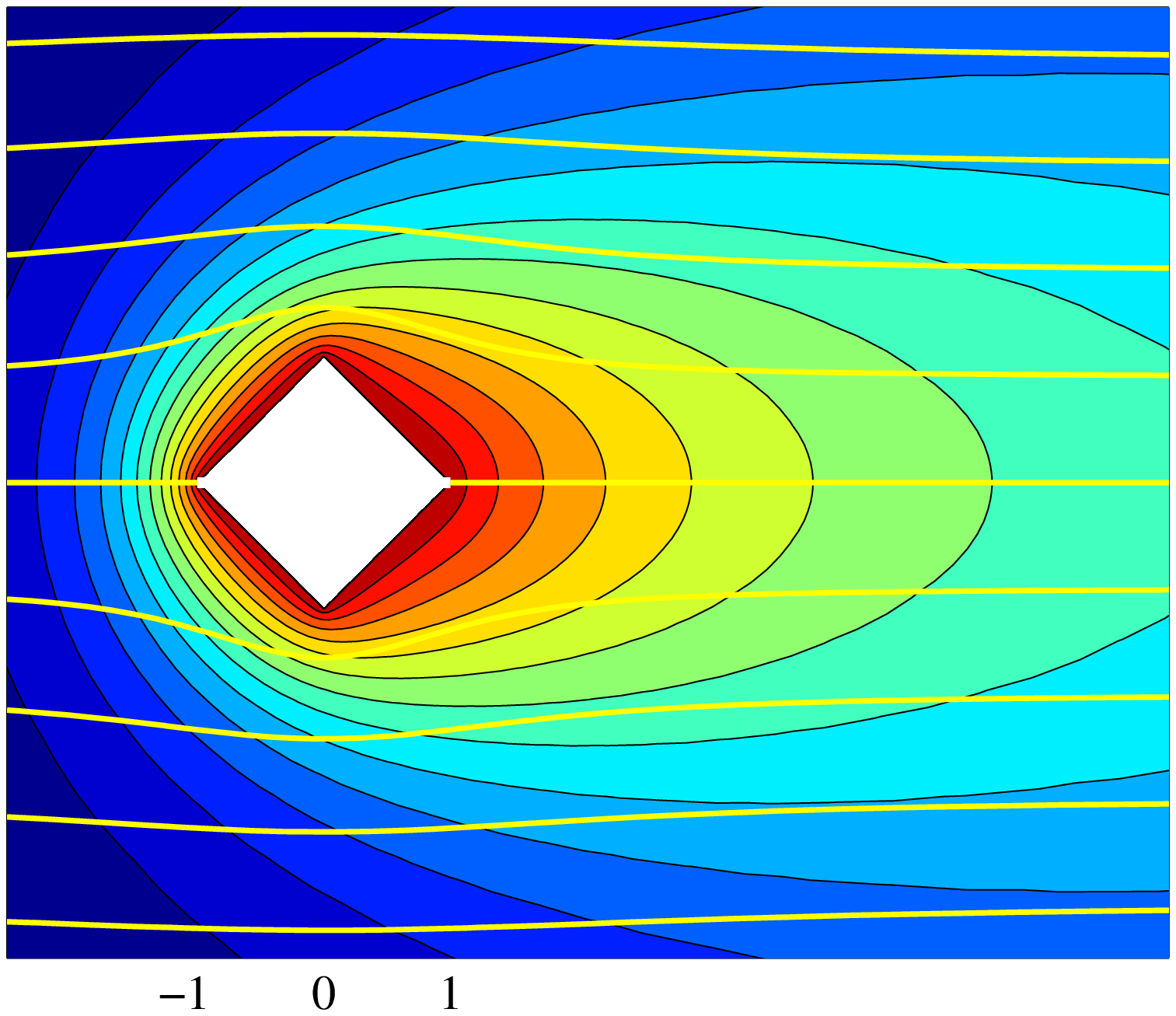} \quad
  \raisebox{1.7in}{(d)} \includegraphics[height=1.9in]{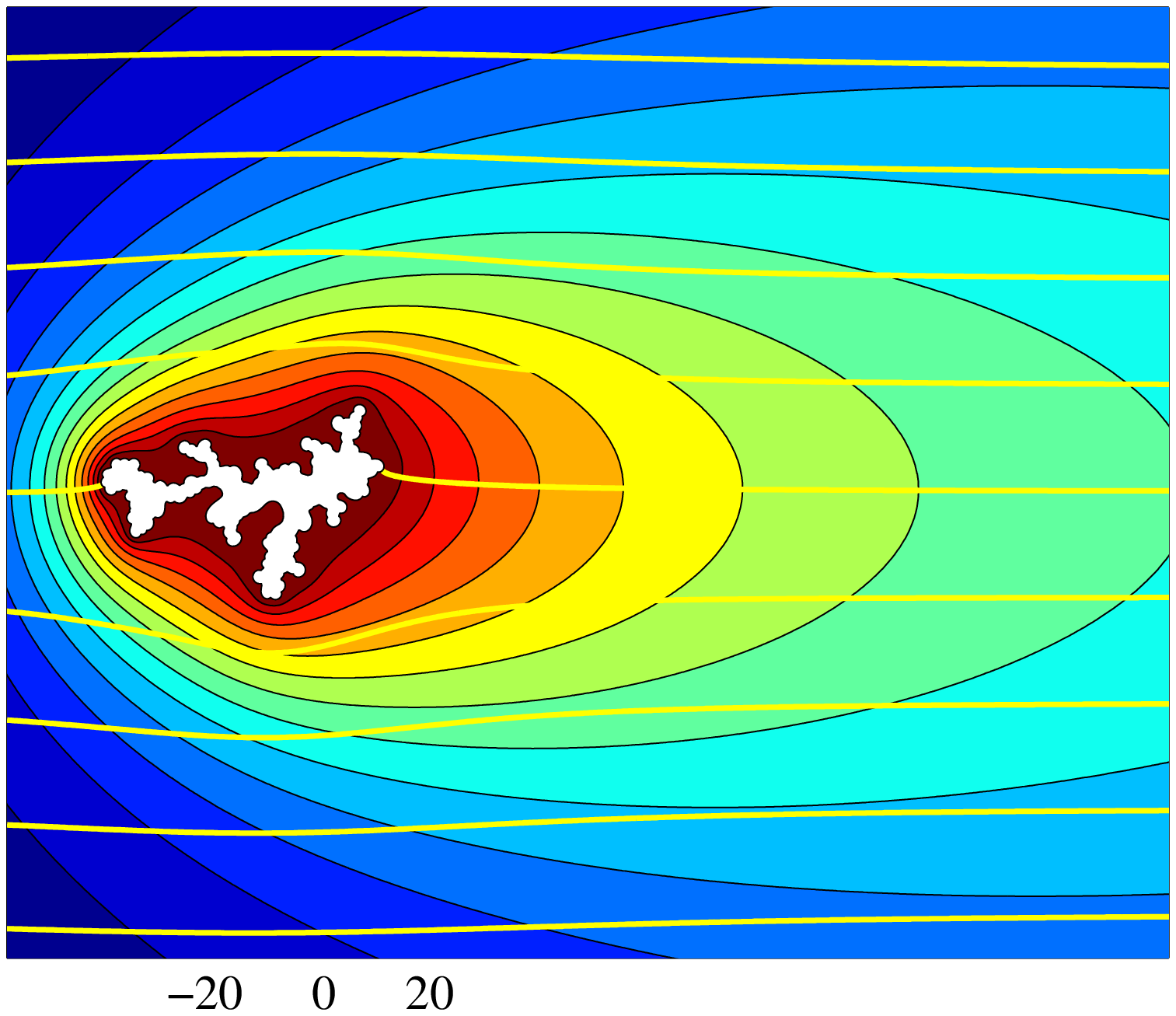}
  \caption{\label{fig:profiles} Numerical solutions using the method
  of Sec.~\ref{sec:numerical_soln} for the concentration
  profile (color contour plot) and streamlines (open yellow curves)
  around different absorbing objects in a uniform background potential
  flow: (a) the unit disk, (b) a finite strip, (c) a square, and (d)
  an ADLA fractal cluster with $\Peo = 1, \; 2,\; 1.2$ and $0.05$
  respectively. The geometries in (b), (c) and (d) are obtained by
  conformal mapping from (a). Case (b) corresponds to streamline
  coordinates; case (c) is obtained by the numerical
  Schwartz-Christoffel mapping (\cite{trefethen86}, 1986); case (d) is
  obtained from a stochastic, iterated conformal map (\cite{bazant03},
  2003).  The horizontal axis is labeled by numerical values of the
  spatial coordinate $x$. However, the distances in the figures
  (a)-(d) are scaled by the ``conformal radius'' $A_1$, see
  (\ref{eq:A1-laurent}) below, which is the characteristic size of the
  physical object. The $A_1$ is chosen so that the renormalized
  P\'eclet number is the same, $\Pe=A_1\Peo=1$, in all cases, which
  explains why the far-field solutions look the same.  }
\end{figure}

Boussinesq's transformation corresponds to a conformal mapping to a
plane of a uniform flow, described by a constant
$\del\phi$. Therefore, an arbitrary domain, $\Omega_z$, as shown in
Fig.~\ref{fig:domain}, is mapped to the exterior of a straight line
segment, or strip, parallel to the streamlines (which is a branch cut
of the inverse map). Some examples are given in
Fig.~\ref{fig:profiles}.  In streamline coordinates, the BCs
(\ref{eq:z_bc1}) and (\ref{eq:z_bc2}) are transformed as follows:
\begin{equation}
  \label{eq:stream_bc}
  c=1,\quad \psi=0,\; -2A<\phi<2A, \quadtext{and} c \to 0 \quadtext[\;\;]{as} \phi^2+\psi^2 \to \infty.
\end{equation}
The constant $A$ is determined so that $4A$ is equal to the difference of the 
flow potential $\phi$ between two stagnant points, for example, 
the points $\mathrm{S_u}$ and $\mathrm{S_d}$ shown in Fig.~\ref{fig:domain}.

\subsection{Formulation as an Integral Equation}
\label{subsec:integral_eqn}

In streamline coordinates, the advection-diffusion process past a
finite strip can be formulated in terms of an integral equation using
the classical method of Green's functions (\cite{stakgold98}, 1998). For
reasons to become clear in Sec.~\ref{subsec:invariance}, we let
$x=\phi$, $y=\psi$, $A=1/2$, and $\Pe = \Peo/2$, so the BVP
(\ref{eq:stream_pde})--(\ref{eq:stream_bc}) takes the form
\begin{gather}
  \label{eq:strip_pde} 2\, \Pe \p{c}{x} = \p[2]{c}{x} + \p[2]{c}{y},\\
  \label{eq:strip_bc} c=1,\quadtext[\;\;]{on} y=0,\; -1<x<1 \quadtext{and}
  c=0,\quadtext[\;\;]{as} x^2+y^2\to \infty.
\end{gather}
Green's function $G(x,y)$ for (\ref{eq:strip_pde}) and (\ref{eq:strip_bc}), which expresses
the concentration profile generated by a unit source flux at the
origin, satisfies the PDE 
$$ 2\, \Pe \p{G}{x} - \p[2]{G}{x} - \p[2]{G}{y} = \delta(x)\delta(y).$$
After removing the first derivative term by a change of variables
and using polar coordinates,  
\begin{equation}
  G(x,y) = e^{\Pe x}F(r,\theta)\;,
\end{equation}
we find that $F$ obeys the Helmholtz equation
\begin{equation}
\del^2 F -\Pe[^2] F = \delta(x)\delta(y)\;,
\end{equation}
whose solution is a modified Bessel function of the second kind, $K_0(\Pe r)$,
where $r=(x^2+y^2)^{1/2}$. Taking
into account the unit normalization, we obtain Green's function,
\begin{equation}
G(x,y)=e^{\Pe x}K_0(\Pe r)/2\pi.\label{eq:Green}
\end{equation}

The concentration profile, $c$, everywhere in streamline
coordinates is obtained by convolving Green's function, $G$, with the flux on
the strip,
\begin{equation}  
c(x,y) = \int_{-1}^1 G(x-x',y)\cdot 2\sigma(x') dx',\label{eq:c-G-sigma}
\end{equation} 
where the factor $2$ is included because the flux has the same value,
$$\sigma(x) = -\frac{\partial c}{\partial y}(x,0^+) = \frac{\partial
c}{\partial y}(x,0^-),$$ 
on the upper and the lower sides of the strip, respectively.
Therefore, the boundary value problem described by (\ref{eq:strip_pde}) and (\ref{eq:strip_bc}) 
is equivalent to finding the $\sigma(x)$ that satisfies the 
integral equation (\cite{pearson57}, 1957;~\cite{wijngaarden66}, 1966)
\begin{equation}
\label{eq:ie} \int_{-1}^{1} e^{\Pe(x-x')} K_0 \left(\Pe|x-x'|\right)
\sigma(x') dx' = \pi ,\quad -1<x<1, 
\end{equation}
which forms the basis for the theory of solidification in flowing
melts (\cite{maksimov76}, 1976). In this context, (\ref{eq:ie}) has
been analyzed for large and small $\Pe$ by Kornev and collaborators,
as cited in the introduction. Below, we will extend these results and
construct an analytical approximation that is uniformly accurate in
both $\Pe$ and $x$.

The reader may worry about the existence and uniqueness of solutions
because (\ref{eq:ie}) is a Fredholm-type equation of the first kind
(with a difference kernel). In the present case, however, the 
symmetrized kernel, $K_0(\Pe |x-x'|)$, is positive definite, and 
thus invertible; see Appendix~\ref{app:invertibility} for an explanation. 

It seems tempting to approach (\ref{eq:ie}) using Fourier-type methods 
because it involves a convolution (\cite{titchmarsh}, 1948), but the kernel is
not a periodic function. We also note that the kernel is singular and not of the classical 
Cauchy type (\cite{musk}, 1992).  It is known that (\ref{eq:ie}) 
admits a solution which can be expanded in terms of Mathieu 
functions~(\cite{rvachev56}, 1956; \cite{protsenko76}, 1976), but 
such representations are impractical for computations over a wide 
range of $\Pe$, and give no insight into the dependence of the 
flux $\sigma$ on $x$ and $\Pe$. The collocation method has been 
used successfully to obtain numerically the solution of (\ref{eq:ie}) (\cite{kornev94}, 1994).


\subsection{The General Principle of Conformal Invariance}
\label{subsec:invariance}

There is a simple way to understand why Boussinesq's transformation
works: The advection-diffusion PDE (\ref{eq:z_pde}) is invariant under
conformal changes of variables, even though its solutions are not
harmonic functions, which also holds more generally for some other
equations (\cite{bazant04}, 2004). As such, the boundary value can be
transformed to any convenient geometry by conformal
mapping. Streamline coordinates is a good choice for asymptotics, but
other choices are better suited for numerical analysis and similarity
solutions.

Here, we exploit this general principle to map the BVP
(\ref{eq:z_pde})--(\ref{eq:z_bc2}) to other useful coordinate
systems. If $\phi_w = \Re \Phi(w)$ and $c_w = F(w,\overline{w})$ 
(where $\overline{w}$ denotes the complex conjugate of $w$) solve
(\ref{eq:z_pde}) in some simple domain, $\Omega_w$, then $\phi_z = \Re
\Phi(f(z))$ and $c_z = F(f(z),\overline{f(z)})$ solve
(\ref{eq:z_pde}) in an arbitrary mapped domain, $\Omega_z =
g(\Omega_z)$, where $z=g(w)=f^{-1}(w)$. The concentration BCs
(\ref{eq:z_bc1}) are conformally invariant, but, since the BC
(\ref{eq:z_bc2}) prescribing the background flow is not, due to the
gradient, care must be taken in transforming the solution.
 
For advection-diffusion-limited growth (\cite{bazant03},
2003), it is natural to let $\Omega_w$ be the exterior of the unit
circle, so our canonical problem is that of a concentrated flow past a
circular absorber, shown in Fig.~\ref{fig:profiles}(a), with the velocity potential 
\begin{equation}
\phi = \Re\left\{w+\frac1w\right\},\quad |w|>1 .  \label{eq:phiw}
\end{equation}
To reach other geometries, the mapping, $z=g(w)$, must be univalent
(conformal and one-to-one), so it has a Laurent series of the form
\begin{equation}
g(w) = A_1 w + A_0 +
\frac{A_{-1}}{w} + \ldots,\quad |w| > 1,\label{eq:A1-laurent}
\end{equation} where $A_1$ is a positive, real constant which defines an
effective diameter of $\Omega_z$ (the ``conformal radius'').  In order
to preserve the BC (\ref{eq:z_bc2}) which sets a unit flow speed at
$z=\infty$, we would need to set the dimensionless flow speed to $A_1$
at $w=\infty$. Instead, we choose to redefine the velocity potential
in $\Omega_w$ to preserve the unit flow speed, as in (\ref{eq:phiw}),
and then define a {\it renormalized} P\'eclet number, $\Pe = A_1\;
\Peo$, analogous to the time-dependent P\'eclet number for ADLA
defined by \cite{bazant03} (2003).

The BVP for the concentration $\Omega_w$ then becomes
\begin{gather}
  \label{eq:w_pde} \Pe \del \phi \cdot \del c = \del^2 c, \quad |w|>1, \\
  \label{eq:w_bc} c=1\ \mbox{on}\ |w|=1, \quadtext{and} c=0 \quadtext[\;\;]{as} |w|\to \infty.
\end{gather}
The physical significance of the renormalized P\'eclet number, $\Pe$, is that
it determines the far-field solution, independent of the absorber's
shape and the bare P\'eclet number.  This point is illustrated in
Fig.~\ref{fig:profiles}, where the concentration and flow field far
away from various objects at $\Pe=1$ looks the same, in spite of
extremely different shapes, ranging from a circle to a fractal ADLA
cluster. Therefore, we view $\Pe$ as the basic parameter in our
analysis, from which we define the bare P\'eclet number, $\Peo =
\Pe/A_1$, for arbitrary domains, in terms of the univalent map from
the exterior of the unit circle.

From this perspective, streamline coordinates are obtained via the
Joukowski transformation (\cite{carrier83}, 1983), $g(w) = (w+1/w)/2$,
which maps the unit circular disk onto the finite strip of length $2$
centered at the origin along the real axis. The BVP (\ref{eq:w_pde})--
(\ref{eq:w_bc}) is then transformed to the form
(\ref{eq:strip_pde})--(\ref{eq:strip_bc}) given above. In general, the
fluxes on the boundaries $\partial\Omega_z$ (the absorber surface) and
$\partial\Omega_w$ (the unit circle, $w=e^{i\theta}$) are related by
\begin{equation}
  \label{eq:sigma_wz}
  \sigma_w(\theta;\Pe) = |g\prime(w)|\;\sigma_z(g(w);\Pe) ,
\end{equation}
where $\sigma_w$ is the flux on $\partial\Omega_w$ and $\sigma_z$ is
the flux on $\partial\Omega_z$.
In the case of streamline coordinates
(\ref{eq:strip_pde})--(\ref{eq:strip_bc}), the flux on the strip,
$\sigma_z(x;\Pe)$, is thus related to the flux on the circle,
$\sigma_w(\theta;\Pe)$, by
\begin{equation}
  \label{eq:sigma_thz}
  \sigma_w(\theta;\Pe) = |\sin\theta|\; \sigma_z(\cos\theta;\Pe).
\end{equation}
For a bounded flux on the circle, $\sigma_w$, the flux on the strip,
$\sigma_z$, always diverges as $O[(1-x^2)^{-1/2}]$ as $x$ approaches
$\pm 1$. Therefore, although we will use the strip geometry,
$\Omega_z$, for asymptotic analysis in Secs.~\ref{sec:highPe_asymptot}
and \ref{sec:lowPe_asymptot}, the circle geometry, $\Omega_w$, is a
better starting point for our numerical analysis in
Sec.~\ref{sec:numerical_soln}. In all cases, however, our goal is to
obtain the flux on the circle, the canonical geometry for a finite
absorber.

\subsection{Similarity Solutions for Semi-Infinite Leading Edges}
\label{subsec:similarity}

Before proceeding with our analysis, we mention a
class of similarity solutions for ``leading edges'' which have
relevance for the high-$\Pe$ limit of our problem. For this section
only, we use the upper half plane, $\{\Im w>0\}$, as our simple
domain, $\Omega_w$. For a straining velocity field, $\phi(w) = \Re
w^2$, bringing fluid toward the plane, there is a classical
similarity solution for the concentration profile, $c(w,\overline{w}) =
\erfc(\sqrt{\Peo}\Im w)$,  for which the 
flux on the real axis is a constant, $\sigma_w =
2\sqrt{\Peo/\pi}$. (As usual, the complementary error function is
defined by $\erfc(z)=(2/\sqrt{\pi})\int_z^{\infty}dt\,e^{-t^2}$.)  

For every conformal map, $w=f(z)$, from the $z$ plane to the upper
half $w$ plane, there corresponds another similarity solution
(\cite{cummings99}, 1999; \cite{bazant04}, 2004),
\begin{equation}
\phi = \Re f(z)^2 \quadtext{and} c=\erfc(\sqrt{\Peo}\Im f(z))
\quadtext[\;\;]{for} \Im w\ge 0.\label{eq:burgers}
\end{equation} 
Note that the boundary condition $\del\phi\sim\boldsymbol{\hat{x}}$ as
$|z|\to\infty$ holds only if $f(z)\sim\sqrt{z}$ as $|z|\to\infty$, so
these solutions correspond to more general flows near stagnation
points. For the purposes of this paper, we discuss two choices for
$f(z)$:
\begin{enumerate}
\item $f(z) = \sqrt{z}$, which maps the entire $z$ plane, with the 
exception of the branch cut $\{\Im z=0, \Re z >0\}$, onto
$\Omega_w$. The ($x,y$) coordinates of the $z$ plane then correspond
to the streamline coordinates, and the flux on the strip is $\sigma(x)
= \sqrt{\Peo/\pi x}$. This well known formula can also be derived via
replacing the upper limit of integration in (\ref{eq:ie}) by $\infty$,
and applying the Wiener-Hopf method of factorization (\cite{krein62},
1962;~\cite{wijngaarden66}, 1966;~\cite{carrier83}, 1983).  This
procedure is carried out systematically to all orders of approximation
in Sec.~\ref{sec:highPe_asymptot} and via a different, rigorous method
by \cite{diochoi04} (2004).

\item  $f(z) = \sqrt{z}+1/\sqrt{z}$, which maps the exterior of the circular
rim, $\{z:|z|>1\quadtext[\;\;]{and} 0<\arg z<2\pi\}$, onto $\Omega_w$
(\cite{bazant04}, 2004). This solution describes
the advection-dominated (high-$\Pe$) fixed point of the ADLA
fractal-growth process (\cite{bazant03}, 2003).  From
(\ref{eq:burgers}) with $z=r e^{i\theta}$ and $r\ge 1$, we find
\begin{equation}
  \label{eq:erfc} c(r,\theta) = \erfc
  \left[\sqrt{\Peo}\right(\sqrt{r}+\frsq1{r}\left)\sin\left(\frac{\theta}{2}\right)\right],
  \quad \sigma(\theta) =
  2\sqrt{\frac{\Peo}{\pi}}\sin\left(\frac{\theta}{2}\right), 
\end{equation}
which is the leading-order solution for our 
BVP (\ref{eq:w_pde})--(\ref{eq:w_bc}) as $\Peo=\Pe\rightarrow \infty$. 
\end{enumerate}

The fact that we consider {\it finite} absorbers leads to significant
analytical and numerical complications, which are the focus of this
paper.

\section{Numerical Solution}
\label{sec:numerical_soln}
\subsection{ Conformal Mapping to Polar Coordinates Inside the Unit Disk }

\begin{figure}
  \centering
  \parbox[t]{0.30\linewidth}{(a)\\
    \includegraphics[height=\linewidth]{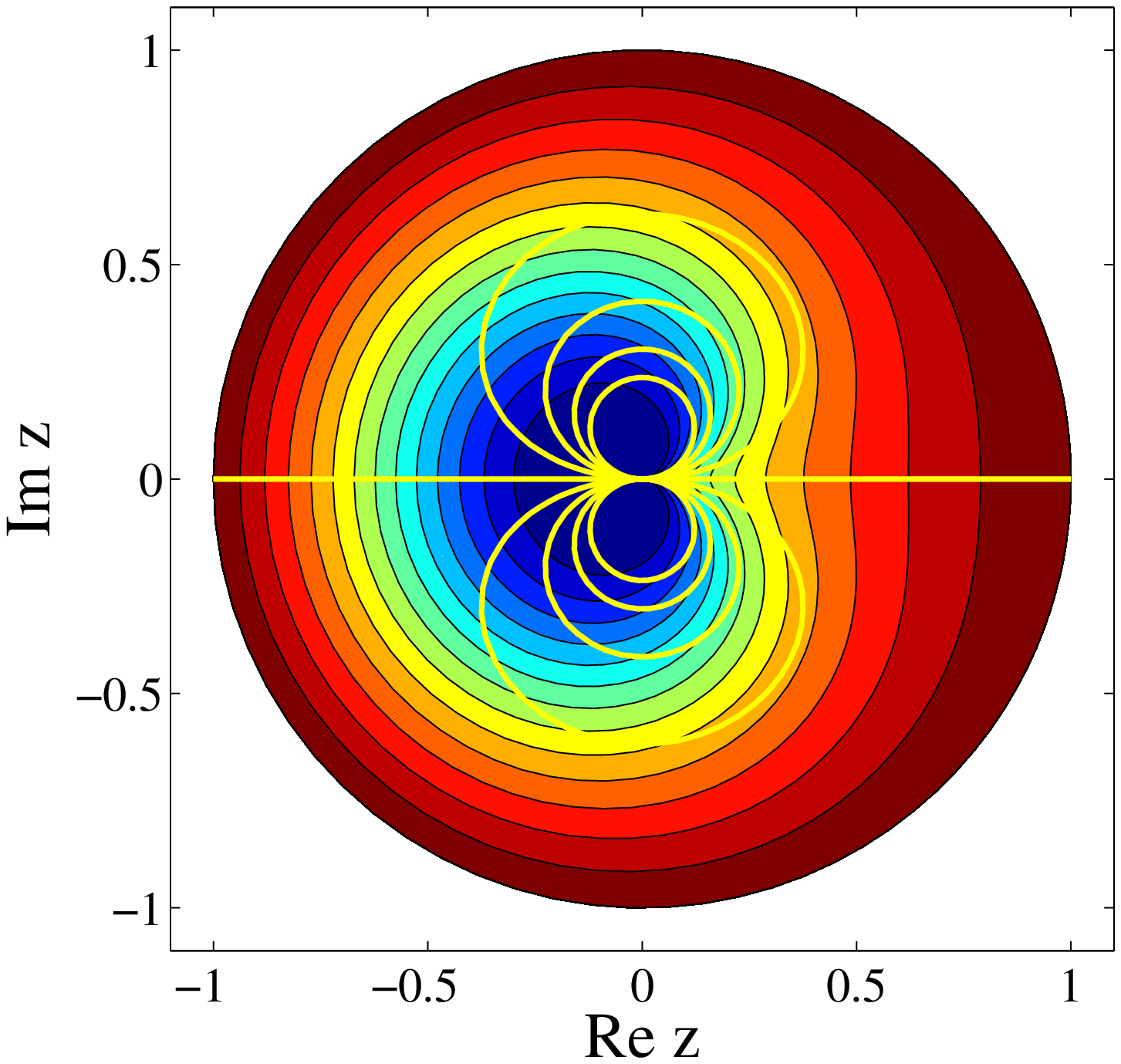}} \quad
  \parbox[t]{0.32\linewidth}{(b)\\
    \includegraphics[height=\linewidth]{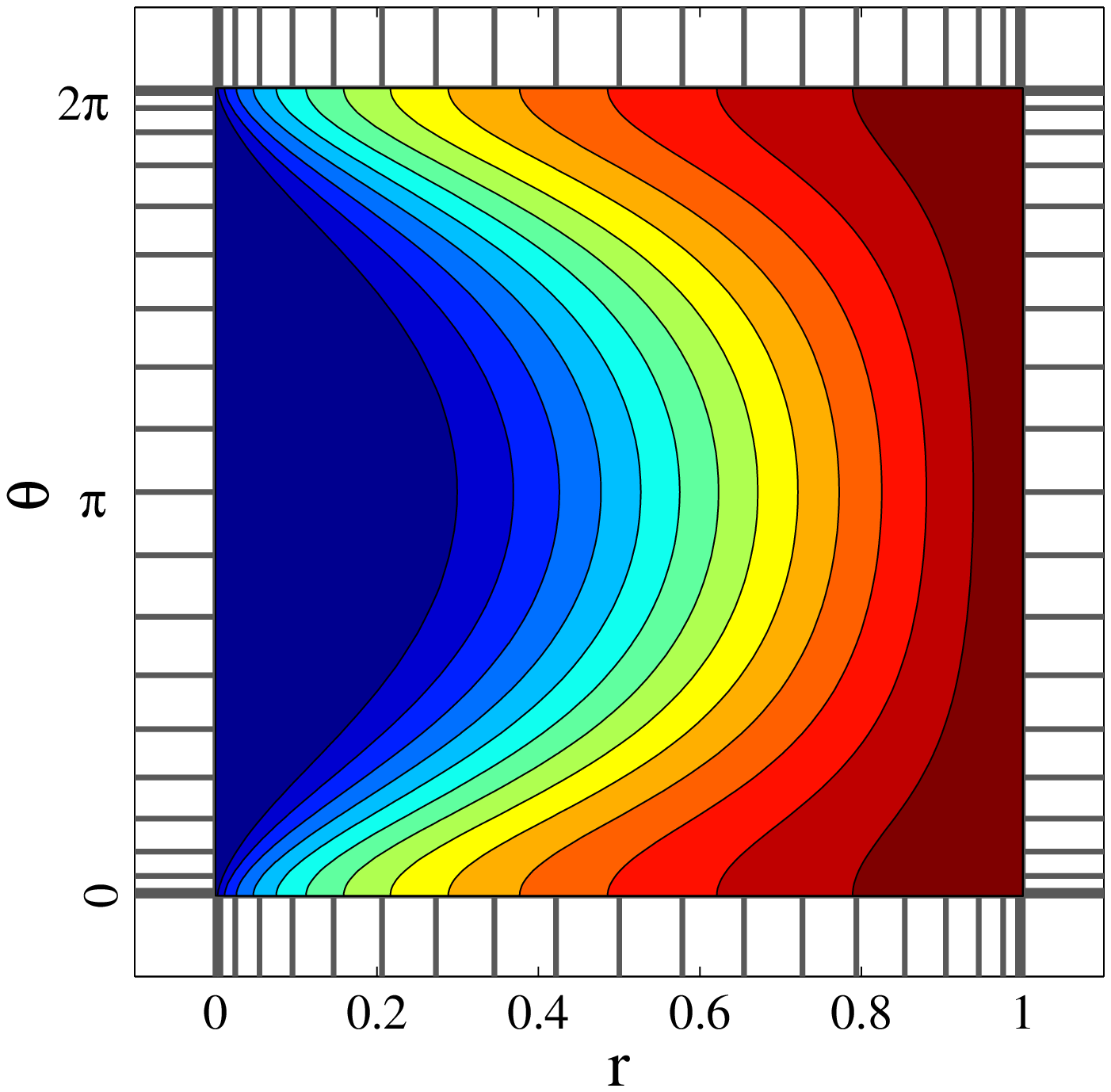}} \quad
  \parbox[t]{0.31\linewidth}{(c)\\
    \includegraphics[height=\linewidth]{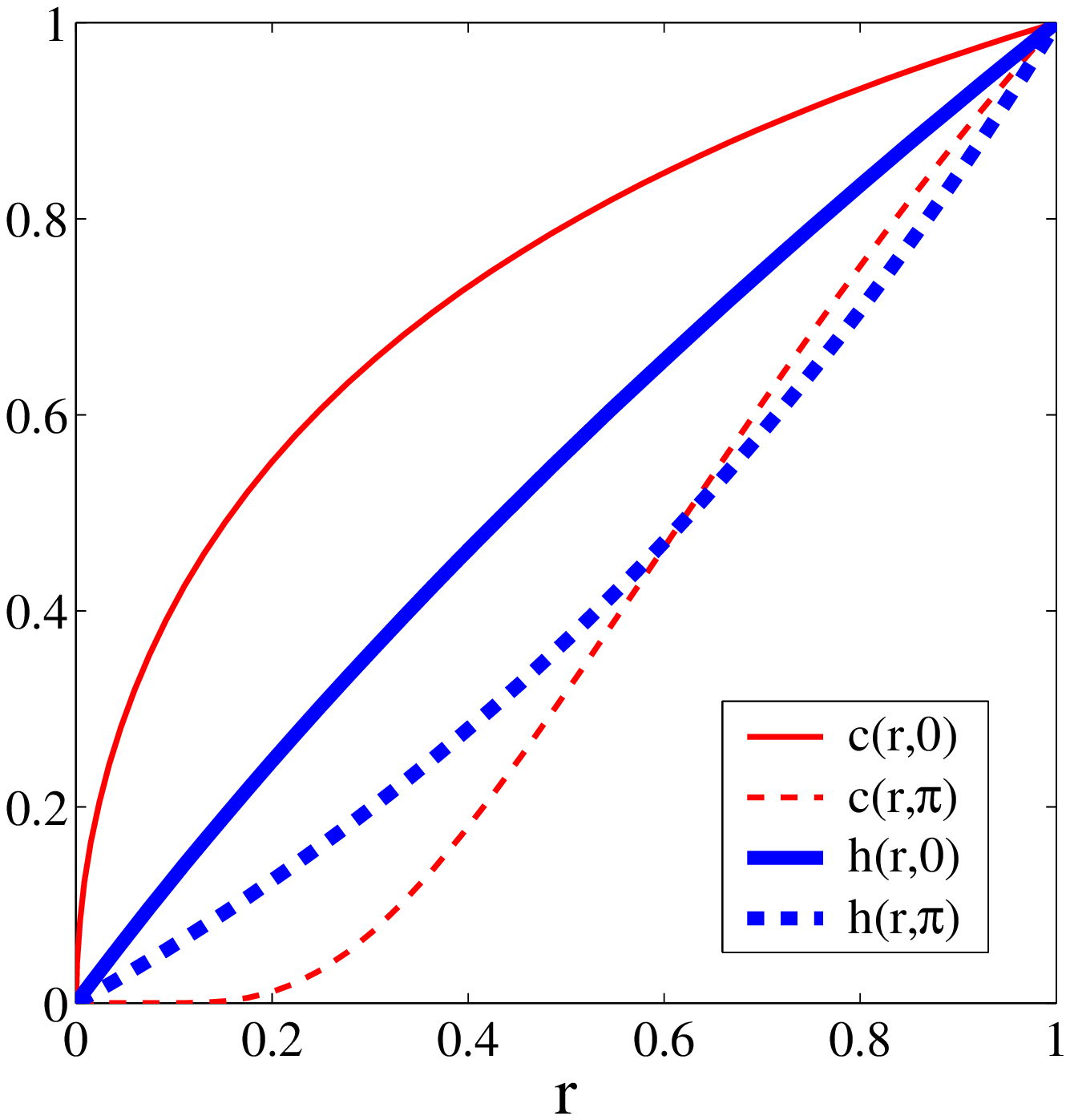}}
  \caption{\label{fig:polar} The concentration profile ($\Pe=1$) calculated
    numerically in the interior of the circular disk is shown (a) in 
    Cartesian coordinates, and (b) in polar coordinates. In (c) the 
    numerically obtained values are shown for $c(r,\theta=0)$ (thin solid 
    line), $c(r,\theta=\pi)$ (thin dashed line), $h(r,\theta=0)$ 
    (thick solid line), and $h(r,\theta=\pi)$ (thick dashed line).}
\end{figure}

In this section we determine the concentration profile $c$ by
numerically solving the BVP (\ref{eq:w_pde})-- (\ref{eq:w_bc}).  One
of the difficulties in applying a numerical method is related to the
fact that the region $\Omega_w$ is unbounded.  By invoking conformal
invariance again, however, we can apply a transformation that leaves
the BVP unchanged yet maps $\Omega_w$ to a bounded region. In
particular, we use inversion, $g(w) = 1/w$, to map the disk
exterior onto its interior. Physically, this corresponds to a dipole
source of concentrated fluid inside an absorbing circular cylinder, as
shown in Fig.~\ref{fig:polar}(a).  
With $g(w)=r e^{i\theta}$ the problem is expressed in the polar
coordinates $(r,\theta)$ by
\begin{subequations}
\label{alleqs:c_pde}
\begin{equation}
  \label{eq:c_pde1} r^3 \p[2]{c}{r} + \{r^2 + \Pe r(1-r^2)\cos\theta\}
  \p{c}{r} + r \p[2]{c}{\theta} + \Pe(1+r^2)\sin\theta \p{c}{\theta} = 0,
\end{equation}
\begin{equation}
\label{eq:c_pde2}
  c = 0,\quadtext[\;\;]{at} r=0, \quadtext{and} c = 1,\quadtext[\;\;]{at}  r=1,
\end{equation}
\end{subequations}
where $r\le 1$ and $0\le\theta< 2\pi$. A solution in the
$(r,\theta)$ plane is shown in Fig.~\ref{fig:polar}(b).

\subsection{ Analytical Treatment of Singularities }

Before we apply any numerical method to (\ref{alleqs:c_pde}) directly,
we note that the concentration profile $c$ as a function of
($r,\,\theta$) exhibits singular behavior as $r$ approaches $0$. We
first need to modify (\ref{eq:c_pde1}) in order to eliminate this
behavior, which undermines the accuracy of our numerical method. From
the similarity solution (\ref{eq:erfc}) and Green's function of
Sec.~\ref{subsec:integral_eqn} we obtain the leading-order behavior
\begin{equation}
  c(r,\theta)=O\left\{\sqrt{r}
  \;\exp\left[\Pe\left(2-\frac1r-r\right)\sin^2\left(\frac{\theta}{2}\right)
    \right]\right\}
  \quadtext{as} r\to 0.
\end{equation}
First, the square-root limit $c(r,\theta=0)=O(\sqrt{r})$ can
hardly be dealt with in numerical methods because of the resulting
diverging derivative near $r=0$. Second, the essential singularity at
$r=0$, $c(r,\theta=\pi)=O\bigl(\sqrt{r}\;e^{-\Pe/r}\bigr)$, forces $c$ to
change drastically near $r=0$.  Especially when $\Pe$ is large, this
limiting behavior is extended even to $r < 1-O(1/\sqrt{Pe})$.  To
avoid this behavior, we define a function $h(r,\theta)$ by factoring
out the leading-order singular behavior of $c(r,\theta)$ as
\begin{equation}
  \label{eq:h_def} \sqrt{r}\;
  \exp\left\{\Pe\left(2-\frac1r-r\right)\sin^2\left(\frac{\theta}{2}\right)
  \right\} h(r,\theta) \equiv r\; c(r,\theta),
\end{equation}
and apply the numerical method shown below directly to this
$h(r,\theta)$. Note that the $c$ is multiplied by $r$ in the
right-hand side of (\ref{eq:h_def}) to ensure that $h(r,\theta)=O(r)$
as $r\to 0$; thus, $h=0$ at $r=0$, which is the same condition as for
$c$. Combining (\ref{eq:c_pde1}) and (\ref{eq:h_def}) we obtain a PDE
for $h(r,\theta)$:
\begin{equation}
  \label{eq:h_pde} r^3 \p[2]{h}{r} + \Pe (r-r^3) \p{h}{r} + r
  \p[2]{h}{\theta} + 2\Pe r\sin\theta \p{h}{\theta} + \left\{\Pe
  (r\cos\theta - 1) + \frac{r}{4}\right\}h = 0.\\
\end{equation}
By comparison of (\ref{eq:c_pde1}) and (\ref{eq:h_pde}), we note that
the coefficients of the derivatives are simplified in
(\ref{eq:h_pde}).  Once $h$ is determined from (\ref{eq:h_pde}), $c$
is simply recovered via (\ref{eq:h_def}), and $\sigma(\theta)$ is
obtained as $$ \sigma(\theta) = \left.\p{c}{r}\right|_{(r=1,\theta)} =
\left.\p{h}{r}\right|_{(r=1,\theta)} -\frac12.$$ Fig.~\ref{fig:polar}(c)
shows how the singular behavior of $c$ is mitigated by introducing the new
variable $h$.

\subsection{ Spectral Method }

The numerical differentiations with respect to the variables $r$ and
$\theta$ are carried out by spectral methods~(\cite{trefethen00},
2000). The spatial nodes, the points ($r_j,\theta_k$) where the
function is evaluated numerically, are determined by $$r_j = \frac12
\left( 1-\cos\frac{j\pi}{N_r} \right) \quadtext{and} \theta_k = \pi
\left( 1-\cos\frac{k\pi}{N_{\theta}}\right), $$ where
$j=1,\ldots,\,N_r$ and $k=0,\ldots,\,N_{\theta}$; the nodes have
higher density at the endpoints as shown in the frame of
Fig.~\ref{fig:polar}(b). Once the function values are given at the nodes, the
derivatives at these points are calculated by interpolation via
Chebyshev's polynomials. This procedure is very efficient, as the
error in the spectral method is known to decrease exponentially in the
number of nodes~(\cite{trefethen00}, 2000). We used $N_r=50$ and 
$N_\theta=100$ (practically $N_\theta=50$ exploiting the symmetry
in $\theta$) for all the numerical results appearing in this paper.

\begin{figure}
  \centering \raisebox{0.8in}{(a)}
  \includegraphics[width=0.6\linewidth]{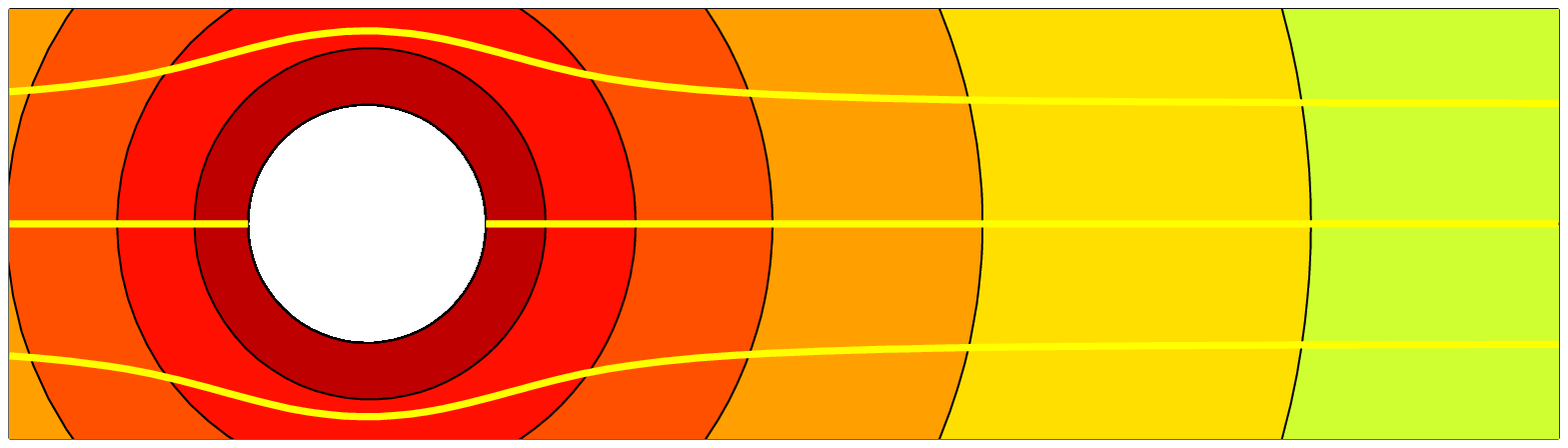}\\
  \raisebox{0.8in}{(b)}
  \includegraphics[width=0.6\linewidth]{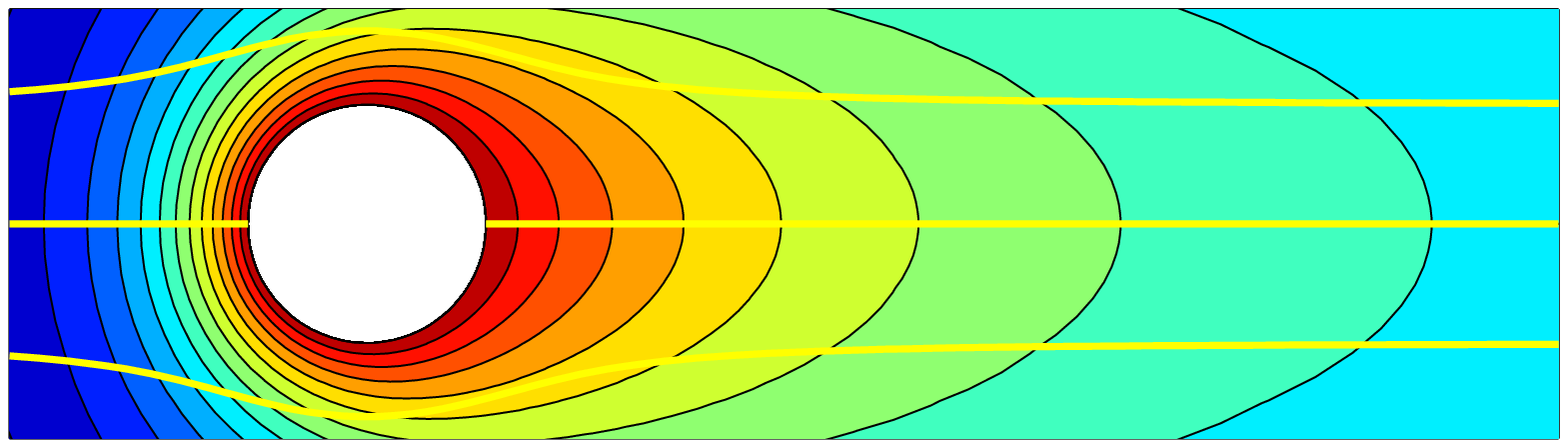}\\
  \raisebox{0.8in}{(c)}
  \includegraphics[width=0.6\linewidth]{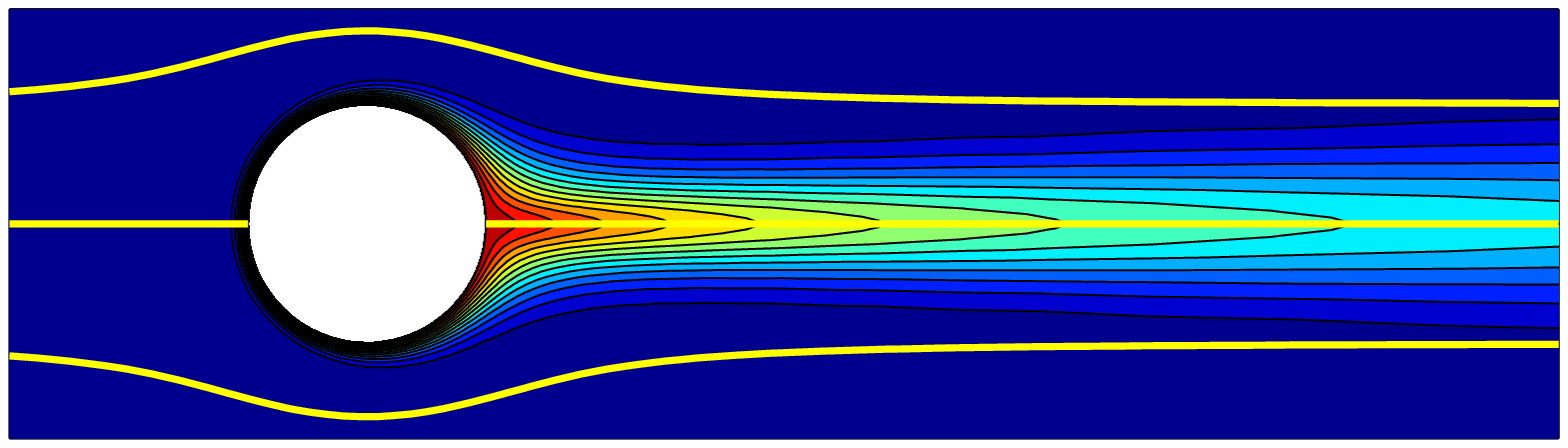}
  \caption{\label{fig:tail} Concentration profiles for adsorption (or
  desorption) around the unit circular disk for (a) $\Pe=0.01$, (b)
  $\Pe=1$, and (c) $\Pe=100$.  }
\end{figure}

Fig.~\ref{fig:tail} shows the concentration profile,
$c(r,\theta)$, for $\Pe$ of different orders of magnitude. As $\Pe$
increases, there is an apparent crossover from a diffusion-dominated
regime (a), where the concentration disturbance looks like  a ``cloud''
extending in all directions, and an advection-dominated regime (c),
where concentration gradients are confined to a narrow boundary layer
which separates into a thin wake downstream.  Understanding the 
crossover regime (b) to (c) is an important part of this paper, revisited below in Sec.~\ref{sec:discussion}, following our analysis of the flux profile.

\begin{figure}
  \centering 
  \parbox[t]{0.57\linewidth}{(a)\\
    \includegraphics[width=\linewidth]{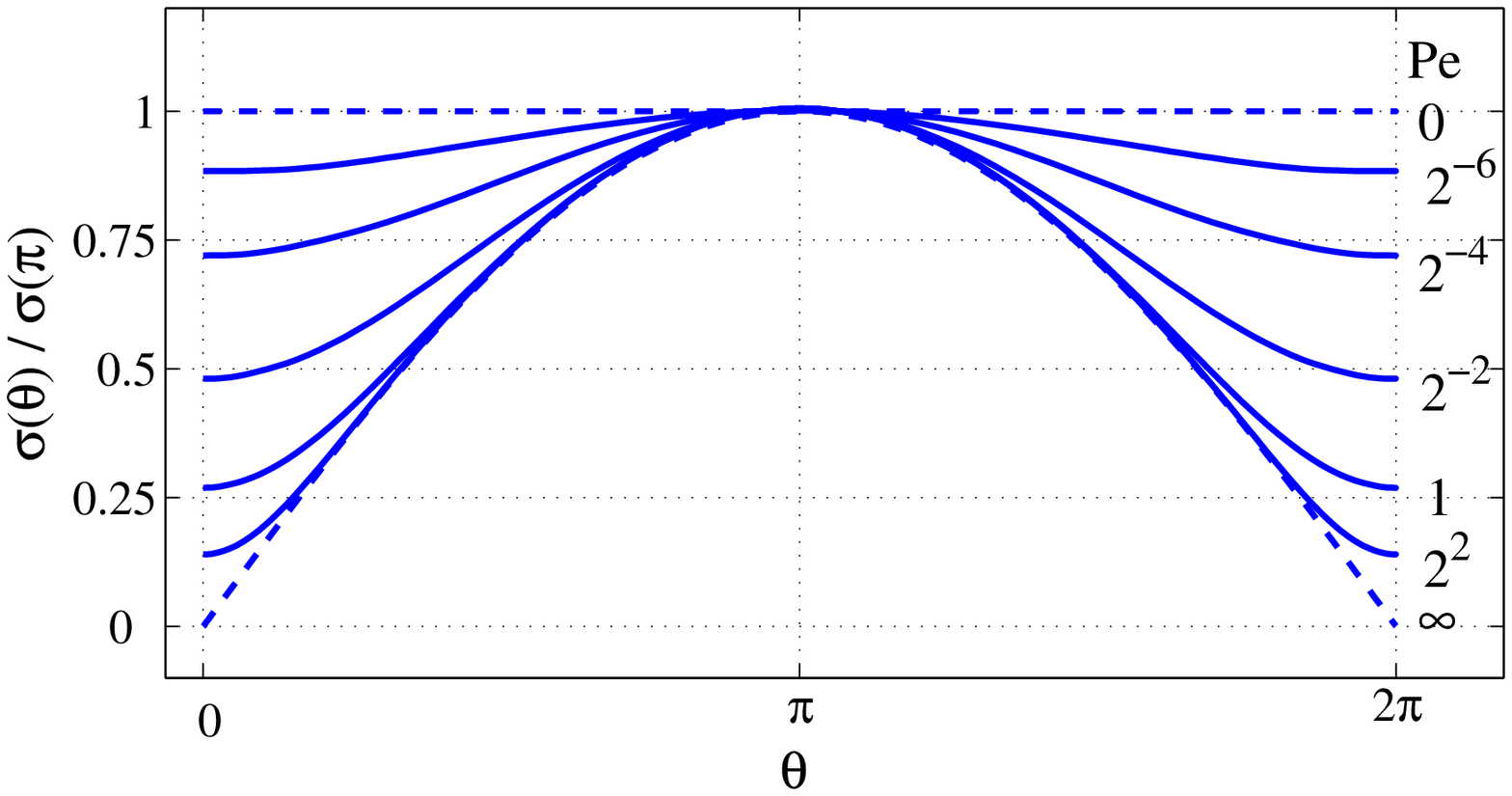}} 
  \hspace{0.03\linewidth}
  \parbox[t]{0.38\linewidth}{(b)\\
    \includegraphics[width=\linewidth]{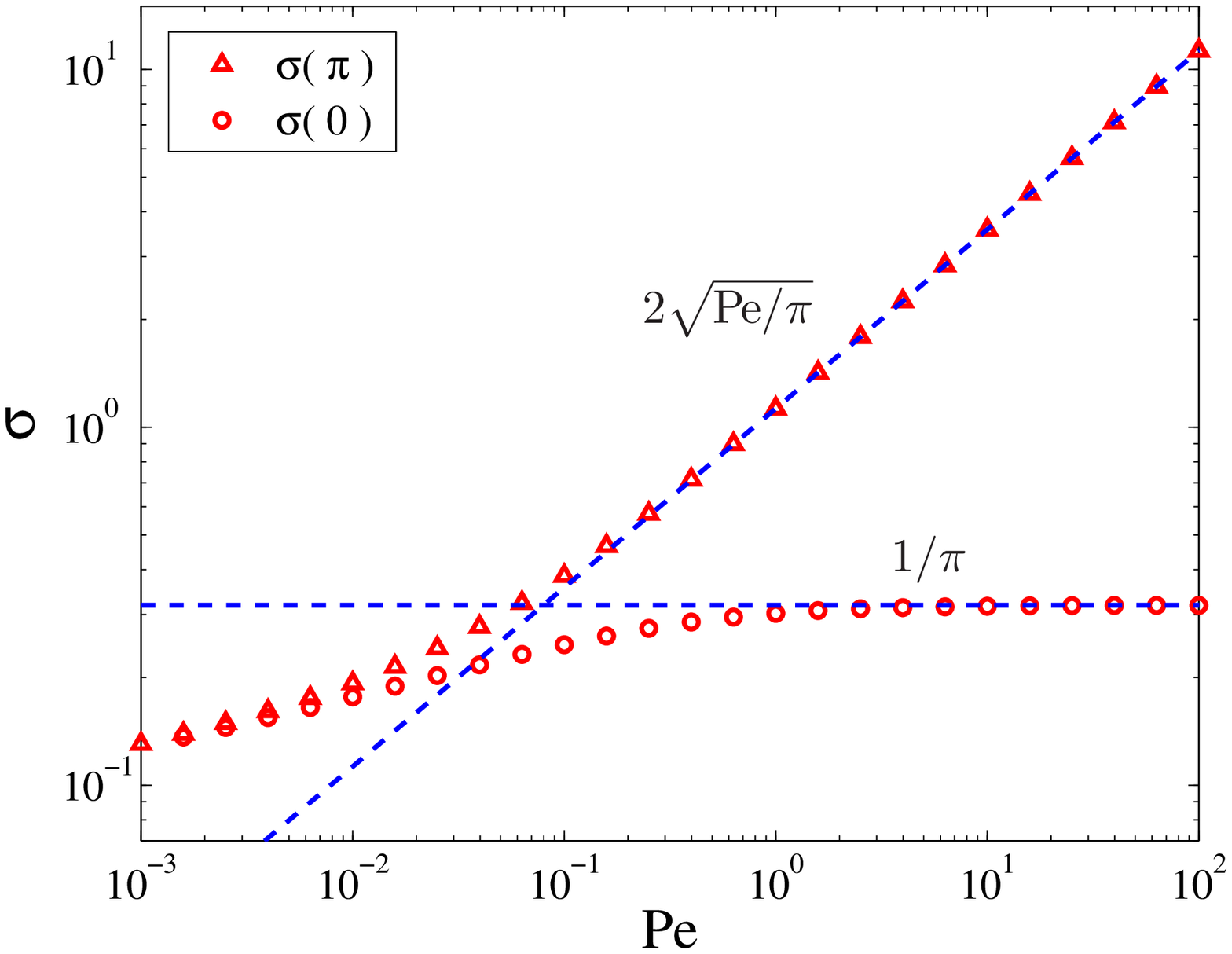}}
  \caption{\label{fig:flux} The flux $\sigma(\theta)$ is plotted around 
    the unit disk for different values of $\Pe$: (a) $\sigma(\theta)/
    \sigma(\pi)$ for $\Pe=0, 2^{-6},\,2^{-4},\,2^{-2},\,1,\,2^2,\,\infty$, 
    and (b) the values $\sigma(\theta=\pi)$ and $\sigma(\theta=0)$ for 
    a range of $\Pe$. }
\end{figure}

Fig.~\ref{fig:flux} shows the flux density on the absorber,
$\sigma(\theta)$, for different values of $\Pe$.  To check the
validity of our numerical method, we compare the result for $\sigma$
with the asymptotic expansion for high $\Pe$ from
Sec.~\ref{sec:highPe_asymptot} below, which can be numerically
evaluated to any order.  For this comparison we used the intermediate
range of P\'eclet numbers $10^{-2}<\Pe<10^2$ for which the series
converge fast enough yet the numerical method is stable.  When $N_r =
50$ and $N_\theta = 100$ nodes are used for the discretization, the
relative error measured as $
||\sigma_\text{num}-\sigma_\text{asym}||\;/\;||\sigma_\text{asym}||$
is of the order of $10^{-5}$ or smaller; here, $\sigma_\text{num}$ and
$\sigma_\text{asym}$ denote the numerical and the asymptotic
solutions, respectively and the norm is defined as $ ||\sigma|| =
\max\{\theta\in[0,2\pi): |\sigma(\theta)|\}$.  

If $\Pe$ lies outside the given intermediate range, care should be 
exercised in using our numerical method as explained in the next
subsection. For such cases, however, the asymptotic, analytical formulae
of Secs.~\ref{sec:highPe_asymptot} and \ref{sec:lowPe_asymptot}
give sufficently accurate solutions. 
The reader is referred to Secs.~\ref{sec:highPe_asymptot} 
and \ref{sec:lowPe_asymptot} for details on the 
formulae and the comparisons with the numerical solution.  

\subsection{ Adaptive Mesh for Very High P\'eclet Numbers }

\begin{figure}
  \centering 
  \raisebox{1.9in}{(a)}\includegraphics[height=2in]{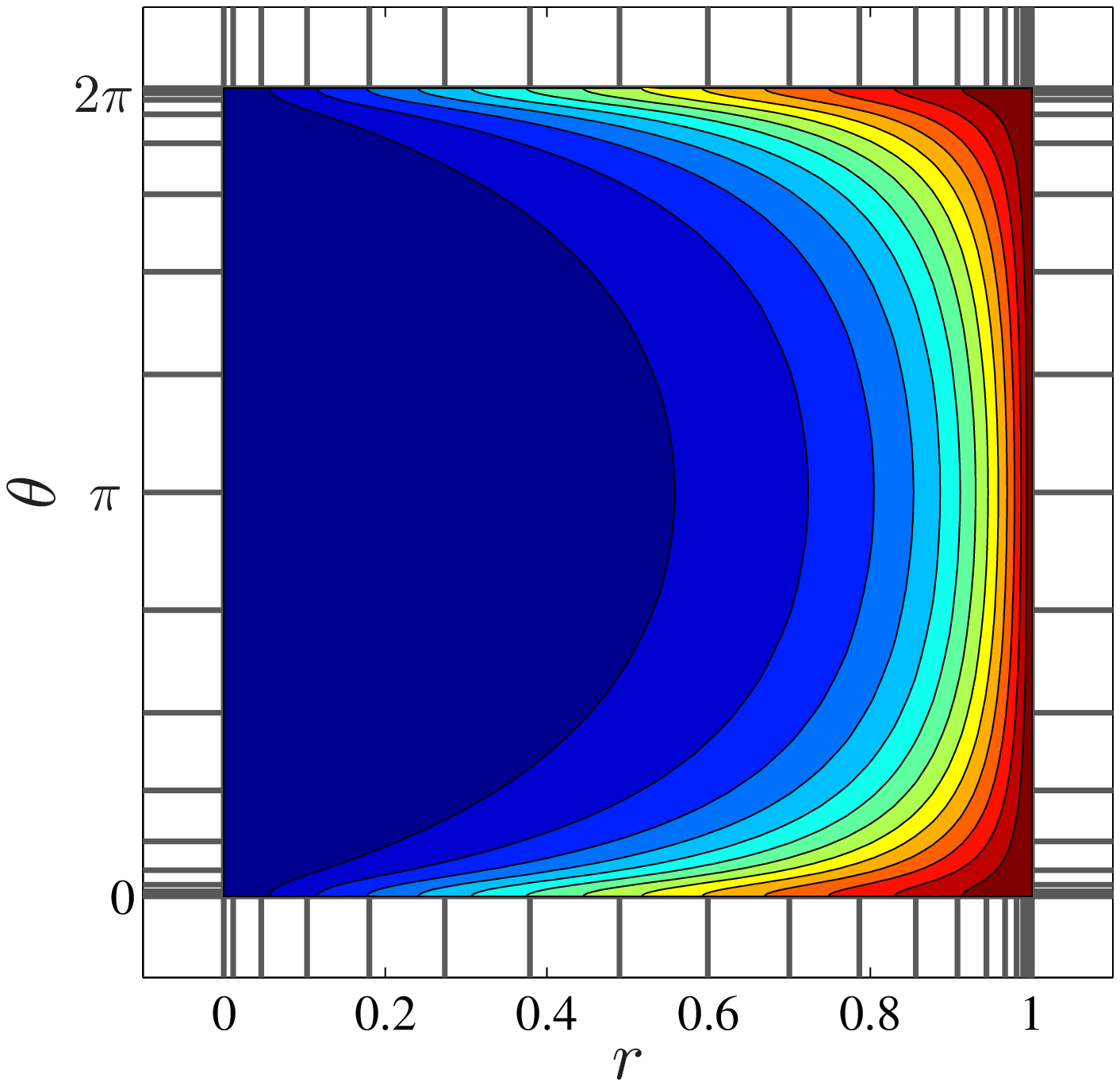} \quad
  \raisebox{1.9in}{(b)}\includegraphics[height=2in]{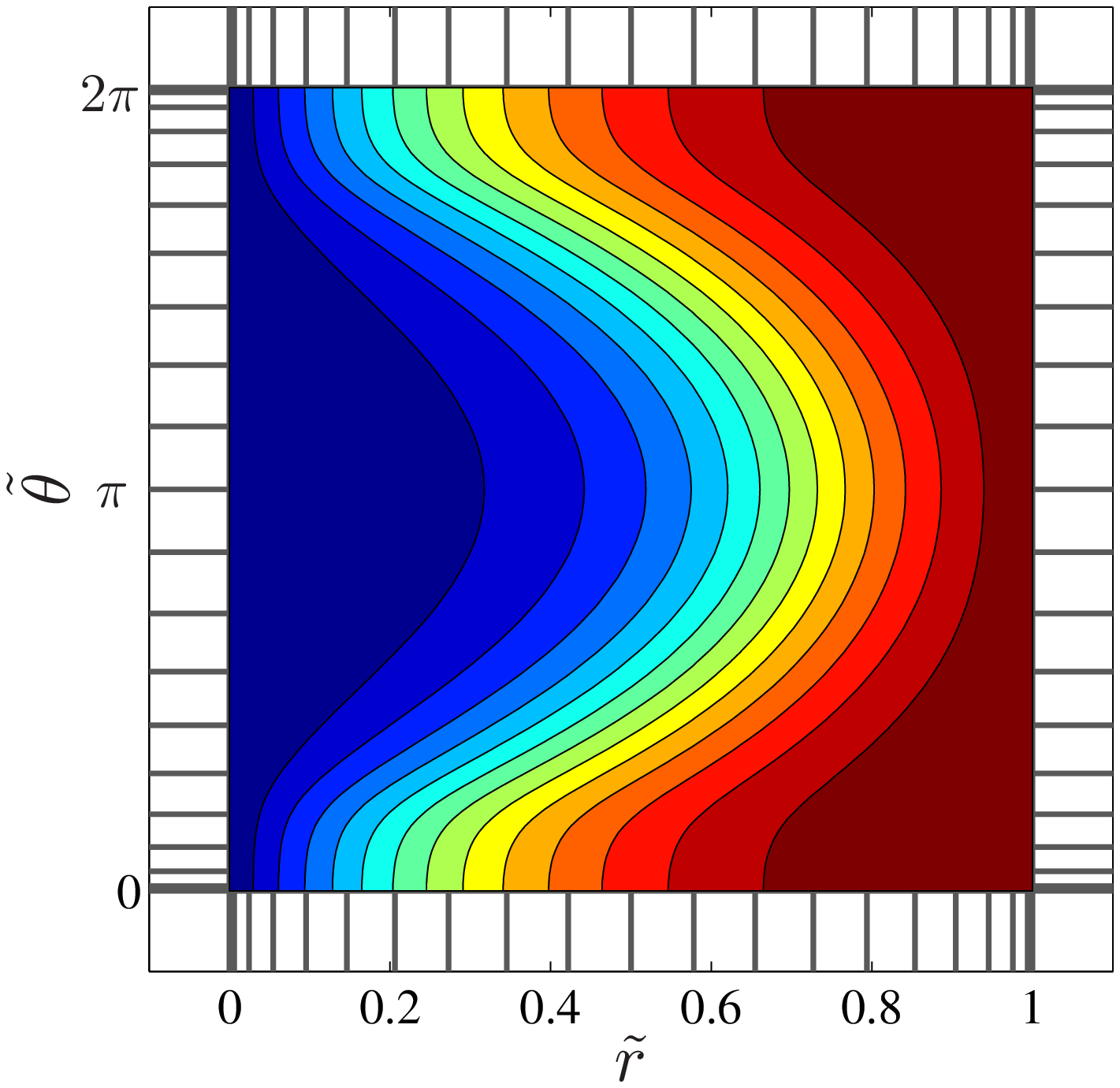}
  \caption{\label{fig:blayer} Contour plots of $h$ for $\Pe=100$ 
    (a) in the $(r,\theta)$ plane, and (b) in the $(\tilde r,\tilde\theta)$
    plane.}
\end{figure}

A feature of the solution $c$ that may undermine the accuracy of our
numerics is the emergence of boundary layers for sufficiently large
$\Pe$. By virtue of (\ref{eq:sig_th_12}) below we
expect that for $\Pe\gg 1$,
\begin{align}
  \left.\p{c}{r}\right|_{(r=1,\theta)} = \sigma(\theta) = 
  \begin{cases} 
    2\sqrt{\Pe/\pi}\;\sin(\theta/2) &\quadtext{when} \theta\gg 
O\bigl(1/\sqrt{\Pe}\bigr) \\
    \quad 1/\pi &\quadtext{when}\theta \le O\bigl(1/\sqrt{\Pe}\bigr).
  \end{cases}
\end{align}
So, $c(r,\theta)$ has boundary layers near $r=1$ and $\theta=0,\,2\pi$ whose
widths are $O(1/\sqrt{\Pe})$, as indicated in Fig.~\ref{fig:blayer}(a). The 
layer at $\theta=0$ corresponds to the ``tail'' shown in Fig.~\ref{fig:tail}(c).
Thus, the numerical method starts to break down when the node spacing 
becomes of the order of $1/\sqrt{\Pe}$.

We next outline a technique to deal with the case of very high $\Pe$ within our
numerical procedure. The idea is to introduce a set of 
independent variables, $\tilde r = \tilde r(r)$ and $\tilde\theta = \tilde\theta(\theta)$,
so that $c$ is a sufficiently smooth function of $\tilde r$ and $\tilde\theta$. 
The following conditions are required for $\tilde r(r)$ and $\tilde\theta(\theta)$:
\begin{gather}
  \tilde r(0) = 0,\; \tilde r(1) = 1,\; \tilde r'(1)=O(\sqrt{\Pe}) \\
  \tilde\theta(0) = 0,\; \tilde\theta(2\pi) = 2\pi,\; \tilde\theta'(0)= O(\sqrt{\Pe}),
\end{gather}
where the prime here denotes differentiation with respect to the
argument.  Once $\tilde r(r)$ and $\tilde\theta(\theta)$ are defined,
we find the PDE for $\tilde h(\tilde r,\tilde\theta)\equiv h(r(\tilde
r),\theta(\tilde\theta))$ from (\ref{eq:h_pde}) by applying the chain
rule for the differentiations with respect to $r$ and $\theta$; for
example,
\begin{equation}
  \p{h}{r} = \frac1{r'(\tilde r)} \p{\tilde h}{\tilde r}, \quad
  \p[2]{h}{r} =
\frac1{r'(\tilde r)^2}\p[2]{\tilde h}{\tilde r} - \frac{r''(\tilde 
r)}{r'(\tilde r)^3}\p{\tilde h}{\tilde r}.\\ 
\end{equation}
We solve the resulting PDE numerically.
A convenient choice for $r(\tilde r)$ and $\theta(\tilde\theta)$ is
\begin{equation}
  \label{eq:adpmesh}
  r(\tilde{r}) = \tilde r + \frac1{\pi} 
\left(1-\frsq1{\Pe}\right)\sin(\pi\tilde{r}),\quad \theta(\tilde\theta) = \tilde\theta - 
\left(1-\frsq1{\Pe}\right) \sin\tilde\theta.
\end{equation}
The advantage of using $\tilde r$ and $\tilde \theta$ instead of $r$
and $\theta$ is illustrated in Fig.~\ref{fig:blayer} for $\Pe=100$.  
The effects of
the boundary layers in $(r,\theta)$ are notably suppressed in the
formulation using $(\tilde r,\tilde\theta)$. In
Secs.~\ref{sec:highPe_asymptot} and \ref{sec:lowPe_asymptot} we
discuss the high- and low-$\Pe$ asymptotics and their comparisons with
the solution determined numerically by the method of this section.

\section{Direct Perturbation Analysis For ``High'' P\'eclet Numbers}
\label{sec:highPe_asymptot}

In this section, we derive an approximate analytical solution to the
integral equation (\ref{eq:ie}) in terms of series expansions produced
via suitable iterations in the coordinate space.  We also obtain
closed-form expressions for the terms of the iteration series as
$\Pe$-dependent multiple integrals. We show that the series is
convergent for $\Pe\ge O(1)$, and that retaining only a few of
its terms produces accurate results even for $\Pe=O(1)$.  
An iterative procedure in the Fourier domain that leads to the same results
is given by \cite{diochoi04} (2004).

\subsection{ Zeroth-order solution via the Wiener-Hopf method }

The starting point of the analysis is the observation that, as
discussed in Sec.~\ref{subsec:similarity}, the solution for the semi-infinite 
strip
$-1<x<\infty$ (in the variable notation of (\ref{eq:ie})) provides
the leading-order term of the high-$\Pe$ asymptotic expansion of the
solution for the finite strip up to a distance $O(1/\sqrt{\Pe})$ from
the endpoint $x=1$.  In order to develop a systematical scheme for
the correction terms, we symmetrize the kernel of (\ref{eq:ie}) and
rescale the independent variable $x$ using $s = \Pe(x+1)$ while we
define $\mu(s)$ by $\sigma(x) = (\sqrt{2}/\pi)\;\Pe e^s
\mu(s)$. (The factor $(\sqrt{2}/\pi)\Pe$ is chosen for
later convenience.) The integral equation (\ref{eq:ie}) thus becomes
\begin{equation}
  \label{eq:ie_h} \int_0^{2\Pe} ds' K_0\left(|s-s'|\right) \mu(s') =
  \frsq{\pi^2 e^{-s}}{2},\quad 0 < s <2\Pe.
\end{equation}
An approximate solution $\mu\sim \mu_0$ that is valid to the leading
order in $\Pe$ is found by taking $\Pe\to\infty$ in the upper limit of
integration in (\ref{eq:ie_h}). The resulting integral equation is
\begin{equation}
  \label{eq:ie_h0} \int_0^\infty ds' K_0\left(|s-s'|\right) \mu_0(s')
  = \frsq{\pi^2 e^{-s}}{2},\quad 0 < s.
\end{equation}

The solution $\mu_0(s)$ is obtained by the Wiener-Hopf technique
(\cite{krein62}, 1962; \cite{noble88}, 1988).  Here we outline the
basic steps of this method, which are also applied to other, similar
integral equations below.  First, we extend the validity of
(\ref{eq:ie_h0}) to $-\infty < s < \infty$ via modifying its right
hand side,
\begin{equation}
  \label{eq:WH1}
  \int_{-\infty}^\infty ds' K_0\left(|s-s'|\right) \mu_0(s')
  = \frsq{\pi^2}{2} \{e^{-s}u(s) + p(s)\} ,\quad -\infty < s < \infty,
\end{equation}
where $\mu_0(s)$ is taken to be zero for $s<0$, $u(s)$ is the Heaviside function
($u(s)=0$ for $s<0$ and $u(s)=1$ for $s>0$),
and $p(s)$ is an unknown function which has non-zero values only for $s< 0$.
Next, we apply the Fourier transform in $s$ to (\ref{eq:WH1}). 
Defining the Fourier transform, $\tilde\mu_0(k)$, of $\mu_0(s)$ as
\begin{equation}
\tilde\mu_0(k)=\int_{-\infty}^{\infty}ds\,\mu_0(s)\,e^{-iks}\quad\mbox{where}
\quad \mu_0(s)=\int_{-\infty}^{\infty}\frac{dk}{2\pi}\,e^{iks}\,\tilde\mu_0(k),\label{eq:mu0-FT}
\end{equation}
(\ref{eq:WH1}) yields
\begin{equation}
  \frsq{\pi\tilde{\mu}_0(k)}{1+k^2} = \frsq{\pi^2}{2} \left[\frac1{1+ik}+\tilde p(k)\right].\label{eq:FT1}
\end{equation}
By simple algebraic manipulations the last equation becomes
\begin{equation}  \label{eq:WH2}
  \frsq{\tilde{\mu}_0(k)}{1+ik} - \frac{\pi}{1+ik} = \frsq{\pi}{2}\left[  
\frac{\sqrt{1-ik}-\sqrt{2}}{1+ik}+\sqrt{1-ik}\,\tilde{p}(k) \right],
\end{equation}
where the left hand side defines a function analytic in the lower half $k$ plane, $\Im k < \varepsilon$ for a 
small positive $\varepsilon$,
and the right hand side defines a function analytic in the upper half $k$ plane, $\Im k > -\varepsilon$; 
each of these functions vanishes as $|k|\to\infty$ in the corresponding half plane.
Thus, the two sides of (\ref{eq:WH2}) together define an entire function of $k$, which is identically
zero by Liouville's theorem (\cite{carrier83}, 1983). It follows that in the region of overlap,
$|\Im k|< \varepsilon$, the solution is $\tilde\mu_0(k)=\pi (1+ik)^{-1/2}$. Inversion of this formula yields
\begin{equation}
  \label{eq:mu0} 
  \mu_0(s) = \sqrt{\frac{\pi}{s}} e^{-s},\quad   0<s<\infty.
\end{equation}

\subsection{Leading-order Uniformly Accurate Approximation }
\label{subsec:high_zero}

The deviation of $\mu_0(s)$ in (\ref{eq:mu0}) from the actual solution 
$\mu(s)$ of the finite strip is interpreted as due to the effect of a fictitious, 
``misplaced'' flux source lying in $2\Pe< s$, which is present in
(\ref{eq:ie_h0}). Therefore a correction term must be found for
$\mu_0(s)$ by placing a ``correction source'' on the
original strip, $0<s<2\Pe$, to compensate for the effect of the
misplaced source.  \cite{diochoi04} (2004) further develop this approach and
place it on a firm mathematical ground using Fourier transforms and a generalization
of the Wiener-Hopf method.  We proceed to calculate the correction to
$\mu_0$ iteratively.  Accordingly, the solution $\mu(s)$ is sought in terms of the series
\begin{equation} 
\label{eq:mu-ser}
\mu = \mu_0 + \mu_1 + \mu_2 + \mu_3 +\;\ldots\;+\mu_n+ \; \ldots,
\end{equation}
where each term, $\mu_n(s)$, corresponds to suitable source corrections as 
described below.

We consider a half-line as the domain of the correction $\mu_1(s)$, as
we did to obtain $\mu_0$, but in the region $-\infty<s<2\Pe$ instead
of the region $0 < s<\infty$. The term $\mu_1$ is determined so that
its effect compensates for the integrated effect of $\mu_0(s)$ in the
region $2\Pe<s$.  Hence, the correction $\mu_1(s)$ satisfies
\begin{equation}
  \label{eq:ie_h1} \int_{-\infty}^{2\Pe}ds' K_0(|s-s'|) \mu_1(s') =
  \int_{2\Pe}^\infty ds' K_0(|s-s'|) \mu_0(s'),\quad s<2\Pe,
\end{equation}
where the right-hand side is known. The next-order corrections can be
formulated and interpreted in a similar way; the correction $\mu_n$
compensates for the effect of the misplaced source corresponding to
$\mu_{n-1}$, where $\mu_{n-1}$ and $\mu_n$ are defined in half-lines
that together cover the entire real axis and overlap only in the
region of the finite strip. In general, $\mu_{n\ge 1}$ satisfy the
recursion relations
\begin{subequations}
\label{eq:ie_mu_n}
\begin{gather}
  \int_0^\infty K_0(|s-s'|) \mu_{n=2k} (s') ds' = \int_{-\infty}^{0} K_0(|s-s'|) \mu_{n-1}(s') ds',\\
  \int_0^\infty K_0(|v-v'|) \mu_{n=2k+1} (2\Pe-v') dv' = \int_{-\infty}^{0} K_0(|v-v'|) \mu_{n-1}(2\Pe-v') dv',
\end{gather}
\end{subequations}
where we made the change of variable from $s$ to $v = 2\Pe-s$ so that
the integral equation for $\mu_{n=2k+1}$ has the same form
as the one for $\mu_{n=2k}$. The variables $s$ and $v$ are both positive
($s>0$ and $v>0$), the left endpoint ($x=-1$) of the strip corresponds
to $s=0$ and the right endpoint ($x=1$) corresponds to $v=0$.

The various $\mu_n$ can be obtained successively, order by order, by
applying the Wiener-Hopf method (\cite{krein62}, 1962; \cite{noble88},
1988) directly to (\ref{eq:ie_mu_n}), but the procedure becomes
increasingly cumbersome with $n$.  Instead, we propose a systematic
procedure that facilitates the derivation of a closed-form expression
for each $\mu_n$. For this purpose, we introduce an operator, $\OP$,
that relates $\mu_{n-1}$ and $\mu_n$ by $\OP[\mu_{n-1}] \equiv
\mu_n$. By (\ref{eq:ie_mu_n}) $\OP$ is linear.  In order to obtain
$\mu_1$, we notice that the leading-order solution $\mu_0(s)$ can be
represented as an integral over a variable, $t_0$,
\begin{equation}
\label{eq:mu0_int}
\mu_0 = \sqrt{\frac{\pi}{s}} e^{-s} = \int_{-\infty}^{\infty}dt_{0}\; e^{-s(1+t_0^2)}.
\end{equation} 
Then $\OP$ acts on $\mu_0$ to yield $\mu_1$ as~\footnote{In the
remaining integrals of this section the range of integration is
understood to be from $-\infty$ to $\infty$ unless it is stated
otherwise.}
\begin{equation}
  \label{eq:mu1_OP_int}
  \mu_1 = \int dt_{0} \;\mathcal{L}[e^{-s(1+t_0^2)}],
\end{equation}
where the order of $\OP$ and integration is safely interchanged. The
advantage of using the $t_0$-representation is that $e^{-s(1+t_0^2)}$,
as a function of $s$, has a Fourier transform simpler than the Fourier
transform of $\mu_0(s)$ itself. The function $\OP[e^{-s(1+t_0^2)}]$ is
found by the Wiener-Hopf method~(\cite{krein62}, 1962) as described in 
Sec.~\ref{subsec:high_zero}, and $\mu_1$ follows by (\ref{eq:mu1_OP_int}):
\begin{align}
  \label{eq:op_mu0}
  \OP[e^{-s(1+t_0^2)}] &= \frac{e^{-2\Pe (1+t_0^2)}}{\pi\sqrt{2+t_0^2}} 
  \left[ \sqrt\frac{\pi}{v}\;e^{-v} - \pi\sqrt{2+t_0^2}\; e^{v(1+t_0^2)}\erfc\sqrt{v(2+t_0^2)} \right],\\
  \label{eq:mu1}
  \mu_1(v) &= K_0(2\Pe)\frsq{e^{-v}}{\pi v} - \intdt{0}\; e^{-(2\Pe-v)(1+t_0^2)}\erfc\sqrt{v(2+t_0^2)}.
\end{align}

Because $\mu_0$ and $\mu_1$ are accurate arbitrarily close to the left
edge ($s=0$) and the right edge ($v=0$) of the strip, respectively,
$\mu_0+\mu_1$ yields a leading-order approximation for $\mu$ as
$\Pe\to\infty$ valid over the entire finite strip.  The corresponding
approximation for the flux $\sigma=(\sqrt{2}/\pi)\Pe e^s\mu(s)$ is
$\sigma\sim \sigma_1+\sigma_2$, which is given by
\begin{equation} 
\label{eq:sig_x_12}
  \begin{split}
    \sigma(x) \sim \sigma^{(\rm hi)}(x) = 2\sqrt{\frac{\Pe}{\pi}} &\left\{\frsq1{2(1+x)} + \frac{K_0(2\Pe)e^{2\Pe x}}{\pi\sqrt{2(1-x)}} \right. \\
    & \left. - \intdt[\tau]{}\; \frsq{e^{-(1+x)\tau^2}}{2\pi}\erfc\sqrt{(2\Pe+\tau^2)(1-x)} \right\}
  \end{split}
\end{equation}
for the geometry of the finite strip, and
\begin{equation} 
\label{eq:sig_th_12}
  \begin{split}
    \sigma(\theta) \sim \sigma^{(\rm hi)}(\theta) = 2\sqrt\frac{\Pe}{\pi} &\left\{|\sin\frac{\theta}{2}| + \frac1{\pi} K_0(2\Pe) e^{2\Pe \cos\theta}  |\cos\frac{\theta}{2}| \right.\\
    & \left. - \frsq{|\sin\theta|}{2\pi} \intdt[\tau]{}\;e^{-(1+\cos\theta)\tau^2}\erfc\sqrt{(2\Pe+\tau^2)(1-\cos\theta)} \right\}
  \end{split}
\end{equation}
for the geometry of the unit circular disk. In the last formula we
changed the variable to $\tau=\sqrt{\Pe}\;t_0$.  
An expansion similar to (\ref{eq:sig_x_12}) has been obtained 
by~\cite{chugunov86} (1986) and \cite{kornev88} (1988) in the physical 
context of artificial freezing
\footnote{
We could not verify whether (\ref{eq:sig_x_12}) is equivalent to
(13) in \cite{chugunov86} (1986) or (3.7) in \cite{kornev88} (1988).};
these authors did not use the operator $\OP$ in their method and apparently did not
obtain higher-order terms. An elaborate mathematical procedure for asymptotic
solutions to the relevant class of integral equations on the basis of 
kernel approximations is described in~(\cite{aleksandrovbelokon68},
1968) and (\cite{aleksandrov99}, 1999).  We note that $\sigma_0(x)$ and
$\sigma_1(x)$ are singular at $x=-1$ and $x=1$, respectively, the
edges of the finite strip; these singularities are properly removed
after the strip is mapped onto the unit disk.

\subsection{ Exact Higher-Order Terms }
\label{subsec:high_high}

In an effort to obtain further insight into the nature of the solution
$\sigma$ of (\ref{eq:ie}), we next derive exact, closed-form
expressions for $\mu_n$ for all $n$ in terms of iterated, multiple
integrals.  For this purpose, we exploit the $\OP$ operator introduced
above.  We observe that (\ref{eq:op_mu0}) has the integral
representation
\begin{equation}
  \label{eq:op_mu0_int}
  \OP[e^{-s(1+t_0^2)}] = \intdt{1} \frac{e^{-2\Pe(1+t_0^2)}}{\pi\sqrt{2+t_0^2}}
  \; \frac{t_1^2}{2+t_0^2+t_1^2} \; e^{-v(1+t_1^2)},
\end{equation}
where the last factor in the integrand has the same form as the
$e^{-s(1+t_0^2)}$ term, on which the $\OP$ acts in
(\ref{eq:mu1_OP_int}), with the $s$ being replaced by $v$ and the
$t_0$ being replaced by $t_1$. It follows that the $\mu_n$ is
expressed as the $n$-th power of $\OP$ acting on a term of the form
$e^{-s(1+t^2)}$, $\OP^n[e^{-s(1+t^2)}]$. Thus, by induction, 
$\mu_n$ is expressed as
an iterated, multiple integral of the independent variable $s$ or $v$,
\begin{equation}
  \label{eq:mu_n} \mu_n(u) = e^{-2n\Pe} \intdt{0} \intdt{1}(Q_0 R_1)
  \intdt{2}(Q_1 R_2)\cdots \intdt{n} (Q_{n-1}R_n) e^{-u(1+t_n^2)},
\end{equation}
where $u\equiv s$ for even $n$ and $u\equiv v$ for odd $n$, and $Q_n$
and $R_n$ are defined by
\begin{equation} 
  \label{eq:QR}
  Q_n \equiv \frac{e^{-2\Pe t_n^2}}{\pi \sqrt{2+t_n^2}},\qquad
  R_n \equiv \frac{t_n^2}{2+t_{n-1}^2+t_n^2}.
\end{equation}

As indicated from (\ref{eq:op_mu0}), $\mu_n(u)$ has has the singularity 
$u^{-1/2}$ at $u=0$ which comes from the last integral of (\ref{eq:mu_n}).
Thus we can single out the singular behavior as $\mu_n(u) = \sqrt{\pi/u}\;
 e^{-2n\Pe-u}\; F_n(u)$ and the singular-free part, $F_n$, is given by
\begin{align}
  \label{eq:Fn-def}
  \begin{split}
    F_n(u) &= \intdt{0} \intdt{1}(Q_0 R_1) \intdt{2}(Q_1 R_2)\cdots \\
    \cdots \intdt{n-1} & (Q_{n-2}R_{n-1}) \left[Q_{n-1} - \sqrt\frac{u}{\pi}\;
      e^{2u - (2\Pe-u) t_{n-1}^2}\erfc\sqrt{u(2+t_{n-1}^2)}\right],
  \end{split}
\end{align}
where (\ref{eq:op_mu0}) and (\ref{eq:op_mu0_int}) are used to evaluate 
the last integral.

Thus, the $n$th term for the flux on the strip, $\sigma_n(x)$, is given by
\begin{subequations}
  \label{eq:sig_x}
  \begin{align}
    \begin{split}
      \sigma_{n=2k}(x) &= 2\sqrt\frac{\Pe}{\pi}\; 
      \frsq{e^{-2n\Pe}}{2(1+x)}\; F_n(\,\Pe(1+x)\,)
    \end{split}\\
    \begin{split}
      \sigma_{n=2k+1}(x) &= 2\sqrt\frac{\Pe}{\pi}\; 
      \frsq{e^{-2(n-x)\Pe}}{2(1+x)}\; F_n(\,\Pe(1-x)\,).
    \end{split}
  \end{align}
\end{subequations}
On the unit circle $\sigma_n(\theta)$ is free of singularities in $\theta$ and is expressed as
\begin{subequations}
\label{eq:sig_th}
\begin{align}
  \begin{split}
    \sigma_{n=2k}(\theta) &= 2\sqrt\frac{\Pe}{\pi} 
    e^{-2n\Pe} |\sin\frac{\theta}{2}|\; F_n(\,\Pe(1+\cos\theta)\,)\,,
  \end{split}\\
  \begin{split}
    \sigma_{n=2k+1}(\theta) &= 2\sqrt\frac{\Pe}{\pi} 
    e^{-2(n-\cos\theta)\Pe} |\cos\frac{\theta}{2}|\; F_n(\,\Pe(1-\cos\theta)\,)\,.
  \end{split}
\end{align}
\end{subequations}
It has not been possible to evaluate $\sigma_n$ from (\ref{eq:sig_th})
in simple closed form, except for $n=1,\; 2$, as described by
(\ref{eq:sig_th_12}).  However, the numerical integrations over the
variables $t_j$ ($j=1,\,2,\,\ldots\,n$) for each $\sigma_n$ can be
carried out efficiently by using recursion.  

\begin{figure}
  \centering
  \parbox[t]{0.48\linewidth}{(a)\\
    \includegraphics[width=\linewidth]{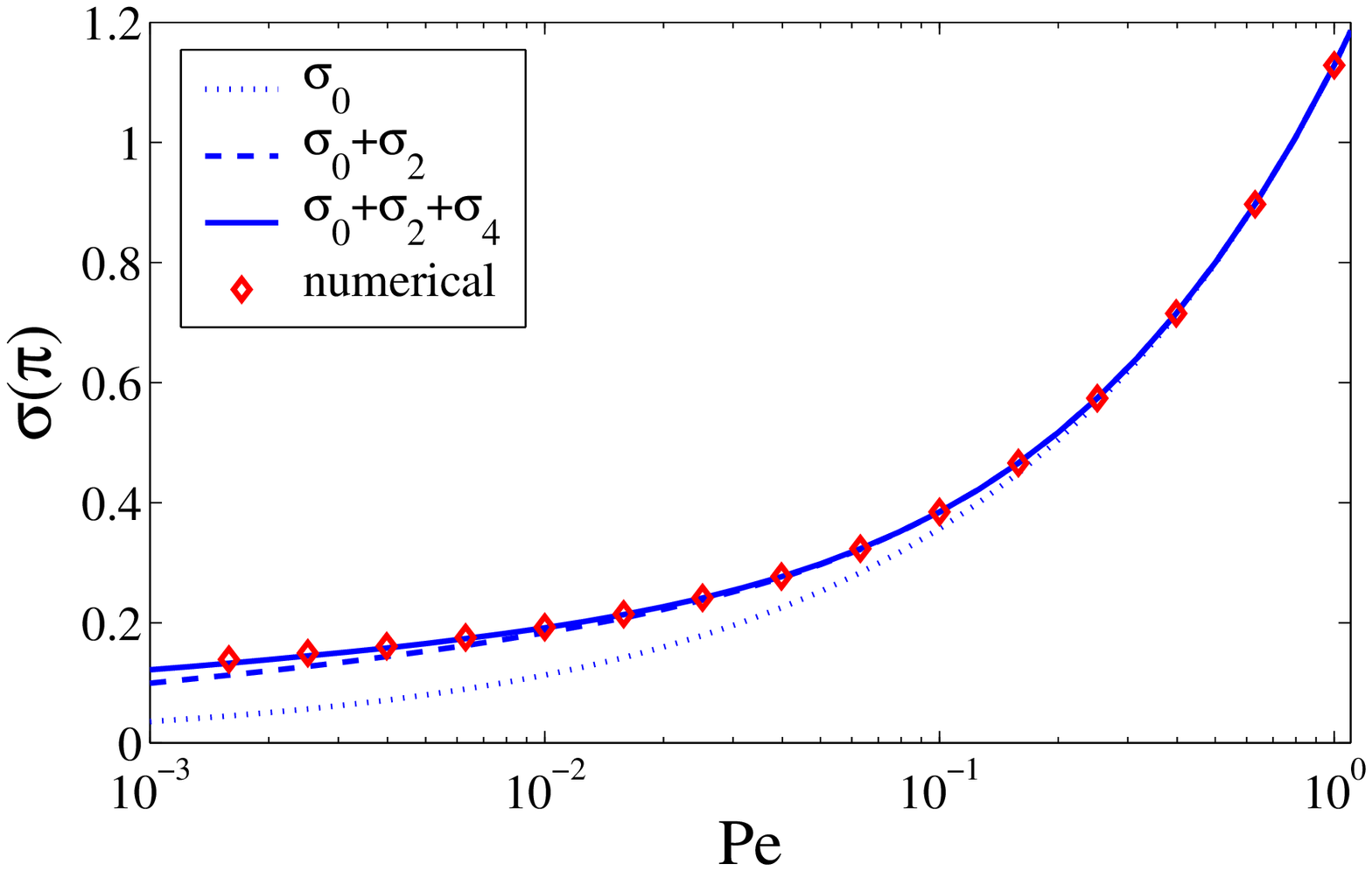}} \hspace{0.02\linewidth}
  \parbox[t]{0.48\linewidth}{(b)\\
    \includegraphics[width=\linewidth]{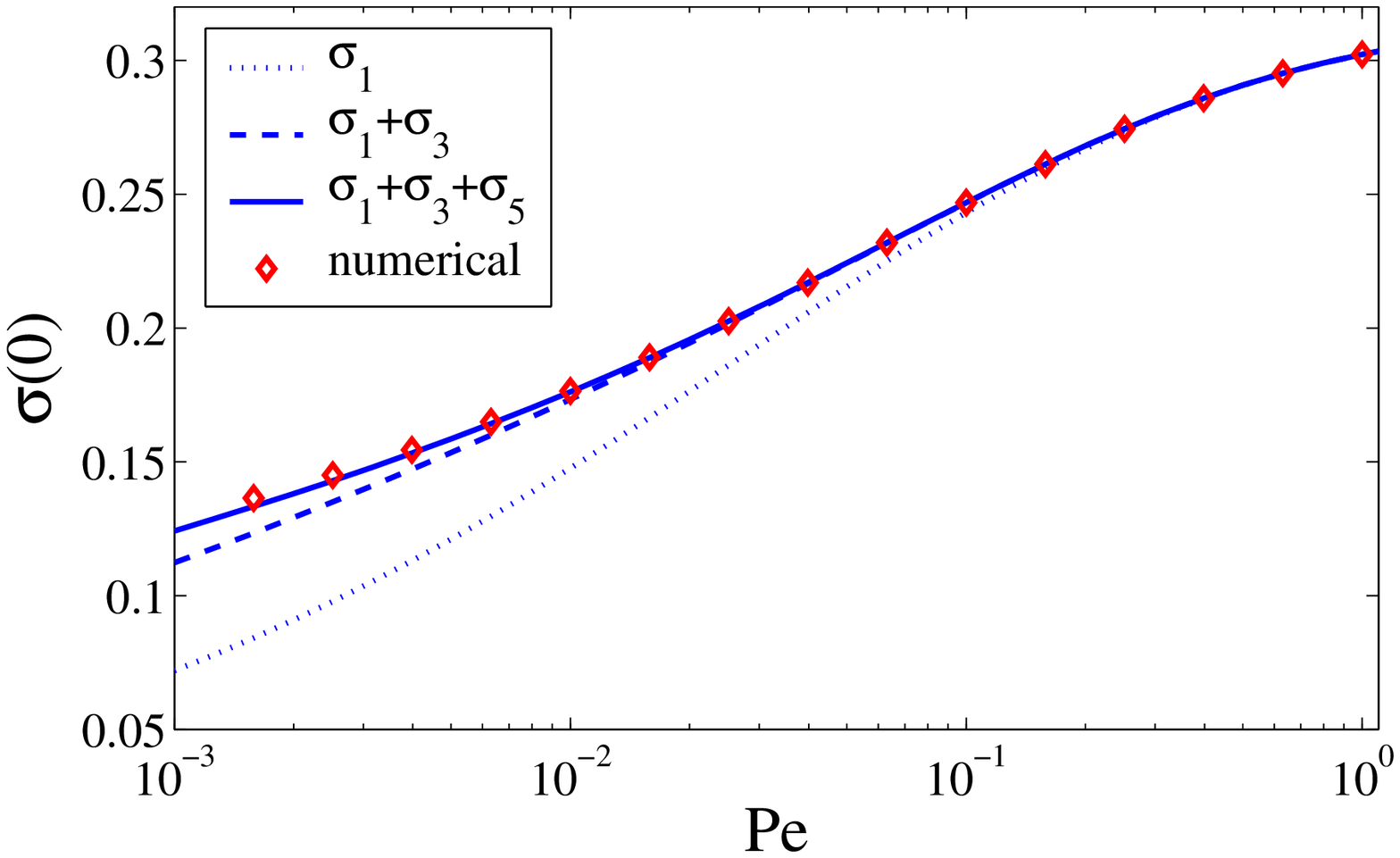}}
  \caption{\label{fig:CompHigh} Asymptotic approximations for high
    $\Pe$ compared with our numerical solution: (a) Upstream flux,
    $\sigma(\theta=\pi;\Pe)$ and (b) downstream flux, $\sigma(\theta=0;\Pe)$.
    The integrals in the asymptotic corrections, $\sigma_n$ ($1\le n\le 5$), 
    were performed numerically.}
\end{figure}

We verify that the sum for the flux,
$\sigma\sim\sigma_0+\sigma_1+\ldots+\sigma_n$, calculated for finite
$n$ via the numerical integration of (\ref{eq:sig_th}), indeed
approaches the numerical solution of Sec.~\ref{sec:numerical_soln}.
In Fig.~\ref{fig:CompHigh}, we show a comparison of the numerical solution 
of Sec.~\ref{sec:numerical_soln} for
$\sigma(0)$ and $\sigma(\pi)$ with formula (\ref{eq:sig_th}) for
different values of $n$.  Because $\sigma_{n=2k}(\theta)$ and
$\sigma_{n=2k+1}(\theta)$ vanish at $\theta=0$ and $\theta=\pi$,
respectively, the term $\sigma_{n=2k}(\theta)$ affects only
$\sigma(\pi)$ whereas the term $\sigma_{n=2k+1}(\theta)$ affects only
$\sigma(0)$.  

Remarkably, with only a few terms, our approximation is uniformly
accurate down to values of $\Pe$ of order $10^{-2}$ or lower, which could
hardly be called ``high'', while some correction terms $\mu_{n\ge 2}$ may not
be small. This behavior suggests that there may be an
intermediate region of overlap between asymptotic approximations for
high and low $\Pe$.  Indeed, by combining such approximations below,  we will
construct a very accurate approximation for all $\theta$ and all $\Pe$.

The closed-form expression of $\sigma_n$ also serves as another 
``numerical method'' for high $\Pe$. The multiple integrals in 
(\ref{eq:Fn-def}) can be numerically evaluated in a recursive way;
the intermediate calculation steps for $F_k$ recur in the calculation
for all $F_{n>k}$. Thus, the computational cost for $\sum_{k=0}^n\sigma_k$
is the same as that for $\sigma_n$, which scales linearly with $n$.

\subsection{ Convergence of the Asymptotic Series  }

We next discuss the convergence properties of the asymptotic series
$\sum_n\sigma_n(\theta)$ for $\sigma$ on the unit circle. Because the
maximum of $\sigma_n(\theta)$ occurs at $\theta=\pi$ for $n=2k$ and
at $\theta=0$ for $n=2k+1$, it is readily shown that
\begin{equation}
\label{eq:sigma-norms}
||\sigma_0|| = 2\sqrt{\frac{\Pe}{\pi}} \quadtext{and}
\left\{\begin{matrix}
||\sigma_{n=2k}|| =  e^{-4\Pe k}\; ||F_{2k}||\;  ||\sigma_0||\\
||\sigma_{n=2k+1}|| = e^{-4\Pe k}\; ||F_{2k+1}||\; ||\sigma_0||,
\end{matrix}\right.
\end{equation}
From (\ref{eq:Fn-def}), $||F_n||$ is simply $F_n(u=0)$.
In particular, for $n=1,\,2$, $||F_n||$ are evaluated in simpler forms:
\begin{equation}
  \label{eq:F01}
  ||F_1|| = \frac{e^{2\Pe}}{\pi}\,K_0(2\Pe), \quad 
  ||F_2||  = \frac{e^{4\Pe}}{\pi^2} \left\{\frac{K_0(2\Pe)^2}{2} - \int_{2\Pe}^\infty dt\; K_0(t)^2 \right\}.
\end{equation}

We have not been able to further simplify the expressions for $||F_{n>2}||$.
We now show that each 
$||F_n||$ is bounded by a function of $\Pe$ that ensures convergence of the series $\sum_n \sigma_n$ for $\Pe\ge O(1)$.
By noting that $Q_n < e^{-2\Pe t_n^2}/\sqrt{2}$ and $R_n < 1/\pi$, we find
\begin{equation}
\label{eq:Fn-bound}
||F_n|| < \pi^{-n/2}\,(4\Pe)^{-n/2},
\end{equation}
for any $\Pe >0$. Hence, by (\ref{eq:sigma-norms}) the series $\sum_n\sigma_n$ is characterized by geometric convergence 
in the parameter $\Pe[^{-1/2}]$ for $\Pe\ge O(1)$.
Finally, for large $\Pe$ and fixed $n$ the asymptotic 
behavior of $||F_n||$ with respect to $\Pe$ is obtained via scaling the original variables 
$t_k$ as $\tau_k = t_k\,\sqrt{2\Pe}$ ($k=0,\ldots,n-1$),
\begin{equation}
\label{eq:Fn-asympt}
||F_{n\ge 1}|| \sim \frac{1}{2\sqrt{\pi\Pe}} \left(\frac1{16\sqrt{\pi}\Pe[^{3/2}] } \right)^{n-1}\quadtext{as}\Pe \to \infty.
\end{equation}Formulae (\ref{eq:Fn-bound}) and (\ref{eq:Fn-asympt}) indicate that the
series $\sum_n \sigma_n$ converges geometrically for a wide range of $\Pe$.
 
\begin{figure}
  \centering
  \parbox[t]{0.48\linewidth}{(a)\\
    \includegraphics[width=\linewidth]{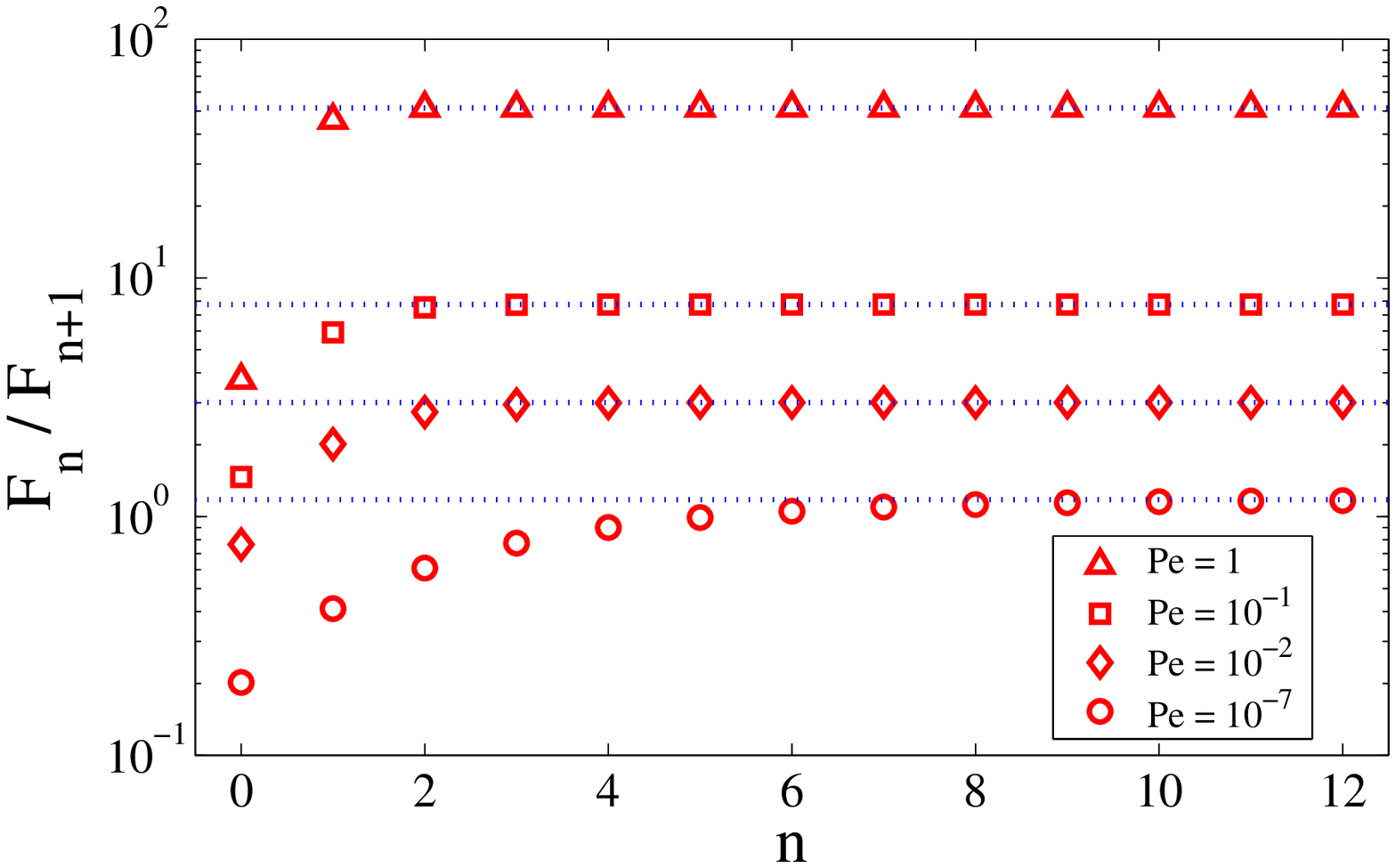}} \hspace{0.02\linewidth}
  \parbox[t]{0.48\linewidth}{(b)\\
    \includegraphics[width=\linewidth]{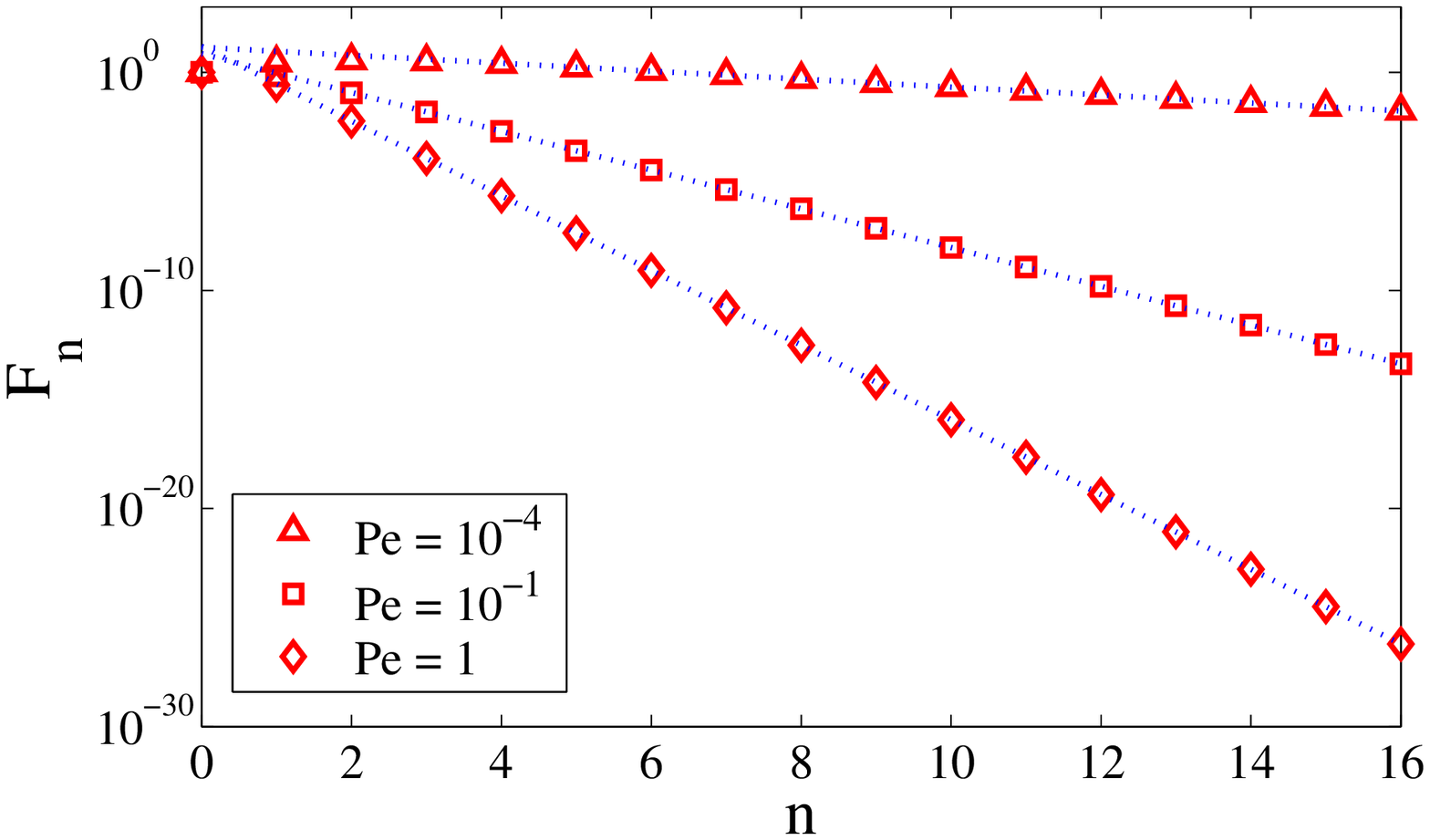}}\\
  \parbox[t]{0.48\linewidth}{(c)\\
    \includegraphics[width=\linewidth]{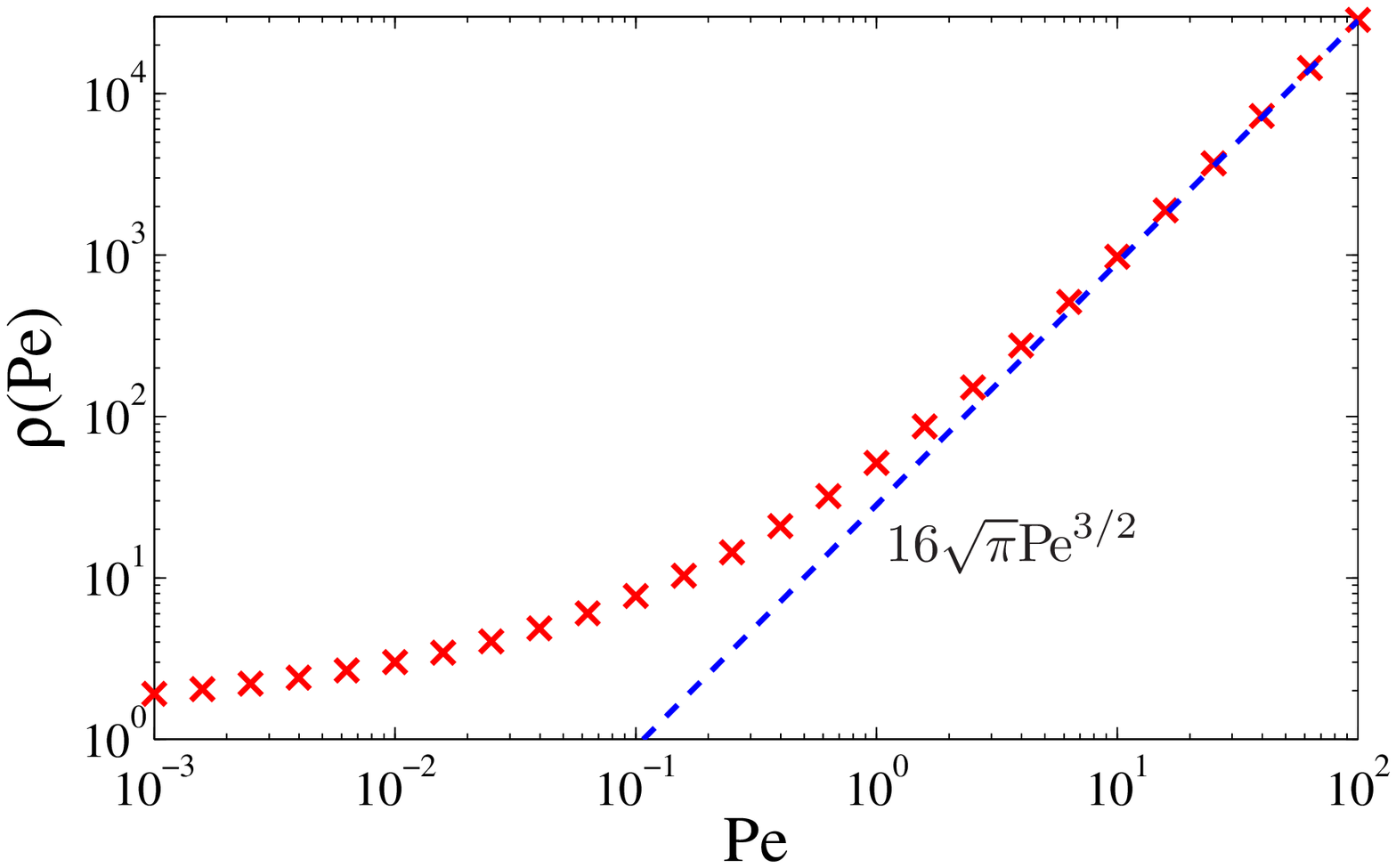}} \hspace{0.02\linewidth}
  \parbox[t]{0.48\linewidth}{(d)\\
    \includegraphics[width=\linewidth]{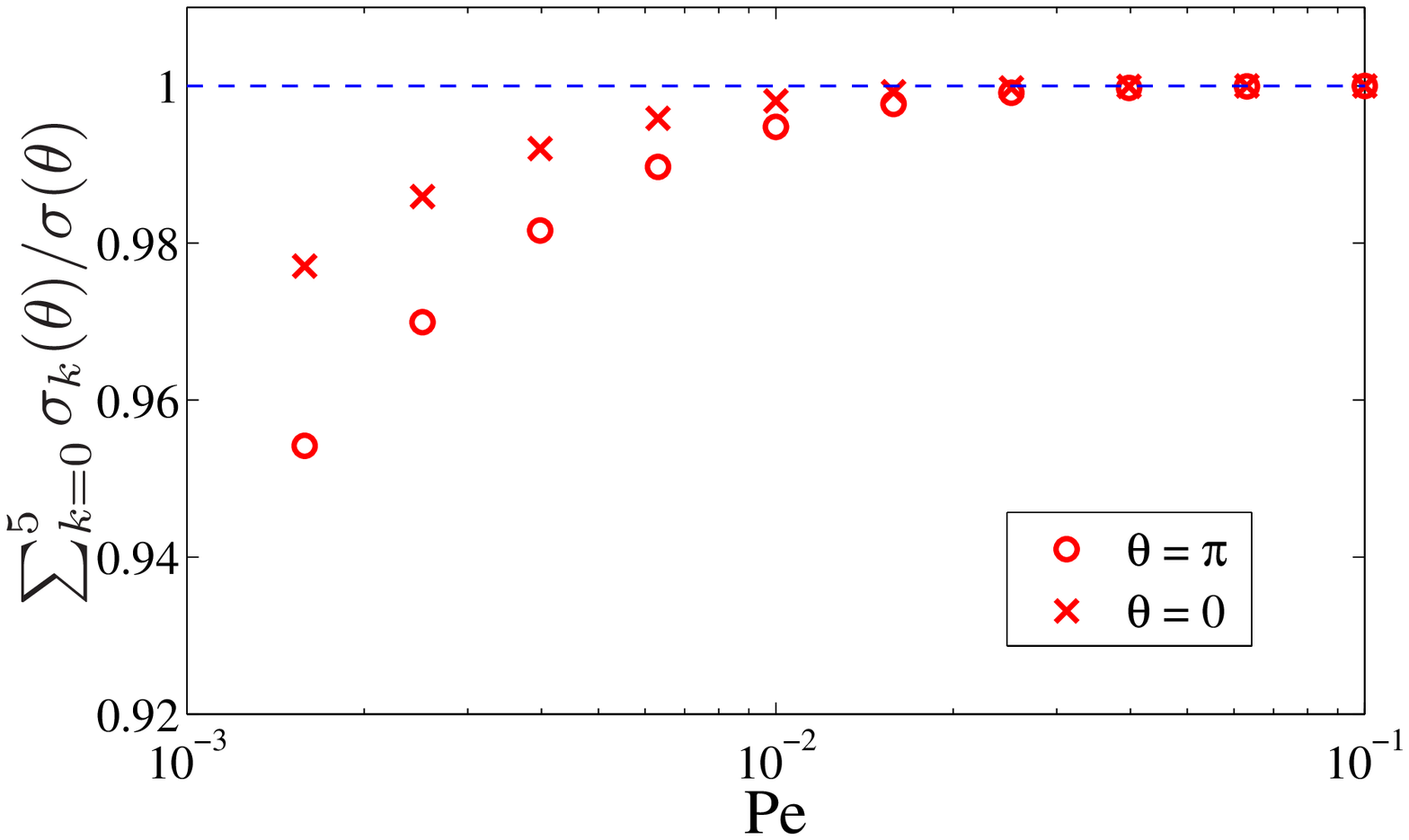}}
  \caption{\label{fig:rho} Numerical evidence for the convergence of the 
    iteration series $\sum_n \sigma_n\sim \sigma$. (a) The ratio $||F_n||/||F_{n+1}||$
    as a function of $n$ for different values of $\Pe$. The convergence of 
    $\sum_n\sigma_n$ is guaranteed if $||F_n||/||F_{n+1}||>1$ by virtue of 
    (\ref{eq:sigma-norms}). (b) The term $||F_n||$ as a function of $n$ for 
    different values of $\Pe$. (c) The decay rate $\rho(\Pe)$ of $||F_n||$ as a 
    function of $\Pe$, where $\rho=\lim_{n\to\infty}(||F_n||/||F_{n+1}||)$.
    The asymptotic behavior (\ref{eq:Fn-asympt}) is shown to be attained for 
    $\Pe> 10$. (d) The ratio of $\sum_{k=1}^5\sigma_k(\theta)$ to 
    $\sigma(\theta)$ as a function of small values of $\Pe$ and $\theta=0,
    \,\pi$, where $\sigma(\theta)$ is evaluated numerically by the method of 
    Sec.~\ref{sec:numerical_soln}.}
\end{figure}

We check numerically that $||F_n||$ decays exponentially in $n$ for a wide range of $\Pe$, $||F_n||\sim \rho^{-n}$
as $n\to\infty$ where $\rho=\rho(\Pe) >0$ is the ``decay rate'' of $||F_n||$, which is independent of $n$. 
For this purpose, we examine the ratio $||F_n||/||F_{n+1}||$
as a function of both $n$ and $\Pe$, expecting that this ratio approaches a constant for fixed $\Pe$ as $n$ becomes
sufficiently large, as shown in Figs.~\ref{fig:rho}(a), (b), (c). 
From Fig.~\ref{fig:rho}(d) 
we find that the relative error in the approximation of $\sigma$ by the 
sum $\sum_{k=1}^5\sigma_k$ becomes higher than 1\% only when $\Pe< 6.5\times 10^{-3}$.

\section{Uniformly Accurate Asymptotics For Low P\'eclet Numbers}
\label{sec:lowPe_asymptot}

In this section we solve approximately the integral equation
(\ref{eq:ie}) for the surface flux $\sigma(x)$, for all $x$ in
$(-1,1)$, when $\Pe$ is sufficiently small, $\Pe < O(1)$. For this
purpose, we define the dependent variable $\varphi(x)=e^{-\Pe x}\,\sigma(x)$.
%
%
%
Equation (\ref{eq:ie}) thus becomes 
\begin{equation}
\int_{-1}^{1}dx'\,K_0(\Pe|x-x'|)\,\varphi(x')=\pi e^{-\Pe x},\qquad -1<x<1.\label{int_eqn}
\end{equation}
Because the argument of the kernel is also sufficiently small, $\Pe
|x-x'|< O(1)$, we invoke the expansion
\begin{equation}
K_0(\Pe|x-x'|)\sim -I_0(\Pe|x-x'|)\,\ln\biggl(\frac{\Pe}{2}|x-x'|\biggr)+\sum_{m=0}^{M}\frac{\psi(m+1)}{(m!)^2}\,
\biggl(\frac{\Pe |x-x'|}{2}\biggr)^{2m},\label{K0_expansion}
\end{equation}
where it is understood that
\begin{equation}
I_0(\Pe|x-x'|)\sim\sum_{m=0}^{M}\frac{2^{-2m}\Pe^{2m} |x-x'|^{2m}}{(m!)^2},\label{I0_exp}
\end{equation} 
and $\psi(z)$ is the logarithmic derivative of the Gamma
function,$\,\psi(z)=\frac{d}{dz}\ln\Gamma(z)$.  It was first pointed
out by \cite{pearson57} (1957) that the resulting integral equation
can be solved exactly for any finite number of terms, $M$, but the
procedure becomes increasingly cumbersome with $M$.

To leading order we consider $M=0$ in (\ref{K0_expansion}) and
(\ref{I0_exp}).  Equation (\ref{int_eqn}) thus reduces to a variant of
Carleman's equation~(\cite{carleman22}, 1922),
\begin{equation}
\int_{-1}^{1}dx'\,\ln(|x-x'|)\,\varphi_0(x')=C_1-\pi e^{-\Pe x},\label{int_eqn2}
\end{equation}
where $\gamma=-\psi(1)=0.577215\cdots$ is the Euler
number,$\,\varphi_0(x)\sim\varphi(x)$ is the corresponding
approximation for $\varphi(x)$, and $C_1$ is the constant
\begin{equation}
  C_1=-[\gamma+\ln(\Pe/2)]\int_{-1}^1 dx'\,\varphi_0(x').\label{C1_def}
\end{equation}
Following~\cite{carrier83} (1983), we introduce the complex function
\begin{equation} 
  \label{Phi_def}
  \Phi(z)=\frac{\sqrt{z^2-1}}{2\pi i}\int_{-1}^1 dx'\ \frac{\varphi_0(x')}{x'-z}
\end{equation}
and single out the limit values
\begin{equation}
  \Phi^{\pm}(x)\equiv \lim_{\varepsilon\to 0} \Phi(x\pm i\varepsilon)
  = \pm\frac{\sqrt{1-x^2}}{2\pi} \lim_{\varepsilon\to 0} 
  \int_{-1\mp i\varepsilon }^{1\mp i\varepsilon}
  dz'\ \frac{\varphi_0(z')}{z'-x},
\end{equation}
by which the integral equation (\ref{int_eqn2}) is equivalent to the equations:
\begin{align} 
  \label{diffPhi+-_eqn}
  \begin{split}
    \Phi^+(x)-\Phi^{-}(x) &= - \frac{\sqrt{1-x^2}}{\pi} \frac{d}{dx} 
   \int_{-1}^{1}dx'\,\ln(|x-x'|)\,\varphi_0(x')
=-\Pe \sqrt{1-x^2}\;e^{-\Pe x},
  \end{split}\\
  \label{sumPhi+-_eqn}
  \Phi^+(x)+\Phi^{-}(x) &= i\sqrt{1-x^2}\; \text{Res}\left\{ 
  \frac{\varphi_0(z')}{z'-x}; z'=x \right\} = i\, \sqrt{1-x^2}\,\varphi_0(x).
\end{align}

First, we find $\Phi(z)$ via applying directly the Mittag-Leffler expansion theorem to
(\ref{diffPhi+-_eqn}) (\cite{carrier83}, 1983): 
\begin{equation}
\Phi(z)=-\frac{1}{2\pi i}\left(\Pe\int_{-1}^{1} dx'\ \frac{\sqrt{1-x'^2}}{x'-z}\,e^{-\Pe x'}+A\right),\label{Fz_eqn}
\end{equation}
where $-(2\pi i)^{-1}\,A$ is the limit as $z\to\infty$ of the function $\Phi(z)$; by inspection of (\ref{Phi_def}),
\begin{equation}
A=\int_{-1}^1 dx'\,\varphi_0(x').\label{A_def}
\end{equation}
We recognize that the constant $C_1$ of (\ref{C1_def}) is $C_1=-[\gamma+\ln(\Pe/2)]A$. 

Next, we obtain the approximate solution $\varphi_0(x)$ in terms of this $A$ by (\ref{diffPhi+-_eqn}):
\begin{equation}
\varphi_0(x)=\frac{1}{\pi\sqrt{1-x^2}}\Biggl[ \Pe\,(P)\!\int_{-1}^1 dx'\ \frac{\sqrt{1-x'^2}}{x'-x}\,e^{-\Pe x'}+A\Biggr],
\label{phi_eqn1}
\end{equation}
where $(P)\int_{-1}^1$ denotes the principal value of the integral.
In order to determine the unknown constant $A$, we multiply both sides of (\ref{int_eqn2}) by $(1-x^2)^{-1/2}$ and integrate
over $(-1,\,1)$ by use of the elementary integral (\cite{carrier83}, 1983)
\begin{equation}
\int_{-1}^{1} dx\ \frac{\ln(|x-x'|)}{\sqrt{1-x^2}}=-\pi\ln 2.\label{ln_integral}
\end{equation}
A few comments on this result are in order. It can be obtained via differentiating the left-hand side
with respect to $x'$, and thus converting the integral to a Cauchy principal value which is found directly to vanish
identically. Hence, the original integral is independent of $x'$ and can be evaluated for $x'=0$ 
by changing the variable to $x=(\xi-1/\xi)/(2i)$, where $\xi$ moves on the unit circle, and
applying the residue theorem (\cite{carrier83}, 1983). 
We thus find
\begin{equation}
A=-\frac{1}{\gamma+\ln(\Pe/4)}\int_{-1}^1 dx\ \frac{e^{-\Pe x}}{\sqrt{1-x^2}},\label{ka_eqn}
\end{equation}
 \begin{equation}
\varphi_0(x)=\frac{1}{\pi\sqrt{1-x^2}}
\Biggl[ \Pe\cdot\,(P)\!\int_{-1}^1 dx'\ \frac{\sqrt{1-x'^2}}{x'-x}\,e^{-\Pe x'}-
\frac{1}{\gamma+\ln(\Pe/4)}\int_{-1}^1 dx'\ \frac{e^{-\Pe x'}}{\sqrt{1-x'^2}}\Biggr].\label{phi_eqn2}
\end{equation}

It is straightforward to carry out the integrations in
(\ref{phi_eqn2}). The second integral on the right-hand side is simply
a modified Bessel function of the first kind:
\begin{equation}
\int_{-1}^1 dx\ \frac{e^{-\Pe x}}{\sqrt{1-x^2}}=\int_{0}^{\pi} dt\, e^{-\Pe\cos t}=\pi J_0(i\Pe)=\pi I_0(\Pe).
\label{integral1}
\end{equation}
The remaining integral can be converted to one that is directly amenable 
to numerical computation for $\Pe\le O(1)$. By defining
\begin{equation}
\mathcal I(\Pe;x)=(P)\!\int_{-1}^1 dx'\ \frac{\sqrt{1-x'^2}}{x'-x}\,e^{-\Pe (x'-x)},\label{calI_def}
\end{equation}
we evaluate the derivative
\begin{equation}
e^{-\Pe x}\p{\mathcal I}{\Pe}=-\int_{-1}^{1} dx'\,\sqrt{1-x'^2}\,e^{-\Pe x'}=
-\pi [I_0(\Pe)-I_0''(\Pe)]=-\pi\,\frac{I_1(\Pe)}{\Pe},\label{deriv_calI}
\end{equation}
where $I_{\nu}$ is the modified Bessel function of the first kind. An expression for the integral
(\ref{calI_def}) then follows by direct integration in $\Pe$ of (\ref{deriv_calI}):
\begin{equation}
\mathcal I(\Pe;x)=\mathcal I(0; x)-\pi\int_0^{\Pe} dt\ e^{t x}\,\frac{I_1(t)}{t},
\label{calI_eqn2}
\end{equation}
where
\begin{equation}
\mathcal I(0; x)= -(1-x^2)\,\frac{d}{dx}\int_{-1}^1 dx'\ \frac{\ln(|x-x'|)}{\sqrt{1-x'^2}}-\pi x=-\pi x.
\label{calI_eqn3}
\end{equation}
Hence,
\begin{equation}
\mathcal I(\Pe;x)=-\pi x-\pi\int_0^{\Pe} dt\ e^{t x}\,\frac{I_1(t)}{t}.
\label{integral2}
\end{equation}
The approximation
\begin{equation}
\gamma+\ln(\Pe/4)\sim - K_0(\Pe/2),\label{replace}
\end{equation}
which becomes useful in Sec.~\ref{sec:connection} where we construct a uniform formula for all $\Pe$ and local
coordinate of the boundary, and
the use of (\ref{integral1}), (\ref{integral2}) and $\sigma(x)=e^{\Pe x}\varphi(x)$ yield a low-$\Pe$ approximation for
the flux on the boundary of the finite strip, $\sigma^{(\rm lo)}=e^{\Pe x}\varphi_0(x)$,
\begin{equation}
\label{eq:sigma0-strip}
\sigma(x) \sim \sigma^{(\rm lo)}(x)=\frac{1}{\sqrt{1-x^2}}
\Biggl\{\frac{I_0(\Pe)}{K_0(\Pe/2)}\,e^{\Pe x}-\Pe\,\left[x+\int_{0}^{\Pe}\hspace{-2pt}dt\; e^{t x}\,\frac{I_1(t)}{t}\right]\Biggr\}.
\end{equation}
Hence, by virtue of (\ref{eq:sigma_thz}), the flux on the unit circle is
\begin{equation}\label{eq:sigma0-cir}
\sigma(\theta) \sim \sigma^{(\rm lo)}(\theta)=\frac{I_0(\Pe)}{K_0(\Pe/2)}\,e^{\Pe\cos\theta}-
\Pe\left[\cos\theta+\int_{0}^{\Pe}\hspace{-2pt}dt\; e^{t \cos\theta}\,\frac{I_1(t)}{t}\right].
\end{equation}

We note in passing that the integral in (\ref{eq:sigma0-strip}) can be
expressed as a powers series in $\Pe$.  With the series expansions
\begin{equation}
e^{t x}=\sum_{n=0}^{\infty}\frac{t^n x^n}{n!},\quad \frac{I_1(t)}{t}=\sum_{m=0}^{\infty}\frac{t^{2m}}{2^{2m+1}}\,
\frac{1}{m!\,(m+1)!},\label{exp_modbess}
\end{equation}
it is straightforward to derive the expansion
\begin{eqnarray}
\int_0^{\Pe} dt\ e^{x t}\,\frac{I_1(t)}{t}&=&
\sum_{l=0}^{\infty} \frac{\Pe^{2l+1}}{2l+1}\sum_{m=0}^l \frac{x^{2m}}{(2m)!\,(l-m)!\,(l-m+1)!}\nonumber\\
\mbox{} &&+\sum_{l=0}^{\infty} \frac{\Pe^{2(l+1)}}{2(l+1)}\sum_{m=0}^l \frac{x^{2m+1}}{(2m+1)!\,(l-m)!\,(l-m+1)!}.
\label{integral2exp}
\end{eqnarray}

\begin{figure}
  \centering
  \parbox[t]{0.48\linewidth}{(a)\\
    \includegraphics[width=\linewidth]{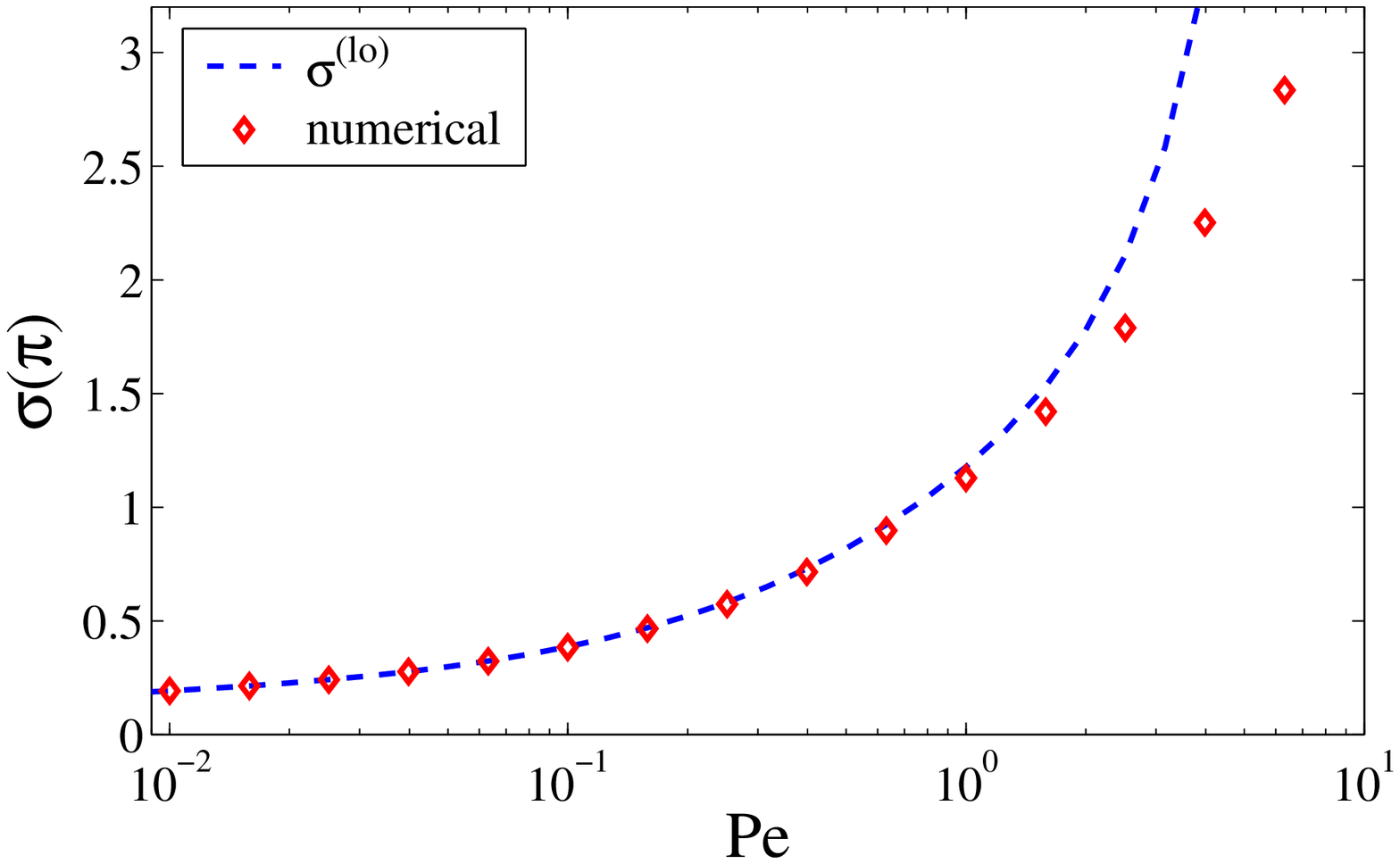}} \hspace{0.02\linewidth}
  \parbox[t]{0.48\linewidth}{(b)\\
    \includegraphics[width=\linewidth]{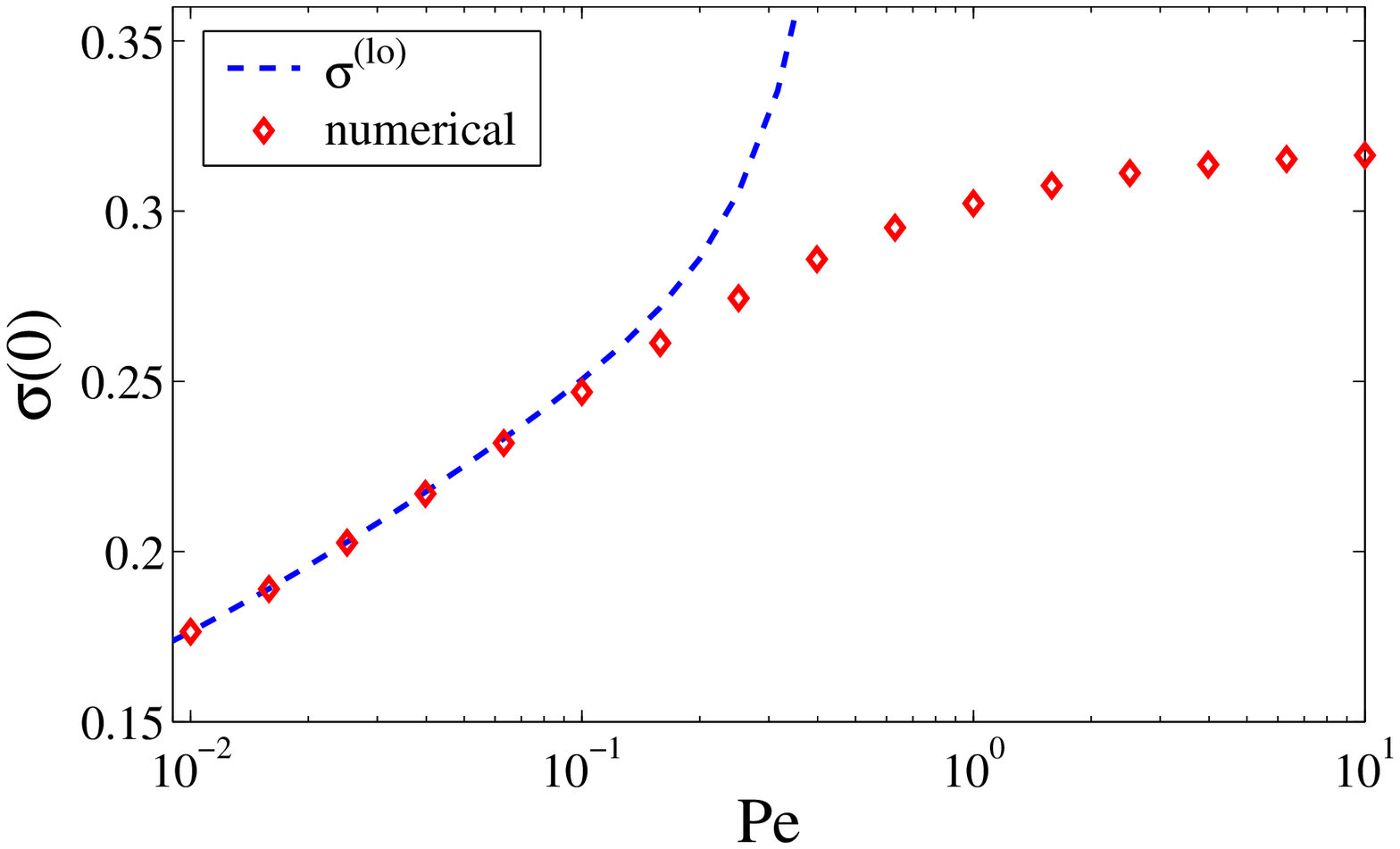}}
  \caption{\label{fig:CompLow} The asymptotic approximation 
    $\sigma^{(\rm lo)}(\theta;\Pe)$ in  (\ref{eq:sigma0-cir}) compared to 
    the numerical solution of Sec.~\ref{sec:numerical_soln} for a range of
    low $\Pe$. (a) Upstream flux, $\sigma^{(\rm lo)}(\theta=\pi;\Pe)$, 
    and (b) downstream flux, $\sigma^{(\rm lo)}(\theta=0;\Pe)$.}
\end{figure}

A few comments on formula (\ref{eq:sigma0-strip}) are in order.~\cite{aleksandrovbelokon67} (1967) expanded to high
orders the kernels of the relevant class of singular integral
equations and derived in generality a more accurate yet complicated
formula for the solution. The procedure here, though being based on
simply taking $M=0$ in (\ref{K0_expansion}) and (\ref{I0_exp}), leaves
intact the right-hand side of (\ref{int_eqn}) and applies
(\ref{replace}); our approximate formula for $\sigma(x)$ turns out to
be accurate for an extended range of low $\Pe$. Fig.~\ref{fig:CompLow} shows a
comparison of (\ref{eq:sigma0-cir}) with the numerical solution of
Sec.~\ref{sec:numerical_soln} for a range of low P\'eclet numbers.

\section{Uniformly Accurate Formula for All Positions and P\'eclet Numbers}
\label{sec:connection}

\subsection{Connecting the High and Low $\Pe$ Approximations for the flux }

\begin{figure}
  \centering 
  \parbox[t]{0.48\linewidth}{(a)\\
    \includegraphics[width=\linewidth]{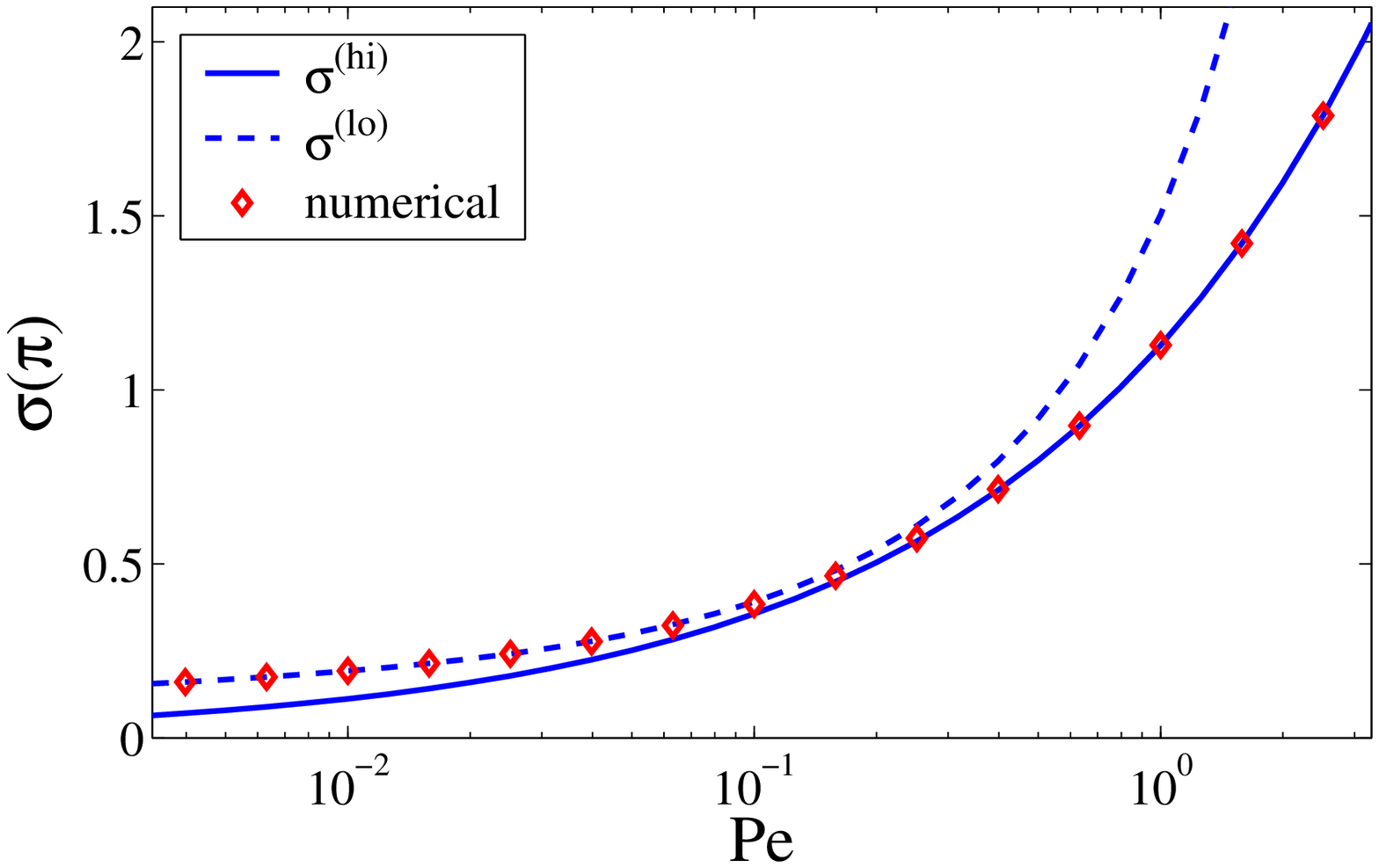}} \hspace{0.02\linewidth}
  \parbox[t]{0.48\linewidth}{(b)\\
    \includegraphics[width=\linewidth]{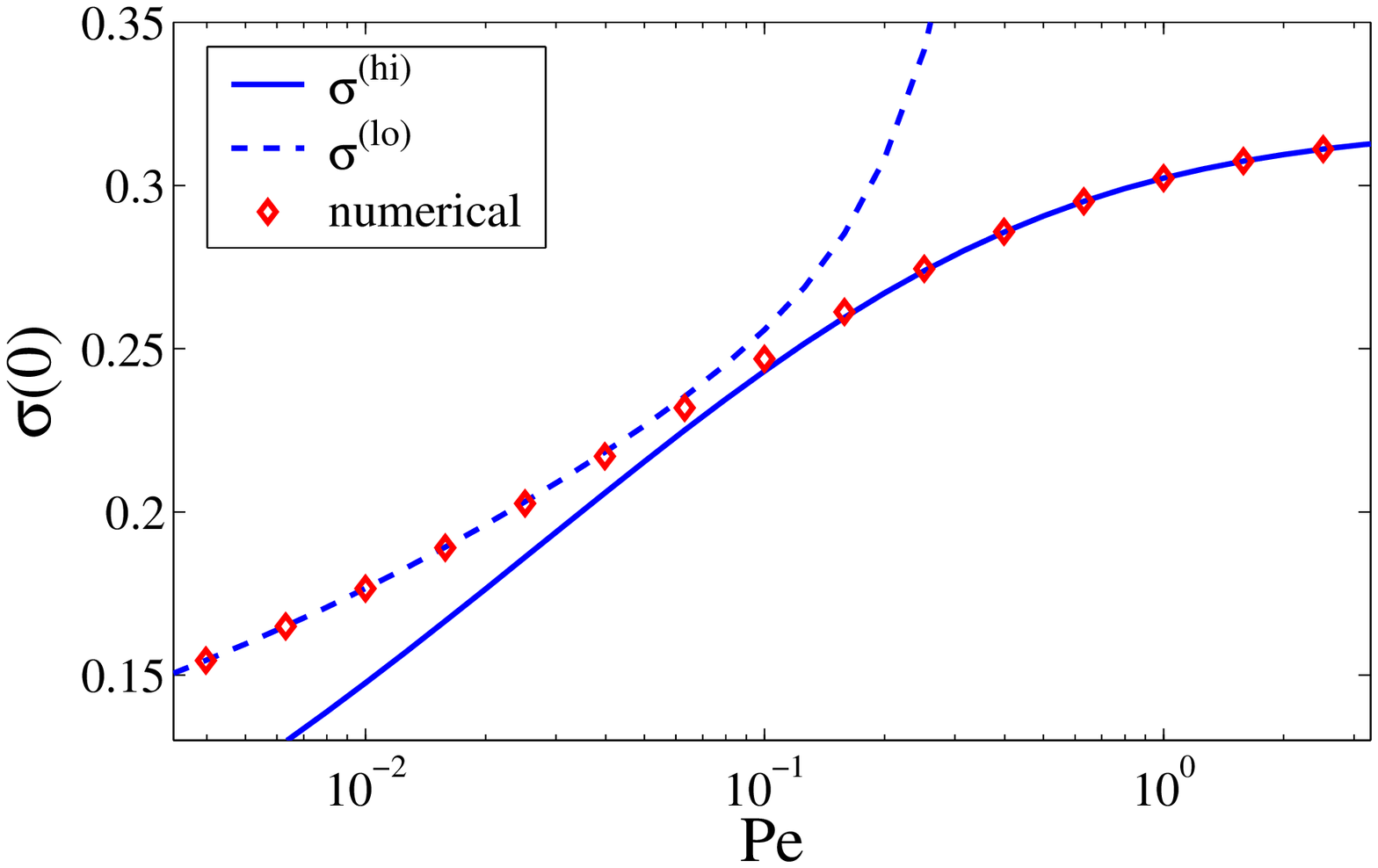}}
  \caption{\label{fig:CompAll} Plots of formulae (\ref{eq:sig_th_12}) for $\sigma^{(\rm hi)}$ and (\ref{eq:sigma0-cir}) for $\sigma^{(\rm lo)}$ versus $\Pe$ in the same graph as the plot for the solution $\sigma$ evaluated numerically by the method of Sec.~\ref{sec:numerical_soln}. (a) Upstream flux, $\sigma(\theta=0;\Pe)$. (b) Downstream flux, $\sigma(\theta=\pi; \Pe)$.}
\end{figure}

In Secs.~\ref{sec:highPe_asymptot} and \ref{sec:lowPe_asymptot} we
derived asymptotic formulae for the surface flux $\sigma$ on the
boundary of the unit circular disk (or finite strip) that are valid
for sufficiently high or sufficiently low $\Pe$; these expressions,
$\sigma^{(\rm hi)}$ and $\sigma^{(\rm lo)}$ in (\ref{eq:sig_th_12}) and
(\ref{eq:sigma0-cir}), respectively, hold for all values of the local
coordinate of the absorber boundary, although we did not analyze to
what extent the approximations are uniformly valid in a rigorous mathematical sense. Comparisons with
the numerical results in Figs.~\ref{fig:CompHigh} and
\ref{fig:CompLow} show that the approximations are comparably accurate
for the rear stagnation point ($\theta=0$) and forward stagnation
point ($\theta=\pi$).
We have also checked that, for fixed $\Pe$, the errors are also
comparable at intermediate values of the local coordinate ($0 < \theta< \pi$).
 In Fig.~\ref{fig:CompAll}, we show that the two approximations,
 $\sigma^{(\rm hi)}$ and $\sigma^{(\rm lo)}$, nearly overlap for some
 range of values $\Pe$ near the value $\Pe=0.1$, while remaining
 remarkably close to the ``exact'' numerical solution from
 Sec.~\ref{sec:numerical_soln}.

The existence of a regime of overlapping accuracy allows us to
construct an analytical formula for $\sigma$, accurate for all values
of $\Pe$ and the local coordinate of the absorbing boundary, $\theta$,
by smoothly connecting $\sigma^{(\rm hi)}$ and $\sigma^{(\rm lo)}$.
A similar overlapping accuracy was found for a three-dimensional problem of heat or mass transfer in a steady shear 
flow by~\cite{phillips90} (1990), 
who combined high- and low-$\Pe$ expansions only for the Nusselt number, $\Nu$, using singular perturbation. 
The dependence of the flux $\sigma$ on $\Pe$ can be described heuristically by a formula of the form
\begin{equation}
\sigma(\theta;\Pe)\sim \sigma^{(\rm conn)}=\sigma^{(\rm hi)}(\theta;\Pe) \,
\mathcal U(\Pe/\rm P_0)+ \sigma^{(\rm lo)}(\theta;\Pe)\,[1-\mathcal U(\Pe/\rm 
  P_0)],
\label{eq:conn-step}
\end{equation}
for the entire range of $\Pe$ and $\theta$; $\mathcal U(\chi)$ is a family of smooth functions
defined for $\chi=\Pe/\rm P_0 >0$ that at least satisfy the conditions
\begin{equation}
\lim_{\chi\to 0^+}\mathcal U(\chi)=0, \qquad \lim_{\chi\to +\infty}\mathcal U(\chi)=1.\label{U-limits}
\end{equation} 
A simple choice for the step-like function, $\mathcal U$, which yields
a rather accurate formula for $\sigma$, is
\begin{equation}
\label{eq:f-alpha}
\mathcal U(\chi;\alpha)=e^{1/(1-e^{\chi^{\alpha}})},\qquad\alpha>0.
\end{equation}
We note that there are two free parameters in (\ref{eq:f-alpha}) with
$\chi=\Pe/\rm P_0$: $\alpha$ and $\rm P_0$.  The parameter $\rm P_0$
corresponds to a value of $\Pe$ in the region of overlap of formulae
(\ref{eq:sig_th_12}) and (\ref{eq:sigma0-cir}); in principle, $\rm
P_0$ may depend on the local coordinate, which is $\theta$ for the
unit circle.  The parameter $\alpha$ determines the steepness of the
curve $\mathcal U(\chi)$ near $\chi=1$.  

\begin{figure}
  \centering
  \parbox[t]{0.48\linewidth}{(a)\\
    \includegraphics[width=\linewidth]{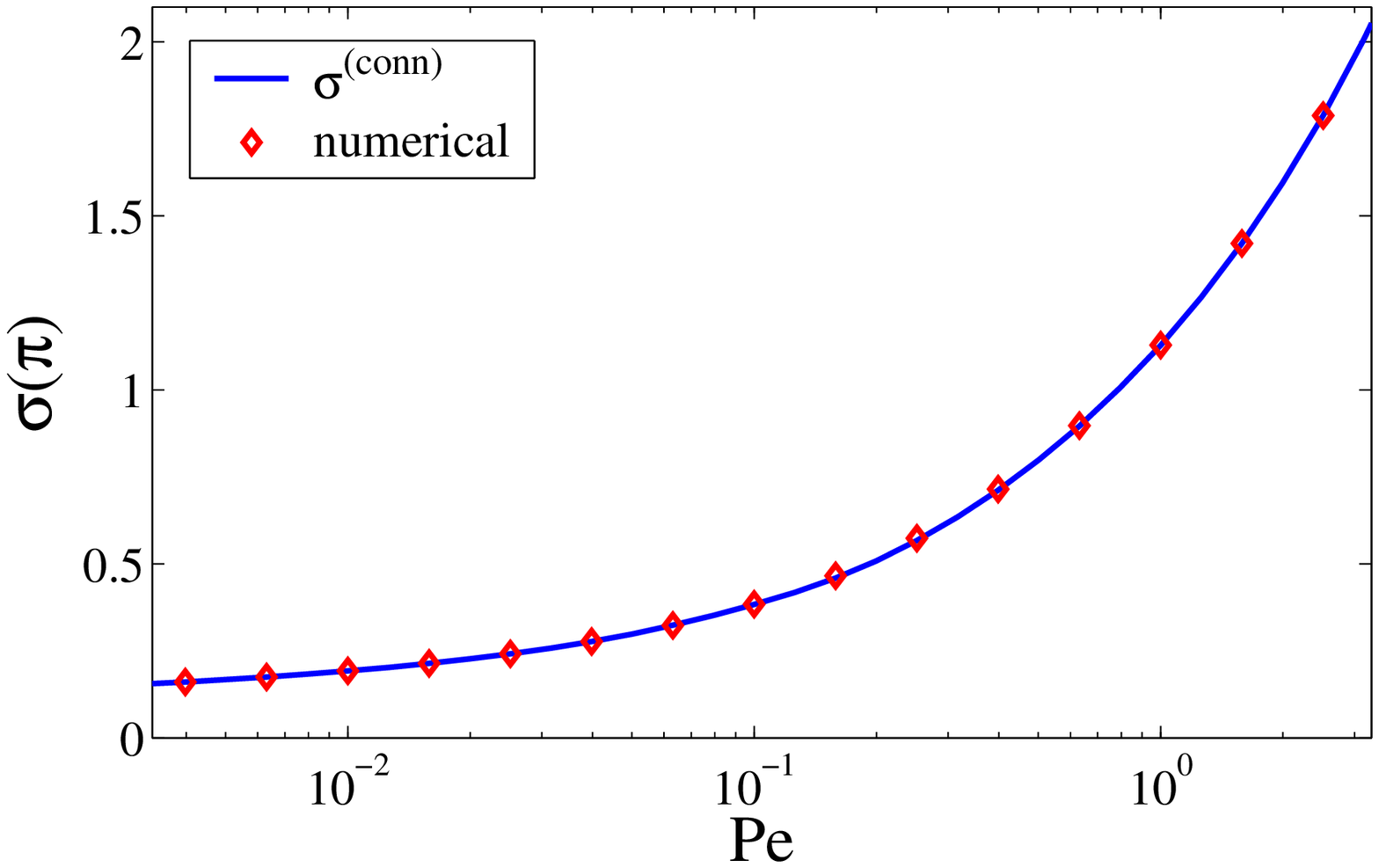}} \hspace{0.02\linewidth}
  \parbox[t]{0.48\linewidth}{(b)\\
    \includegraphics[width=\linewidth]{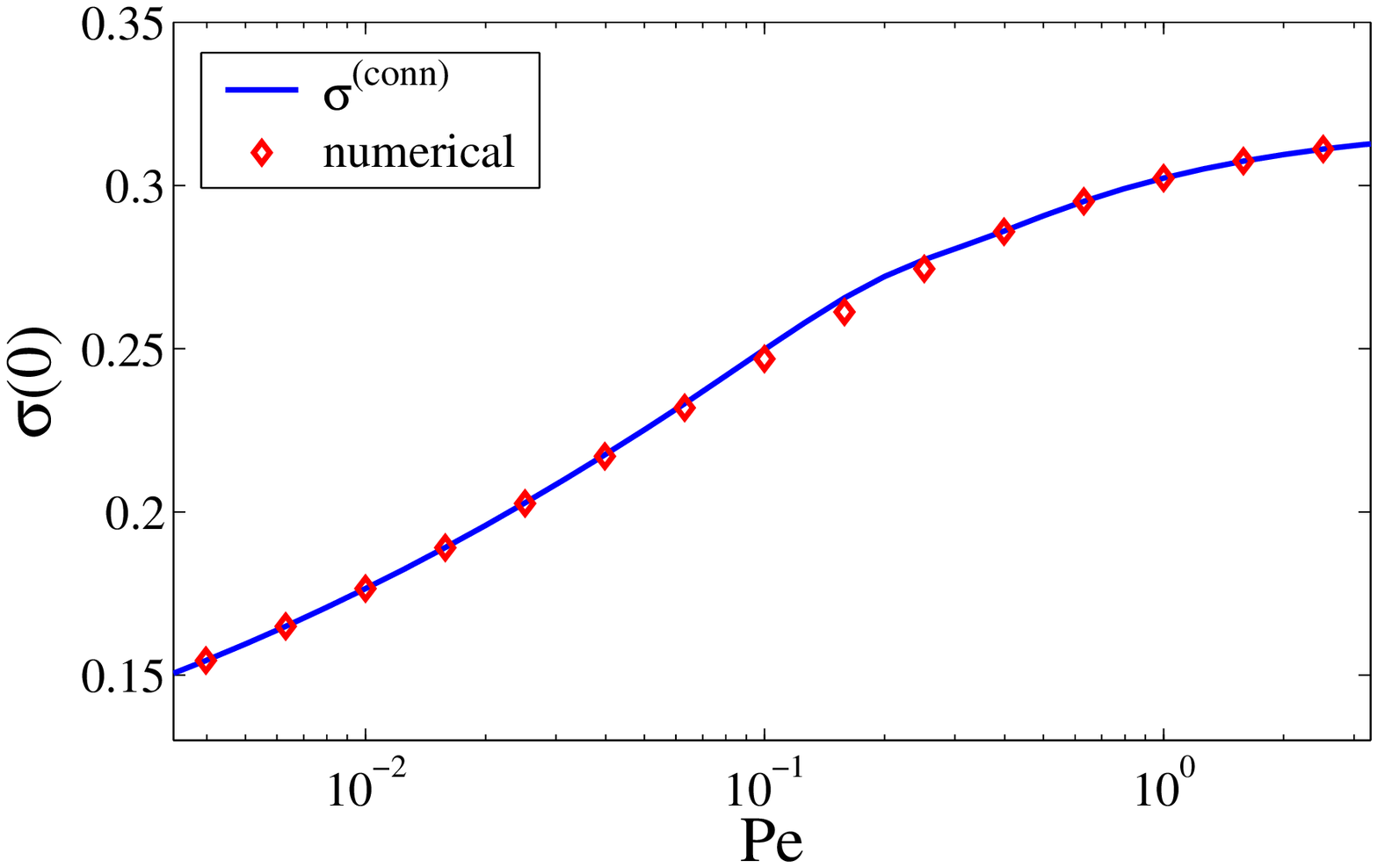}}
  \caption{\label{fig:CompConn}  Plots of the connection formula, $\sigma^{(\rm conn)}$, from
Eqs.~(\ref{eq:conn-step}) and (\ref{eq:f-alpha}), in comparison to the numerical solution
of Sec.~\ref{sec:numerical_soln}:
(a) Upstream flux, $\sigma(\theta=\pi;\Pe)$, and (b) downstream flux,
$\sigma(\theta=\pi; \Pe)$. The relative error of $\sigma^{(\rm conn)}$
compared to the numerical solution is less than $1.75\%$ for all
values of $\Pe$ and $\theta$, and it becomes 
negligibly small at high and low $\Pe$ for all $\theta$.}
\end{figure}

We find that a good fit with the numerical solution is achieved for
$\alpha=2$ and $\rm P_0=1/6$, as shown in Fig.~\ref{fig:CompConn}
where $\sigma^{(\rm conn)}(\theta)$ is plotted versus $\Pe$ for
$\theta=\pi$ (upstream flux) and $\theta=0$ (downstream flux). The
relative error is less than $1.75\%$ for all values of $\Pe$ and
$\theta$. Because we have the exact Green's function (\ref{eq:Green}),
this uniform accuracy also carries over to the solution of the entire
concentration field, $c(x,y;\Pe)$, obtained from the integral
(\ref{eq:c-G-sigma}). 

Our analytical approximation also describes an absorber of arbitrary
shape obtained by conformal mapping, $z=g(w)$ ($w=e^{i\theta}$).  The
flux distribution on the unit circle, $\sigma^{(\rm conn)}(\theta)$, is
transformed to the new surface using (\ref{eq:sigma_wz}). Because the
flux is proportional to a gradient, it is locally amplified by a
factor of $|g^\prime(w)|^{-1}$, which may cause relative errors larger
than 1.75\% in some locations, e.g. near a cusp, where conformality
breaks down ($g^\prime = 0$).  For a well behaved univalent mapping,
however, the approximation should remain very accurate for all
positions, $g(e^{i\theta})$ and $\Pe$, so the general BVP may be
considered effectively solved.

\subsection{ The Total Flux to the Absorber }
\begin{figure}
  \centering
  \parbox[t]{0.48\linewidth}{(a)\\
    \includegraphics[width=\linewidth]{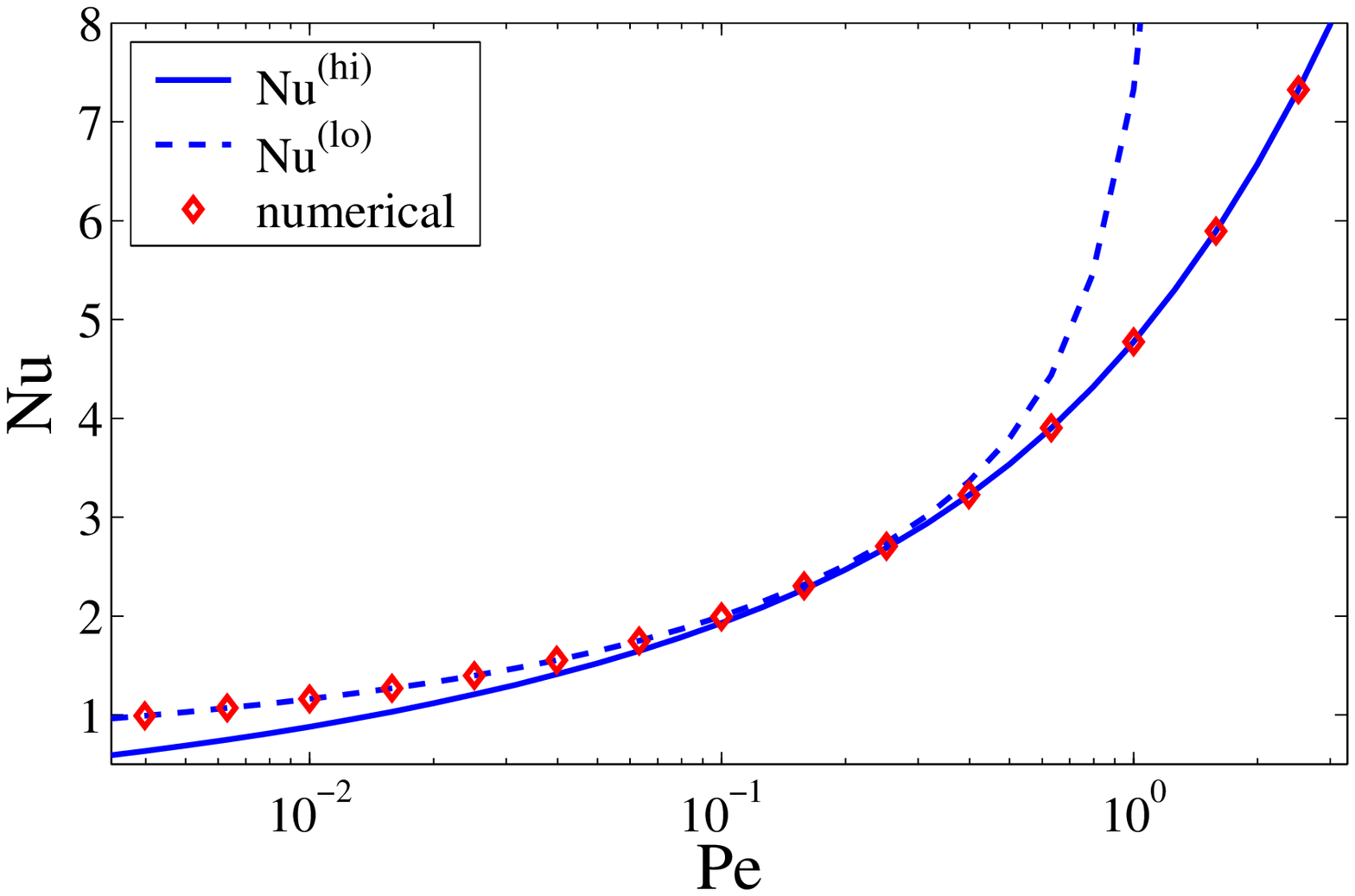}} \hspace{0.02\linewidth}
  \parbox[t]{0.48\linewidth}{(b)\\
    \includegraphics[width=\linewidth]{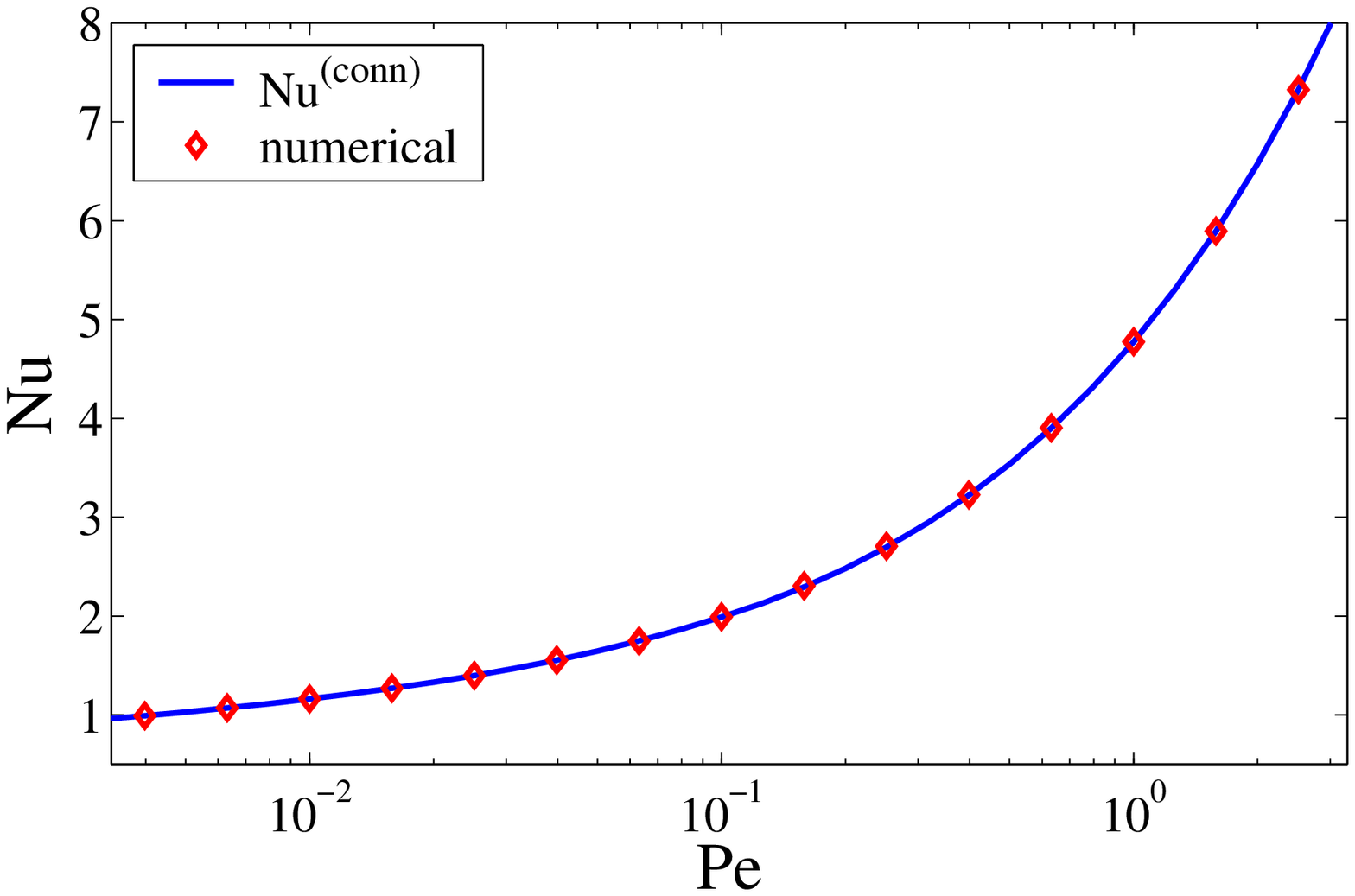}}
  \caption{\label{fig:Nu} The Nusselt number, $\Nu$, which gives the
    total flux to the absorber, versus the P\'eclet number, quantifying
    the importance of advection compared to diffusion. In (a)
    the asymptotic expressions (\ref{eq:NuAsym}) overlap for $0.1<\Pe<0.4$.
    In (b) the analytical connection formula (\ref{eq:NuUniform}) 
    compares very well with the
    ``exact'' numerical result from Sec.~\ref{sec:numerical_soln}.
    The results hold for absorbers of arbitrary shape, as long as $\Pe$ is
    the renormalized P\'eclet number, $\Pe = A_1\, \Peo$, where $A_1$ is
    the conformal radius and $\Peo$ is the bare P\'eclet number for the
    unit circle.}
\end{figure}

It is straightforward to obtain a uniformly accurate approximation of
the Nusselt number by integrating the flux on the unit circle or the 
finite strip:
\begin{equation}
  \label{eq:Nu}
  \Nu(\Pe) = \int_{0}^{2\pi}d\theta\, \sigma(\theta;\Pe) 
  = 2\int_{-1}^1 dx\, \sigma(x;\Pe).
\end{equation} 
The leading-order expressions of $\Nu(\Pe)$ in the high and low $\Pe$ limits are thus obtained by integrating 
(\ref{eq:sig_x_12}) and (\ref{eq:sigma0-strip}) using integration
by parts and the identity $xI_0(x)=[xI_1(x)]'$ (\cite{gradshteyn80}; 1980),
\begin{subequations}
  \label{eq:NuAsym}
  \begin{align}
    \Nu^{(\rm hi)}(\Pe) &= \frac{8}{\pi}\left\{ \sqrt{\frac{\Pe}{\pi}} 
    e^{-2\Pe} K_0(2\Pe) + \Pe e^{2\Pe}
    \erf(2\sqrt{\Pe})[K_0(2\Pe)+K_1(2\Pe)]\right\}, \\
    \Nu^{(\rm lo)}(\Pe) &= 2\pi\left\{ \frac{I_0(\Pe)^2}{K_0(\Pe/2)} 
    + (\Pe)^2[I_1(\Pe)^2-I_0(\Pe)^2] + \Pe I_0(\Pe)\,I_1(\Pe)  \right\}.
  \end{align}
\end{subequations}
The uniform analytical approximation for $\Nu(\Pe)$ follows as
\begin{equation}
  \label{eq:NuUniform}
  \Nu^{(\rm conn)}(\Pe) = \Nu^{(\rm hi)}(\Pe)\,e^{1/(1-e^{36\Pe[^2]})}
  + \Nu^{(\rm lo)}(\Pe)\,[1 - e^{1/(1-e^{36\Pe[^2]})}].
\end{equation}
As shown in Fig.~\ref{fig:Nu}, this analytical result is quite 
accurate over the entire range of P\'eclet numbers, and it becomes
exact as $\Pe\rightarrow 0$ and $\Pe\rightarrow \infty$. The maximum
relative error is found to be 0.53\%.
We are not aware of any such analytical formula for the
complete Nusselt-P\'eclet relation of a finite absorber.

It may come as a surprise that the same result (\ref{eq:NuAsym}) also
holds for an absorber of arbitrary shape, obtained by conformal
mapping of the unit circle, $z= g(w)$, without any additional error. The
reason is that the total flux through any contour is preserved {\it
exactly} under every conformal mapping of a conformally invariant BVP
(\cite{bazant04}, 2004).  When computing $\Nu$ from (\ref{eq:Nu}) for
another shape, therefore, one must simply use the renormalized
P\'eclet number, $\Pe = A_1 \Peo$, equal to the conformal radius,
$A_1$, times the bare P\'eclet number, $\Peo$, for the unit circle.


\section{Conclusion}
\label{sec:discussion}

\subsection{ Summary of Results }

We have performed a detailed study of the BVP
(\ref{eq:z_pde})--(\ref{eq:z_bc2}) for a finite absorber of arbitrary
cross section in a steady two-dimensional potential flow. Our focus has been the
flux profile on the boundary, $\sigma$, from which the concentration
can be obtain everywhere in the plane by convolution with the known
Green's function. We have explicitly considered several simple cases,
notably the canonical problem of uniform flow past a circular
cylinder, which can be mapped to arbitrary geometries by conformal
mapping, as described in Sec.~\ref{sec:background}.
 
In Sec.~\ref{sec:numerical_soln}, we presented an efficient numerical
method to solve the BVP. We 
began by mapping the BVP to the inside of the unit circle in order to
work with a bounded domain. We then eliminated some ``far-field''
singularities (at the origin) using exact asymptotics and applied a
spectral method in polar coordinates, with exponential convergence in
the number of nodes. We also used an adaptive mesh to deal with
boundary layers at very high P\'eclet numbers.  The results, which are
illustrated in Figs.~\ref{fig:profiles} and \ref{fig:tail}, provided
reliable tests of our analytical approximations.

In Secs.~\ref{sec:highPe_asymptot} and \ref{sec:lowPe_asymptot}, we
derived distinct asymptotic expansions for the flux on the boundary of
the finite strip for high and low $\Pe$, respectively, starting from a
well known integral equation in streamline coordinates
(\cite{wijngaarden66}, 1966), which has been studied extensively in
the theory of solidification and freezing in a flowing melt
(\cite{maksimov76}, 1976;  \cite{chugunov86}, 1986; \cite{kornev88}, 1988;
\cite{kornev94}, 1994; \cite{alimovetal94}, 1994; \cite{alimovetal98}, 1998;
\cite{cummingskornev99}, 1999; \cite{cummings99}, 1999). We used some
original variations on classical techniques from the theory of
singular integral equations, including the Wiener-Hopf method of 
factorization, to obtain improvements on previous approximations.

Noteworthy features of our expansions for high $\Pe$ are: (i) The
summands admit closed-form expressions in terms of multiple integrals
(\ref{eq:sig_th}), which are straightforward to evaluate
numerically; (ii) the expansions converge for all distances along the
strip for a wide range of $\Pe$, $\Pe\ge O(10^{-2})$; and (iii)
only the small number of terms in (\ref{eq:sig_th_12}) can be retained
for reasonable accuracy though the number increases as $\Pe$ decreases.
On the other hand, our asymptotic formula (\ref{eq:sigma0-cir}) for low
$\Pe$ is accurate for $\Pe\le 10^{-1}$, which renders it possible to
have a region where the two expansions overlap. 

Therefore, we were able to construct an {\it ad hoc} analytical
connection formula (\ref{eq:conn-step}), which reproduces our
numerical results for the flux to a circular absorber with less than
1.75\% relative error for all angles and P\'eclet numbers, as shown in
Fig.~\ref{fig:CompConn}. We also predicted the $\Nu(\Pe)$ relation
(\ref{eq:Nu}) with comparable accuracy, as shown in
Fig.~\ref{fig:Nu}. These results constitute a nearly exact, analytical
solution to the BVP~(\ref{eq:z_pde})--(\ref{eq:z_bc2}).

The main contribution of our work is a unified description of the
crossover regime, intepolating between the well known asymptotic
limits of high and low $\Pe$. As such, we can draw some mathematical
and physical conclusions about the transition in the following sections.

\subsection{ Mathematical Discussion  }

{\it A posteriori}, we may try to understand why perturbation
methods, which ostensibly require extreme values of $\Pe$, produce a
very accurate analytical solution for {\it all} values of $\Pe$.  One
technical reason is that our high-$\Pe$ approximation, $\sigma^{(\rm
hi)}$, is not the usual asymptotic series of singular perturbation,
since we have essentially ``summed'' parts of a naive expansion
exactly in the Bessel function terms of (\ref{eq:sig_th_12}). This
allows the approximation to be more accurate than a simple
power-series type of expansion, presumably extending its validity to
somewhat higher $\Pe$. Still, the neglected higher-order terms involve
powers of $\Pe$, which become important as $\Pe$ becomes small.  

As we have mentioned throughout the paper, similarity solutions govern
the asymptotic limits.  For $\Pe \rightarrow 0$, we are perturbing
around the trivial similarity solution with uniform diffusive flux
from infinity, $\sigma =$constant and $c(r,\theta)
\propto \ln r$. As shown in Fig~\ref{fig:tail}(a) for $\Pe=0.01$, a
small amount of fluid flow changes this picture only slightly near the
disk, by favoring flux to one side compared to the other.  When
$\Pe=1$, as in Fig.~\ref{fig:tail}(b), the region of depleted
concentration begins to be swept significantly downstream by the
flow. This causes the low-$\Pe$ approximation to break down near the
rear stagnation point, while remaining fairly accurate near the
forward stagnation point, as shown in Fig.~\ref{fig:CompLow}.

The high-$\Pe$ approximation is derived by perturbing around a
different similarity solution (\ref{eq:erfc}) for an absorbing
circular rim on an absorbing flat plate (\cite{bazant04}, 2004). In
this advection-dominated regime, there is a thin diffusion layer of
width, $O(1/\sqrt{\Pe})$, around the disk, as shown in
Fig~\ref{fig:tail}(c), which provides the first term in the
approximation (\ref{eq:sigma0-cir}). The asymptotic corrections in
Sec.~\ref{sec:highPe_asymptot} are obtained by effectively removing
the ``false plate'' from the similarity solution, downstream from the
disk.

To understand the influence of downstream perturbations, consider the
Green's function (\ref{eq:Green}), which has the asymptotic form
\begin{equation}
 G(x,y) \sim \frac{e^{\Pe(x-r)}}{\sqrt{8\pi \Pe r }}\ \ \
\mbox{as}\ \ r = \sqrt{x^2+y^2}\rightarrow \infty.  
\end{equation}
The Green's function decays exponentially at the scale of $\Pe$ in all
directions, except precisely downstream, where it is long-ranged:
$G(x,0) \sim 1/\sqrt{8\pi\Pe x}$ as $x\rightarrow \infty$ for $y=0$.
Therefore, all corrections to the leading-order similarity solution
(\ref{eq:erfc}) decay exponentially upstream beyond an $O(\Pe)$
distance from the rear stagnation point. Our
high-$\Pe$ approximation, $\sigma^{(\rm hi)}$, in (\ref{eq:sig_th_12})
captures the first such correction, which is needed for uniform
accuracy over the disk. Further corrections are $O(e^{-4\Pe})$, as is
clear from formulae (\ref{eq:sigma-norms}).

The fact that the approximation breaks down when $e^{4\Pe}$, rather
than $\Pe$, gets close to unity explains the fortuitous accuracy of
$\sigma^{(\rm hi)}$ down to $\Pe = 0.1$.  Higher-order terms further
extend the region of accuracy by orders of magnitude, e.g., to
$\Pe=O(10^{-3})$ for five terms, as shown in Fig.~\ref{fig:CompHigh}.
Because the approximation is valid for a wide range of $\Pe$ that
would not be considered ``high'', it overlaps with the low-$\Pe$
approximation. 

We also mention some directions for further analysis.  The analytical
treatment on the basis of an integral equation essentially avoids the
complications of boundary-layer theory applied directly to the BVP
(\cite{hinch}, 1991).
%
%
In that sense, it resembles renormalization-group (RG) methods for
PDEs (\cite{goldenfeld}, 1992; \cite{chen96}, 1996), which provide a
general means of deriving asymptotic expansions in place of
traditional problem-specific singular-perturbation methods. An
attractive feature of RG methods is that they promise to produce
globally valid approximations {\it directly}, without the need to
combine distinct overlapping expansions, as we have done. It would be
interesting to see if this could possibly be accomplished for our BVP, and, if
so, whether the resulting approximations are any simpler or more
accurate than ours. We leave this as an open challenge to RG
aficionados.

From the point of view of mathematical methods, another interesting
observation is that the BVP (\ref{eq:z_pde})--(\ref{eq:z_bc2}) can be
formulated in analogous way to the problem of wave scattering by a
finite strip~(\cite{myers65}, 1965) in acoustics or electromagnetics,
which has some variants with other geometries such as the scattering
by a broken corner~(\cite{myers84}, 1984).  Such problems, where the
requisite PDEs are or can be reduced to Helmholtz-type equations by
simple transformations, can be described alternatively by finite sets
of nonlinear ODEs in which the length of the strip, or the P\'eclet
number $\Pe$ in the present case, is the independent variable.  The
underlying method is an improvement over Latta's method
(\cite{latta56}, 1956) for the solution of a class of singular
integral equations via their exact conversion to ODEs. 

Another technical question is how to place the method of high-$\Pe$
expansion pursued in Sec.~\ref{sec:highPe_asymptot} on a more firm
mathematical ground. To address this issue, two of us have developed
an equivalent approach based on a generalization of the Wiener-Hopf
technique applied to the BVP~(\ref{eq:z_pde})--(\ref{eq:z_bc2}) in the
Fourier domain, which yields the same results as the iteration scheme
of Sec.~\ref{sec:highPe_asymptot} (\cite{diochoi04}, 2004).  Since the
Wiener-Hopf method is well known in this context for a semi-infinite
strip (\cite{carrier83}, 1983), it would be interesting to ask how
the method might be extended for the present case of a finite strip,
or perhaps more general situations, such as multiple strips, or
three-dimensional objects. 

A more careful comparison should be made of our numerical solutions
with the works by \cite{kornev94} (1994). Our method in
Sec.~\ref{sec:numerical_soln} seems to require more computational
resources to obtain the same level of accuracy because it solves a
two-dimensional BVP, whereas the other method solves a one-dimensional
integral equation.  Aspects of the technique in Sec.~\ref{subsec:high_high}
also need to
be studied further, especially the number of the required terms versus
$\Pe$, the accuracy for $\Pe\ll 1$ and the propagation of error
through the requisite multiple integrals.

\subsection{ Physical Discussion }

Our analysis provides quantitative information about the 
crossover between diffusion-dominated and advection-dominated
adsorption, which is important in applications.  As described
above, the former regime, at low $\Pe$, is characterized by a broad
diffusion layer extending in all directions like a ``cloud'', with the
flow causing only a minor broken symmetry, as shown in
Fig.~\ref{fig:tail}(a). In contrast, at high $\Pe$, concentration
gradients are confined to a thin boundary layer around the object,
which separates into a long thin ``wake'' near the rear stagnation
point, as shown in Fig.~\ref{fig:tail}(c).

What is the critical $\Pe$ for the cloud-to-wake transition? Of
course, the answer depends on one's definition, but the shape of the
object also plays a role.  In streamline coordinates, the transition
is not so obvious since the object corresponds to an infinitely thin
strip, which cannot ``shield'' a thinner, finite-sized wake.  Objects of
finite thickness, however, do show a clear transition.  A uniformly
valid formula for $c(x,y,\Pe)$ in any geometry may be obtained by
inserting our expression for the flux to the strip
(\ref{eq:conn-step}) into the convolution integral
(\ref{eq:c-G-sigma}) and conformally mapping from the desired shape to
the strip, $w=f(z)$.

\begin{figure}
  \centering \includegraphics[width=0.7\linewidth]{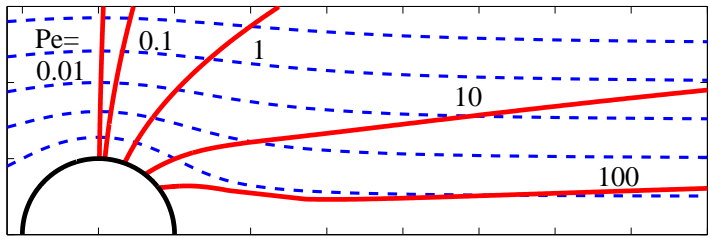} 
 \caption{\label{fig:maxc} The locus of points, for various values of
  $\Pe$, (solid lines) where the concentration $c$ attains its maximum
  along the streamlines (dashed lines) around a circular disk for
  desorption into an unconcentrated fluid. (In the equivalent problem
  of absorption from a concentrated fluid, the solid curves give the
  minimum concentration along streamlines for different $\Pe$. ) }
\end{figure}

The cloud-to-wake transition is related to $f(z)$, as we illustrate
for the case of the disk, $f(z)=z+1/z$. The analytical approximation
is nearly indistinguishable from the numerical solution in
Fig.~\ref{fig:tail}, so we consider the latter. In
Fig.~\ref{fig:maxc}, we show curves where, for different values of
$\Pe$, the concentration change, relative to the background, is
maximal along streamlines. Far from the disk, they approach parabolae,
$\Pe y^2 = 4x$, due to balance between diffusion in the $y$-direction
scaling as $y\sim 2\sqrt{t}$ and advection in the $x$-direction
scaling as $x\sim \Pe t$.  Near the object, the map causes a
significant distortion of the curves, which allows us to define the 
critical value, $\Pe = 60$, above which they become
non-monotonic functions of $x$. At larger values of $\Pe$, the curves
get sucked back into a thinner wake region behind the disk, signifying
the dominance of advection.

It is interesting to note the limiting wake structure as $\Pe
\rightarrow \infty$. As the concentration boundary layer wraps all the
way around the disk, the downstream disturbance begins to look like
that of a Green's function source located at the rear stagnation
point. If one defines the ``wake'' as the region behind the disk
enclosed by a given iso-concentration line, $c=c_0$, then it is easy
to see that the wake tends to a finite length as
$\Pe\rightarrow\infty$, even though its thickness tends to zero, like
$1/\sqrt{\Pe}$.  Physically, the (dimensionless) length $L = O(l_0^2
\Pe) = O(1)$ is the distance traveled in the flow downstream in $x$
during the time for diffusion across the initial wake thickness, $l_0
= O(1/\sqrt{\Pe})$.  For example, the $c=0.5$ contour ends roughly
$2.3$ disk diameters downstream from the rear stagnation point, as
$\Pe \rightarrow \infty$.

These results have relevance for more complicated, physical
situations, such as the coating of fibers from a gas flow, where
quasi-steady advection-diffusion in a two-dimensional potential flow
is a reasonable model for the growth dynamics (K. G. Kornev, private
communication). It is well known that the
P\'eclet number, based on the fiber diameter, controls the total
growth rate, as well as the uniformity of the coating's thickness, for
a {\it single} fiber.  For $\Pe \leq 0.1$, the flux to the disk (or
fiber cross-section), $\sigma(\theta,\Pe)$, is nearly uniform, apart
from a minor asymmetry due to the flow, which is accurately described
by (\ref{eq:sigma0-cir}). For $\Pe \geq 0.01$, the high-$\Pe$
expansion (\ref{eq:sig_th_12}) shows how the flux profile becomes
increasingly asymmetric and approaches the advection-dominated limit,
$\sigma \sim 2\sqrt{\Pe/\pi} \, \sin \theta/2$ as $\Pe \rightarrow
\infty$.  Our work provides an analytical description of how this
crossover occurs for the flux profile (\ref{eq:conn-step}), and
the total flux  (\ref{eq:Nu}), which may be useful in future
analytical or numerical studies of the coating process.

By quantifying the crossover in the concentration field, we also
provide some insight into the possible effect of interactions between
{\it multiple} nearby fibers during real coating process. In the
``cloud'' regime at low $\Pe$, the concentration field approximately
satisfies Laplace's equation, which means that other fibers in all
directions can strongly influence the local flux, due to the
long-range decay of the concentration. In the `wake' regime at high
$\Pe$, the concentration remains uniform (outside a thin boundary
layer) in all directions except directly downstream, where a thin wake
forms of $O(1/\sqrt{\Pe})$ thickness and $O(1)$ extent (at the scale
of a few particle diameters). When the mean fiber spacing is larger
than the typical wake length, there are negligible interactions, but,
whenever a fiber ends up in the wake of another fiber, its coating
becomes much thinner in a localized region, which can be
undesirable. Our analysis shows that this important crossover occurs
at a critical value of $\Pe \approx 60$.

Another physical application of our results is to
advection-diffusion-limited aggregation (\cite{bazant03}, 2003), the
stochastic analog of the fiber coating problem, which involves
discrete, diffusing particles collecting on a seed in a potential
flow. The ADLA model may have relevance for the dendritic growth of
mineral oxide deposits in very thin cracks in rocks, where Hele-Shaw
flow, laden with oxygen diffusing from the surface, is believed to
occur over geological time scales (G. R. Rossman, private
communication).  To describe ADLA growth from a finite seed in a
uniform background flow, the probability measure for adding particles
on a growing aggregate in the $z$ plane is given by the flux to the
unit disk for the same problem in the $w$ plane,
\begin{equation}
p(z,t)|dz| = p(\theta,t)d\theta \propto \sigma_w(\theta;\Pe(t)), 
\end{equation}
where the renormalized P\'eclet number, $\Pe(t) = A_1(t) \Peo$, is
given by the time-dependent conformal radius. Our accurate numerical
solution for $\sigma_w(\theta,\Pe)$ has already been used as the
growth measure in the simulations of \cite{bazant03} (2003). Future
simulations or mathematical analysis could start from our nearly exact
formula, $\sigma^{(\rm conn)}$ from (\ref{eq:conn-step}), which
captures the full crossover between diffusion-dominated and
advection-dominated dynamical fixed points with increasing
$\Pe(t)$. 

\section*{Acknowledgments }
TMS and MZB would like to thank \'Ecole Sup\'erieure de Physique et
Chimie Industrielles (Laboratoire de Physico-chimie Th\'eorique) for
hospitality and partial support.  The authors also wish to thank
Konstantin Kornev, John Myers, Howard Stone, Jacob White, and Tai Tsun
Wu for references and helpful discussions.

\appendix

\section{ Invertibility of the Kernel }
\label{app:invertibility}

In this appendix we show that the symmetrized kernel,
$K_0(\Pe|x-x'|)$, of the integral equation (\ref{eq:ie}) is positive
definite. Using the relation
\begin{equation}
\label{eq:B1}
K_0(x) = \frac{1}{2}\int_{-\infty}^\infty dt\,\frac{\cos(x t)}{\sqrt{1+t^2}},
\end{equation}
the kernel is recast to the form
\begin{equation}
\label{eq:B2}
\int_{-\infty}^{\infty}dt\ \frac{\cos(\varsigma t)\cos(\varsigma_0t)+\sin(\varsigma t)\sin(\varsigma_0t)}{\sqrt{1+t^2}}.
\end{equation}
It follows that, for $\beta>0$,
\begin{equation}
\label{eq:B3}
\int_{-\beta}^\beta d\varsigma_0\!\! \int_{-\beta}^\beta d\varsigma\  
\int_{-\infty}^{\infty}dt\ \frac{\cos(\varsigma t)\cos(\varsigma_0 t)+\sin(\varsigma t)\sin(\varsigma_0 t)}
{\sqrt{1+t^2}} f(\varsigma_0)f(\varsigma)  = \int_{-\infty}^{\infty}dt\ \frac{g(t)^2+h(t)^2}{\sqrt{1+t^2}},
\end{equation}
by which the kernel is positive definite. The functions $g(t)$ and $h(t)$ are real, and given by
\begin{equation}
\label{eq:B4}
g(t)+i h(t) = \int_{-\beta}^\beta f(u)e^{i u t}du.
\end{equation}



\end{document}